\documentclass[nohyper]{JHEP3}

\usepackage{amsmath}
\usepackage{amssymb}
\usepackage{graphicx}
\usepackage{latexsym,epsfig,cite}
\usepackage{ulem}
\usepackage{lineno}

\def\beq{\begin{equation}}
\def\eeq{\end{equation}}
\def\bea{\begin{eqnarray}}
\def\eea{\end{eqnarray}}
\def\barr{\begin{array}}
\def\earr{\end{array}}

\def\lsim{\mathrel{\rlap{\raise 2.5pt \hbox{$<$}}\lower 2.5pt\hbox{$\sim$}}}
\def\gsim{\mathrel{\rlap{\raise 2.5pt \hbox{$>$}}\lower 2.5pt\hbox{$\sim$}}}

\newcommand{\half}{{\textstyle\frac{1}{2}}}

\renewcommand{\Re}{{\rm Re\thinspace}}
\renewcommand{\Im}{{\rm Im\thinspace}}

\newcommand{\hc}{\hbox {h.c.}}

\allowdisplaybreaks 

\title{Exploring the CP-Violating Inert-Doublet Model}
\author{B. Grzadkowski,\\
Institute of Theoretical Physics, Faculty of Physics, University of Warsaw, \\ Ho\.za 69, 
PL-00-681 Warsaw, Poland\\
E-mail: \email{Bohdan.Grzadkowski@fuw.edu.pl}}
\author{O. M. Ogreid,\\
Bergen University College, Bergen, Norway \\
E-mail: \email{omo@hib.no}}
\author{P. Osland\\
Department of Physics,
University of Bergen, Postboks 7803, N-5020 Bergen, Norway\\
E-mail: \email{Per.Osland@ift.uib.no}}
\author{A. Pukhov\\
Skobeltsyn Inst.\ of Nuclear Physics, Moscow State Univ., Moscow 119991, Russia\\
E-mail: \email{pukhov@lapp.in2p3.fr}}
\author{M. Purmohammadi\\
Department of Physics,
University of Bergen, Postboks 7803, N-5020 Bergen, Norway\\
E-mail: \email{Mahdi.PurMohammadi@ift.uib.no}}

\abstract{
We have explored properties of an extension of the Inert Doublet Model by the addition of an extra non-inert scalar doublet. The model offers a possibility of CP violation in the scalar sector and a candidate for the Dark Matter. Allowed regions in the plane spanned by the mass of the Dark-Matter particle and the lightest neutral Higgs particle have been identified, and constraints from direct-detection experiments have been studied. For favorable parameter regions one may observe long-lived charged particles produced at the LHC.}


\preprint{}

\begin{document}

\section{Introduction}
The Inert Doublet Model (IDM) was introduced to accommodate or explain
neutrino masses \cite{Deshpande:1977rw} and independently, to alleviate the little
hierarchy problem while also providing a dark matter (DM) candidate
\cite{Barbieri:2006dq}.  The model represents a very minimal extension of
the Standard Model (SM), it just contains an extra weak scalar
doublet, which is odd under an unbroken $Z_2$ symmetry, rendering the
lightest member stable. The other members of this doublet are another neutral particle
and a pair of charged ones.  These particles can all be produced at
colliders via their couplings to electroweak gauge bosons, subject to
the constraint of the $Z_2$ symmetry.  The collider phenomenology has
been explored in \cite{Cao:2007rm,Lundstrom:2008ai} and the Early
Universe phenomenology has been studied in some detail in
\cite{LopezHonorez:2006gr} and \cite{Hambye:2009pw}.

While the IDM has many attractive features, simplicity being an
important one, it was felt that the introduction of CP violation in
the scalar sector would make the model more attractive, therefore an
extension to a Two-Higgs-Doublet Model (2HDM) plus an inert doublet
model was proposed \cite{Grzadkowski:2009bt}. This also allows for an
alleviation of the little hierarchy problem. We shall refer to the
resulting model as IDM2.

It has been found that the IDM permits a DM particle with a mass in
one of three regions: light ($m\ll m_W$)
\cite{Andreas:2008xy,Hambye:2007vf}, medium ($m\sim m_W$)
\cite{Barbieri:2006dq,LopezHonorez:2006gr} or heavy
($m\gsim535~\text{GeV}$)
\cite{Cirelli:2005uq,LopezHonorez:2006gr,Hambye:2009pw}. Two of these
mass regions ($m\sim m_W$ and $m\gsim 500~\text{GeV}$) were also found
to yield solutions for the IDM2 \cite{Grzadkowski:2009bt}. The aim of
the present paper is to explore the IDM2 in more detail, determine
the allowed mass regions for the DM particle, its dominant annihilation
channels, and the corresponding mass regions for the lightest Higgs boson
$H_1$. Furthermore, we will confront the model with constraints from direct-detection experiments, and briefly comment on possible signals in LHC experiments.

The paper is organized as follows. In Sec.~\ref{Sec:def-model} we review the model, and in Sec.~\ref{sect:constraints} we discuss the theoretical and experimental constraints. In Sec.~\ref{sect:annihilation} we consider various annihilation channels and in Sec.~\ref{Sec:-scan-strategy} we present the scan strategy adopted to search for allowed regions in the parameter space.  In Sec.~\ref{sec:overview} we give an overview of allowed regions of DM particle masses, whereas in Secs.~\ref{sec:med-mass} and \ref{sec:high-mass} we explore in more detail parameters that are compatible with all the constraints in the low and high DM mass regions, respectively. Then, in Sec.~\ref{Sec:direct-detection} we discuss constraints from direct detection experiments, in Sec.~\ref{Sec:LHC} we briefly explore relevant LHC phenomenology, and in Sec.~\ref{Sec:summary} we summarize. The relevant couplings of the model are given in an appendix.

\section{IDM2---model and notation}
\label{Sec:def-model}
\setcounter{equation}{0}
\subsection{Fields and potential}
We denote the doublets of the 2HDM as
\begin{equation}
\Phi_1=\left(
\begin{array}{c}\varphi_1^+\\ (v_1+\eta_1+i\chi_1)/\sqrt{2}
\end{array}\right), \quad
\Phi_2=\left(
\begin{array}{c}
\varphi_2^+\\ (v_2+\eta_2+i\chi_2)/\sqrt{2}
\end{array}
\right),
\label{Eq:basis}
\end{equation}
where $v^2=v_1^2+v_2^2$ and $\tan\beta=v_2/v_1$.
The inert doublet is decomposed as
\begin{equation}
\eta = \left(
\begin{array}{c}
 \eta^+ \\ (S+iA)/\sqrt{2}
\end{array}
\right),
\label{eta}
\end{equation}
it transforms under an unbroken $Z_2$ symmetry as $\eta
\to -\eta$ which ensures that $\eta$ couples only bilinearly to other
scalars and to the gauge sector. All other fields remain neutral under
this transformation.

The potential reads
\begin{equation} \label{Eq:fullpot}
V(\Phi_1,\Phi_2,\eta)
= V_{12}(\Phi_1,\Phi_2) + V_3(\eta) + V_{123}(\Phi_1,\Phi_2,\eta)
\end{equation}
where the 2HDM and inert-sector potentials read
\begin{align}
V_{12}(\Phi_1,\Phi_2) &= -\frac12\left\{m_{11}^2\Phi_1^\dagger\Phi_1
+ m_{22}^2\Phi_2^\dagger\Phi_2 + \left[m_{12}^2 \Phi_1^\dagger \Phi_2
+ \hc\right]\right\} \nonumber \\
& + \frac{\lambda_1}{2}(\Phi_1^\dagger\Phi_1)^2
+ \frac{\lambda_2}{2}(\Phi_2^\dagger\Phi_2)^2
+ \lambda_3(\Phi_1^\dagger\Phi_1)(\Phi_2^\dagger\Phi_2) \nonumber \\
&+ \lambda_4(\Phi_1^\dagger\Phi_2)(\Phi_2^\dagger\Phi_1)
+ \frac12\left[\lambda_5(\Phi_1^\dagger\Phi_2)^2 + \hc\right],
\label{v12} \\
V_3(\eta) &= m_\eta^2\eta^\dagger \eta + \frac{\lambda_\eta}{2}
(\eta^\dagger \eta)^2,
\label{v3}
\end{align}
whereas their mutual couplings, bilinear in the $Z_2$-odd field $\eta$, are given by
\begin{align}
V_{123}(\Phi_1,\Phi_2,\eta)
&=
\lambda_{1133} (\Phi_1^\dagger\Phi_1)(\eta^\dagger \eta)
+\lambda_{2233} (\Phi_2^\dagger\Phi_2)(\eta^\dagger \eta) \nonumber  \\
& +\lambda_{1331}(\Phi_1^\dagger\eta)(\eta^\dagger\Phi_1)
+\lambda_{2332}(\Phi_2^\dagger\eta)(\eta^\dagger\Phi_2) \nonumber  \\
&
+\half\left[\lambda_{1313}(\Phi_1^\dagger\eta)^2 +\hc \right]
+\half\left[\lambda_{2323}(\Phi_2^\dagger\eta)^2 +\hc \right].
\label{v123}
\end{align}
Here, $\lambda_{1133}$, $\lambda_{2233}$, $\lambda_{1331}$ and
$\lambda_{2332}$ are real, whereas $\lambda_{1313}$ and
$\lambda_{2323}$ can be complex.  It has been assumed that the total Lagrangian is symmetric under $Z'_2$ which acts as $\Phi_1\to -\Phi_1$ and $u_R\to -u_R$ (all other fields are neutral) and therefore the FCNC are eliminated at  the 
tree level. In order to allow for CP-violation in the scalar sector we allow for soft $Z'_2$ breaking by the mass $m_{12}^2 \Phi_1^\dagger \Phi_2
+ \hc$ (so that the renormalizability is preserved).
Note that, as a consequence of the
unbroken $Z_2$ associated with the inert doublet, there is no mixing in mass terms between
$\Phi_{1,2}$ and $\eta$\cite{Grzadkowski:2009bt}.

The model we discuss here bears some similarity to the three-Higgs-doublet Weinberg model \cite{Weinberg:1976hu,Branco:1980sz}. The Weinberg model was constructed as the minimal model that accommodates CP violation in the scalar sector together with natural flavour conservation. In fact, the symmetries of the Weinberg model are exactly the same as those we impose here: $Z_2\times Z'_2$, where $Z_2$ is responsible for flavour conserving Yukawa couplings of $\Phi_{1}$ and $\Phi_{2}$ while $Z'_2$ decouples $\eta$ from fermions. One difference between the Weinberg model and the present one is that we allow for soft $Z_2$ violation by the mass term
$m_{12}^2 \Phi_1^\dagger \Phi_2 + \hc$ The second is that, in the Weinberg model, all three doublets develop non-zero vacuum expectation values, while in our case $\langle\eta\rangle=0$. 
That implies that there is no mixing in the mass matrix between $\Phi_{1,2}$ and $\eta$;
in particular, $\eta^\pm$ decouples from $G^\pm$ and $H^\pm$, so that there is no
CP-violation mediated by charged scalars. However, since $m_{12}^2\neq 0$, CP
is violated in the neutral non-inert scalar sector in the same way as in the 2HDM.

In the absence of the potential (\ref{v123}), one would have to require $m_\eta^2>0$ in order to ensure $\langle\eta\rangle=0$. However, the non-zero expectation values of the other fields, $v_1/\sqrt2$ and $v_2/\sqrt2$, lead to an overall coefficient of the term that is bilinear in $\eta$ of the form
\begin{equation}
\label{Eq:eta-no-vev}
m_\eta^2+(\lambda_{1133}+\lambda_{1331}+\Re\lambda_{1313})\frac{v_1^2}{2}
+(\lambda_{2233}+\lambda_{2332}+\Re\lambda_{2323})\frac{v_2^2}{2}.
\end{equation}
This must be positive, but $m_\eta^2$ is not necessarily positive.
\subsection{Mass eigenstates of the IDM2}
The neutral states of $\Phi_1,\Phi_2$ will in general mix to form three neutral states $H_1, H_2, H_3$.
These are linear combinations of $\eta_1$, $\eta_2$, and $\eta_3$,
\begin{equation}
\begin{pmatrix}
H_1 \\ H_2 \\ H_3
\end{pmatrix}
=R
\begin{pmatrix}
\eta_1 \\ \eta_2 \\ \eta_3
\end{pmatrix},
\end{equation}
where $\eta_3\equiv-\sin\beta\chi_1+\cos\beta\chi_2$ is orthogonal to the
neutral Goldstone boson $G^0=\cos\beta\chi_1+\sin\beta\chi_2$ and the
rotation matrix $R$ is parametrized in terms of three angles
$\alpha_1$, $\alpha_2$ and $\alpha_3$ according to the convention of
\cite{Accomando:2006ga}.

For the quartic couplings describing the interaction between $\eta$ and
$\Phi_1$ and $\Phi_2$, we adopt for simplicity the ``dark democracy'':
\begin{align} \label{Eq:DarkDemocracy}
\lambda_a\equiv \lambda_{1133}&=\lambda_{2233}, \nonumber \\
\lambda_b\equiv \lambda_{1331}&=\lambda_{2332}, \nonumber \\
\lambda_c\equiv \lambda_{1313}&=\lambda_{2323} \text{ (real)},
\end{align}
then $V_{123}$ is invariant under $\Phi_1 \leftrightarrow \Phi_2$.
The dark-sector masses can be written as:
\begin{align}
M^2_{\eta^\pm}
&=m_\eta^2+\half\lambda_a\,v^2, \nonumber \\
M^2_S
&=m_\eta^2+\half(\lambda_a+\lambda_b+\lambda_c)v^2
=M^2_{\eta^\pm}+\half(\lambda_b+\lambda_c)v^2, \nonumber \\
M^2_A
&=m_\eta^2+\half(\lambda_a+\lambda_b-\lambda_c)v^2
=M^2_{\eta^\pm}+\half(\lambda_b-\lambda_c)v^2,
\label{inmass}
\end{align}
where $m_\eta$ is a mass parameter of the $\eta$ potential (\ref{v3}).
We shall take the scalar, $S$, to be the DM
particle, i.e., $M_S<M_A$.  The other choice would simply correspond
to $\lambda_c\to-\lambda_c$, without any modification of the phenomenology described in the following.

It is instructive to invert the relations (\ref{inmass}):
\begin{subequations} 
\label{Eq:lambda-vs-splitting}
\begin{align}
\lambda_a&=\frac{2}{v^2}
\left(M^2_{\eta^\pm}-m_\eta^2\right), \\
\lambda_b&=\frac{1}{v^2}\left(M^2_S+M^2_A-2M^2_{\eta^\pm}\right), \\
\lambda_c&=\frac{1}{v^2}\left(M^2_S-M^2_A\right).
\end{align}
\end{subequations}
Thus, these couplings of the inert doublet to the non-inert Higgs sector can be expressed in terms of the mass splittings (including also the soft mass parameter $m_\eta$).

It is convenient to introduce the abbreviation
\begin{equation} \label{Eq:lambda_L}
\lambda_L\equiv \half(\lambda_a+\lambda_b+\lambda_c)
=\frac{M_S^2-m_\eta^2}{v^2},
\end{equation}
From Eq.~(\ref{Eq:eta-no-vev}), the condition $\langle\eta\rangle=0$ can now be written as
\begin{equation}
m_\eta^2+\lambda_L v^2=M_S^2>0,
\end{equation}
which is automatically satisfied by our choice of input parameters.
\section{Theoretical and experimental constraints}
\label{sect:constraints}
\setcounter{equation}{0}

We here present a summary of the constraints imposed on the model. Some of the theoretical ones (positivity, in particular) are {\it absolute}, whereas the experimental ones are quantitative, and subject to experimental precision.
\subsection{Theoretical constraints}
\label{subsec:th-constraints}

\begin{itemize}
\item \textbf{CP violation} 

We do {\it not} impose CP conservation on the neutral Higgs sector.
The amount of CP violation that remains after all constraints are imposed is determined afterwards.
For a detailed discussion of the conditions for CP to be violated in this model, 
see Appendix~B of ref.~\cite{Grzadkowski:2009bt}.

\item \textbf{Stability or positivity} 

  The potential should be bounded from below for any values of the
  fields $\Phi_1$, $\Phi_2$ and $\eta$. This condition is rather
  involved for the potential (\ref{Eq:fullpot}). The full set of
  conditions are given in Appendix~A of \cite{Grzadkowski:2009bt}. For
  the somewhat simpler case of dark democracy considered here, we must
  impose
\begin{gather}
\lambda_1>0,\quad \lambda_2>0,\quad \lambda_\eta>0,\\
\lambda_x>-\sqrt{\lambda_1\lambda_2},
\quad \lambda_y>-\sqrt{\lambda_1\lambda_\eta},\quad
\lambda_y>-\sqrt{\lambda_2\lambda_\eta},
\label{constraints}
\end{gather}
\begin{equation} \label{Eq:positivity-cond.2}
\lambda_y\geq0 \vee \left(\lambda_\eta\lambda_x-\lambda_y^2
>-\sqrt{(\lambda_\eta\lambda_1-\lambda_y^2)
(\lambda_\eta\lambda_2-\lambda_y^2)}\right),
\end{equation}
where
\begin{subequations}
\begin{align}
\lambda_x&=\lambda_3+\min\left(0,\lambda_4-|\lambda_5|\right),\\
\lambda_y&=\lambda_{a}+\min\left(0,\lambda_{b}-|\lambda_{c}|\right).
\end{align}
\end{subequations}
With $M_S<\min(M_A,M_{\eta^\pm})$, we have 
\begin{equation}
\lambda_y=\frac{2}{v^2}\left(M_S^2-m_\eta^2\right).
\end{equation}

In \cite{Grzadkowski:2009bt} we constrained the potential further, by requiring
$V_{12}$, $V_{3}$ and $V_{123}$ individually to satisfy
positivity. The condition (\ref{Eq:positivity-cond.2}) was then replaced by
\begin{equation}
\lambda_a\geq \max (0,- 2\lambda_b, - \lambda_b \pm \lambda_c).
\label{posit}
\end{equation}
In terms of masses, this means for the two cases:
\begin{subequations} \label{Eq:positivity-simple}
\begin{alignat}{2}
&M_S<M_A<M_{\eta^\pm}:
&\quad &m_\eta^2+M_{\eta^\pm}^2-M_A^2\leq M_S^2, \\
&M_S<M_{\eta^\pm}<M_A:
&\quad &m_\eta^2\leq M_S^2.
\end{alignat}
\end{subequations}

In the present study we go beyond the domain of parameters allowed by (\ref{posit}) or, in terms of masses, (\ref{Eq:positivity-simple}).  The latter condition allows for checking positivity ``once and for all'', for a given set of inert-sector parameters.  The
full condition (\ref{Eq:positivity-cond.2}) depends also on the parameters of the non-inert sector, and must thus be checked for each point. The ``reward'' is that less parameter space will be excluded.

\item \textbf{Electroweak symmetry breaking}  

In order to break the electroweak symmetry spontaneously, the vacuum expectation values of $\Phi_1$ and $\Phi_2$ should be non-zero. The most general form of the vacuum can always be written in the form
\begin{equation}
\langle\Phi_1\rangle = \frac{1}{\sqrt{2}}\left( 
\begin{array}{c}
0 \\
v_1
\end{array} \right),
\quad
\langle\Phi_2\rangle = \frac{1}{\sqrt{2}}\left( 
\begin{array}{c}
u \\
v_2+i\delta
\end{array} \right) 
\end{equation}
where $v_1>0$ and $u$, $v_2$ and $\delta$ are real numbers. 
Non-zero $u$ would imply spontaneous $U(1)_{\rm EM}$ violation, so electric charge non-conservation.
However, it has been shown in \cite{Ferreira:2004yd} that if a local charge-conserving minimum exists, then there can be no charge-breaking minima (there may exist a stationary point with $u\neq 0$, which is a saddle point). Therefore from here on we assume $u=0$.
Nevertheless, the potential of the 2HDM can have more than one charge conserving minimum. In that case it is important to make sure that the theory is expanded around the global one, so that the issue of tunneling to the lowest one does not appear. In our approach we start out by assuming $\delta=0$~\footnote{Note that for a given minima it is always possible to make its location real by performing an appropriate global phase rotation.} and choosing a value of $\tan\beta$. 
Then the parameters of the potential are adjusted so that $v_1=v\cos\beta$ and $v_2=v\sin\beta$ satisfy
the stationary-point equations. Since all scalar mass squares that we consider are positive the stationary point must be a minimum.
However we do not beforehand know if our starting minimum is the global minimum of the potential since the 2HDM allows for more than one minimum. 
In our scans, we look for points (``good") in the parameters space that satisfy all theoretical and experimental constraints. 
For each ``good" point we also check if our starting minimum ($v_1,v_2$) is the global minimum. If there exists a deeper minimum we discard the point. Of all the ``good'' points that satisfied all other restrictions, approximately 7\% were thus discarded because a deeper global minimum exists.
Note that if the Universe was indeed in a state corresponding to a false vacuum, then the tunneling to the true vacuum would in principle be possible. If the tunneling time was shorter than the Universe age that could have  important cosmological consequences. This, however lies beyond the scope of the present study. Therefore we restrict ourselves to the case of global minimum only. Note that the presence of 
the third doublet $\eta$ does not influence the above arguments since $ \langle \eta \rangle =0 $.

\item \textbf{Unitarity and perturbativity}  

  We impose unitarity on the non-inert Higgs-Higgs-scattering sector
  \cite{Kanemura:1993hm,Akeroyd:2000wc,Ginzburg:2003fe}.  At large values of $\tan\beta$, the soft mass parameter $\mu$ is rather constrained, $\mu\sim  M_2\sim M_{H^\pm}$, as discussed in Refs.~\cite{Kaffas:2007ei,WahabElKaffas:2007xd}. Furthermore, perturbativity is imposed, in the form
  \begin{equation}
    \lambda_i, \frac{\sqrt{2} m_t}{v}|a_j|, \frac{\sqrt{2} m_t}{v}|\tilde a_j|,
    \frac{ m_t}{\sqrt{2}v}\cot\beta, \lambda_a, \lambda_b, \lambda_c < 4 \pi.
    \label{pert}
  \end{equation}
  Here, $a_j$ and $\tilde a_j$ are coefficients of the CP-even and odd
  parts of the Yukawa couplings \cite{ElKaffas:2006nt}. For the
  couplings to $t$ ($b$) quarks, they are given by
  $a_j=R_{j2}/\sin\beta$ and $\tilde a_j=-R_{j3}/\tan\beta$
  ($a_j=R_{j1}/\cos\beta$ and $\tilde
  a_j=-R_{j3}\tan\beta$). 

\item \textbf{The little hierarchy} 

In order to avoid excessive  computational requirements, and in distinction from the approach of \cite{Grzadkowski:2009bt}, we will not a priori impose an alleviation of the little hierarchy. However, it turns out that for $M_S\simeq75~\text{GeV}$, the masses of the non-inert Higgs sector can be lifted to rather high values, as will be discussed in Sec.~\ref{sec:med-mass}. This provides for a considerable alleviation of the little hierarchy problem around $M_S\simeq75~\text{GeV}$.

\end{itemize}

\subsection{Experimental constraints}
\label{subsec:expt-constraints}

We impose a variety of relevant experimental constraints. These can
be grouped as follows:
\begin{itemize}
\item \textbf{Charged-Higgs sector}  

  The charged-Higgs sector is constrained by several observables. The
  $B\to X_s \gamma$ data constrain low values of $\tan\beta$ and low
  $M_{H^\pm}$, the details of which depend on QCD effects
  \cite{Chetyrkin:1996vx,Borzumati:1998tg,Misiak:2004ew,Bauer:1997fe}.
  Likewise, the $B_0-\bar{B}_0$ mixing constrains low values of
  $\tan\beta$ and low $M_{H^\pm}$
  \cite{Abbott:1979dt,Inami:1980fz,Urban:1997gw}, whereas $B\to D\tau
  \bar\nu_\tau$ and $B\to \tau \bar\nu_\tau X$ constrain low values of
  $M_{H^\pm}$ and high values of $\tan\beta$
  \cite{Krawczyk:1987zj,Abbiendi:2001fi,Ikado:2006un,Hou:1992sy,
    Aubert:2007dsa,Nierste:2008qe}.

\item \textbf{Neutral-Higgs sector} 

  The LEP2 Higgs boson searches have given limits on the coupling of
  the lightest Higgs to the $Z$ and to $b\bar b$
  \cite{Abdallah:2004wy}.  At low values of $\tan\beta$, and low
  $M_{H^\pm}$, the well-measured $\Gamma(Z\to b\bar b)$ decay rate
  also constrains charged-Higgs contributions, and to a much lesser
  extent, neutral-Higgs couplings \cite{Denner:1991ie,Cheung:2003pw}.
  There are also bounds stemming from the electroweak precision data, in terms of the $T$ and $S$ parameters
  \cite{Grimus:2007if,Grimus:2008nb,Amsler:2008zz}.  
  Among these, the most serious one is the constraint on $T$, which basically is a constraint on the mass splitting of {\it pairs} of scalars: A pair of neutral scalars or a pair of charged scalars of different masses both give a {\it positive} contribution $\Delta T(M_i^2,M_j^2)>0$, whereas a neutral--charged pair gives a {\it negative} contribution, $\Delta T(M_i^2,M_j^2)<0$. Such contributions must roughly cancel, in order not to violate the electroweak precision data.
  
  Furthermore, at
  large values of $\tan\beta$, the model is constrained by the
  electron electric dipole moment
  \cite{Regan:2002ta,Pilaftsis:2002fe,Barr:1990vd}, for which we
  adopt the bound:
\begin{equation} \label{Eq:edm-bound}
|d_e|\lsim1\times10^{-27} [e\,\text{cm}],
\end{equation}
at the 1-$\sigma$ level. This is calculated directly from the neutral-Higgs-sector 
mixing matrix \cite{Pilaftsis:2002fe,Grzadkowski:2009bt}.
  The muon anomalous magnetic moment
  \cite{Cheung:2003pw,Barr:1990vd,Bennett:2004pv} has
  however little impact, since the large $\tan\beta$ region tends to
  be excluded by the unitarity constraint \cite{Cheung:2003pw}. For
  the relevant loop calculations, we use the {\tt LoopTools} package
  \cite{Hahn:1998yk,vanOldenborgh:1989wn}.

\item \textbf{Inert-sector constraints} 

The amount of dark matter has now been measured to  an impressive precision \cite{Hinshaw:2008kr}
\begin{equation}
\Omega_\text{DM} h^2=0.1131\pm0.0034
\end{equation}
We estimate the model prediction of the amount of dark matter from an
implementation of {\tt micrOMEGAs}
\cite{Belanger:2006is,Belanger:2008sj}.

For the heavier, neutral member of the inert sector, we adopt the bound obtained from a re-analysis of LEP data \cite{Lundstrom:2008ai}, approximated as $M_A>110~\text{GeV}$. For the charged member, we adopt the LEP bound on the chargino mass \cite{Pierce:2007ut}, $M_{\eta^\pm}>70~\text{GeV}$. This is slightly more conservative than the bound on charged Higgs bosons, $M_{H^\pm} > 79.3~\text{GeV}$, adopted by Ref.~\cite{LopezHonorez:2006gr}.

\end{itemize}

Since some of these constraints are correlated, we do not accumulate
their ``penalties'' in the form of an overall $\chi^2$ measure, but rather
demand that each of them be satisfied to within $2\sigma$.
\section{Annihilation mechanisms}
\label{sect:annihilation}
\setcounter{equation}{0}

In order not to over-produce dark matter in the Early Universe, annihilation channels must be kinematically open. These are of different kinds, depending on the mass scales involved. ``External'' reference mass scales are the $W$ and the lightest Higgs mass scales, $m_W$ and $M_1$. If the DM mass is low compared to $m_W$ it will annihilate via the lightest (but off-shell) Higgs particle which then decays to $b\bar b$ or $c\bar c$. If the mass is comparable to $m_W$, it can pair-annihilate to $W^+W^-$ or $ZZ$. In the higher mass range, the neutral and charged members of the inert doublet will be near-degenerate, and several channels will be open.
We here review the different annihilation mechanisms, relevant in
different mass ranges.

\subsection{DM couplings}
The gauge and scalar couplings involving inert-sector fields are
collected in Appendix~A. Here we list some of the
most relevant ones.

The DM particles can annihilate via the gauge coupling:
\begin{subequations}
\begin{alignat}{2}
&SSW^+W^-:&\quad &\frac{ig^2}{2}, \\
&SSZZ:&\quad &\frac{ig^2}{2\cos^2\theta_W},
\end{alignat}
\end{subequations}
or to non-inert scalars via the following trilinear or quartic couplings:
\begin{subequations}
\begin{alignat}{2}
\label{Eq:SShiggs}
&SSH_j: &\quad &-2iF_{SSj}\lambda_Lv, \quad
\text{with} \quad
F_{SSj}=\cos\beta R_{j1}+\sin\beta R_{j2}, \\
&SSH_jH_j:&\quad &-2i(\lambda_L-\lambda_c R_{j3}^2), \\
&SSH_jH_k:&\quad &2i\lambda_c R_{j3}R_{k3}, \\
&SSH^+H^-:&\quad &-i\lambda_a,
\end{alignat}
\end{subequations}
where $\lambda_L$ is defined in Eq.~(\ref{Eq:lambda_L}) and the pre-factor in (\ref{Eq:SShiggs}) satisfies $|F_{SSj}|\leq1$,
since $R$ is unitary. In particular,
$F_{SS1}=\cos(\beta-\alpha_1)\cos\alpha_2$.  Also, we note that the
splitting $M_S^2-m_\eta^2$ controls the strength of the important
trilinear coupling $\lambda_L$ to a neutral Higgs field. Likewise, the
couplings $\lambda_a$ and $\lambda_c$ are related to mass splittings,
$\lambda_a\sim(M_{\eta^\pm}^2-m_\eta^2)$ and
$\lambda_c\sim(M_{S}^2-M_{A}^2)$, see
Eq.~(\ref{Eq:lambda-vs-splitting}).

\subsection{Representative branching ratios}
In the medium-mass region, the early-universe abundance is controlled by
$SS$ annihilation, typically to $b\bar b$ or $W^+W^-$. In the
high-mass region, in part due to the high degree of mass degeneracy of the different states, there are also
significant losses due to annihilations of $A$ and $\eta^\pm$. The
different losses are provided by {\tt micrOMEGAs} as fractions of
$1/(\Omega_\text{DM} h^2)$ (see ref.~\cite{Belanger:2006is}). We refer to this variable as ``loss channel''.  Some
representative values are given below, where we consider separately the low- and medium-mass region, and the high-mass region.

\subsubsection{Low- and medium-mass region}
In the low and medium-mass region, the annihilations mostly proceed via an intermediate Higgs boson, or via a $W^+W^-$ pair or a $ZZ$ pair:
\begin{itemize}
\item
For $M_S=40~\text{GeV}$ and $M_1=120~\text{GeV}$, representative loss channels are:
\begin{equation}
SS\to
\begin{cases}
b\bar b \quad &(86-99\%), \\
c\bar c \quad &(1-9\%),
\end{cases}
\end{equation}
\item
For $M_S=80~\text{GeV}$ and $M_1=120~\text{GeV}$, representative loss channels are:
\begin{equation}
\label{Eq:80-annihilate}
SS\to
\begin{cases}
W^+W^- \quad &(19-96\%), \\
b\bar b \quad &(2-91\%),
\end{cases}
\end{equation}
\end{itemize}
where the ranges relate to the scans over $M_{\eta^\pm}$, $m_\eta$, $\tan\beta$, $M_{H^\pm}$ and the $\alpha_i$, defined in Sec.~\ref{Sec:-scan-strategy}.
\subsubsection{High-mass region}
The inert scalar masses are generated by $m_\eta$, $v_1$ and $v_2$ such that $m_\eta$ contributes universally to all the masses while the splitting between them is 
controlled by the interaction terms ($\lambda_a$, $\lambda_b$ and $\lambda_c$) between the 2HDM and the inert sector. 
As the DM mass $M_S$ increases, annihilation channels to pairs of vector bosons opens (with fixed gauge coupling constant strength). Therefore annihilation through intermediate $H_i$ must be suppressed  in order not to introduce a too small value of $\Omega_\text{DM}$, that implies small $\lambda_a$, $\lambda_b$ and $\lambda_c$.
In other words, the inert sector masses must be
similar. For some representative mass parameters, loss channels  are given below:
\begin{itemize}
\item
For $M_S=550~\text{GeV}$ and $M_1=120~\text{GeV}$, representative loss channels are:
\begin{subequations} \label{Eq:low-edge-of-high-region}
\begin{equation}
SS\to
\begin{cases}
W^+W^-  \quad &(14-17\%), \\
ZZ \quad &(11-14\%),
\end{cases}
\end{equation}
\begin{equation}
\eta^+\eta^-\to 
\begin{cases}
\gamma Z \quad &(14-17\%), \\
W^+W^- \quad &(13-16\%), \\
\gamma \gamma \quad &(5-6\%),
\end{cases}
\end{equation}
\begin{equation}
AA\to 
\begin{cases}
W^+W^- \quad &(8-9\%), \\
ZZ \quad &(6-7\%),
\end{cases}
\end{equation}
\begin{equation}
S\eta^\pm, A\eta^\pm\to
\gamma W^\pm \quad (6-8\%).
\end{equation}
\end{subequations}
\item
For $M_S=3000~\text{GeV}$ and $M_1=120~\text{GeV}$, representative loss channels are:
\begin{subequations} \label{Eq:high-edge-of-high-region}
\begin{equation}
SS\to
\begin{cases}
W^+W^- \quad &(4-12\%), \\
H^+H^-  \quad &(3-11\%), \\
ZZ \quad &(1-4\%), \\
H_j H_j  \quad &(1-3\%), \quad \text{each }j,
\end{cases}
\end{equation}
\begin{equation}
\eta^+\eta^-\to 
\begin{cases}
W^+W^- \quad &(1-9\%), \\
H^+H^- \quad &(1-9\%), \\
ZZ \quad &(2-7\%), \\
H_j H_j  \quad &(2-7\%), \quad \text{each }j,
\end{cases}
\end{equation}
\begin{equation}
AA\to 
\begin{cases}
W^+W^- \quad &(3-10\%), \\
H^+H^- \quad &(3-9\%), \\
ZZ \quad &(1-2\%), \\
H_j H_j  \quad &(1-3\%), \quad \text{each }j,
\end{cases}
\end{equation}
\begin{equation}
S\eta^\pm, A\eta^\pm\to 
\begin{cases}
ZW^\pm \quad &(1-9\%), \\
H_j W^\pm  \quad &(1-10\%), \quad \text{each }j, \\
H_j H^\pm  \quad &(1-10\%), \quad \text{each }j,
\end{cases}
\end{equation}
\end{subequations}
\end{itemize}
where again the ranges relate to the scans over $M_{\eta^\pm}$, $m_\eta$, $\tan\beta$, $M_{H^\pm}$ and the $\alpha_i$, defined in Sec.~\ref{Sec:-scan-strategy}.

\section{Parameters and Scan Strategy}
\label{Sec:-scan-strategy} 
\setcounter{equation}{0}
\subsection{Model parameters}
\label{Sec:-scan-strategy-parameters} 

The model contains a total of 13 parameters defining the spectrum and
the couplings. Among these, the inert-sector self-coupling, $\lambda_\eta$, plays no role and is kept fixed.
We choose the remaining ones to be:
\begin{enumerate}
\item
$M_S$, $M_1$ (lowest physical masses of inert and 2HDM sectors, fixed)
\item 
$M_A$, $M_{\eta^\pm}$ (inert sector, physical masses, fixed).  In the
  high $M_S$ regime ($M_S>500~\text{GeV}$), $M_A$ and $M_{\eta^\pm}$
  should be rather close to $M_S$, in order to prevent $\lambda_b$ and $\lambda_c$ 
  from becoming large (and thus lead to
  too much annihilation of DM in the Early Universe).  In the low $M_S$ regime
  ($M_S<100~\text{GeV}$), a wider range of values is possible, but
  LEP2 data constrain $M_A\gsim110~\text{GeV}$ \cite{Lundstrom:2008ai}.
  For $M_{\eta^\pm}$, we impose the constraint $M_{\eta^\pm}>70~\text{GeV}$, adopted from the LEP searches for charginos \cite{Pierce:2007ut}.
\item
$M_2$, $\mu$ (2HDM sector parameters)
\item $m_\eta$ (inert sector, soft mass parameter, fixed). Then $\lambda_a, \lambda_b,\lambda_c$ are all fixed. In the high $M_S$
  regime ($M_S>500~\text{GeV}$), $m_\eta$ should be rather close to
  $M_S$, in order to prevent $\lambda_L$ from
  becoming large (and thus lead to too much annihilation of DM in the Early Universe). In the
  low $M_S$ regime ($M_S<100~\text{GeV}$), a wider range of values is
  possible.
\item
$\tan\beta$, $M_{H^\pm}$ (2HDM sector). We allow for the following variation: $0.5\leq\tan\beta\leq50$ and
$300~\text{GeV}\leq M_{H^\pm}\leq700~\text{GeV}$. We consider a logarithmic grid
in $\tan\beta$, and linear in $M_{H^\pm}$, typically 30 points in each parameter.
Representative, allowed regions are shown in Sec.~\ref{Sec-CPV}.
\item $\alpha_1$, $\alpha_2$, $\alpha_3$ (2HDM sector). The allowed
  range of variation is $-\pi/2\leq\alpha_{1,2}\leq\pi/2$, and
  $0\leq\alpha_{3}\leq\pi/2$. A random set of 1000 points in this
  three-dimensional space is typically adequate.
  Representative, allowed regions are shown in Sec.~\ref{Sec-CPV}.
\end{enumerate}
From this input, the value of $M_3$ and all $\lambda$'s of the 2HDM
can be reconstructed \cite{Khater:2003wq}.  For the inert sector, we
take $\lambda_\eta=0.2$ (in the notation of the IDM
\cite{Barbieri:2006dq,LopezHonorez:2006gr}
$\lambda_\eta=2\lambda_2$). This parameter has little influence on the
model \cite{LopezHonorez:2006gr}.  From the chosen mass input,
together with $\lambda_\eta$, the $\lambda_a$, $\lambda_b$ and
$\lambda_c$ of Eq.~(\ref{Eq:DarkDemocracy}) can be determined.
\subsection{General scanning strategy}
\label{subsec:-scan-strategy} 

We scan over the parameters in a hierarchical fashion.
In the notation
of Sec.~\ref{Sec:-scan-strategy-parameters}, we hold parameter $p_{i-1}$
fixed while scanning over $p_{i}$. For each parameter $p_i$, we have
two options: 
\begin{itemize}
\item[(i)] 
if we find an acceptable solution, proceed to the
next value at the higher level, $p_{i-1}$, or 
\item[(ii)] 
exhaust a
predefined range from $p_i^\text{min}$ to $p_i^\text{max}$. 
\end{itemize}
The latter
condition is adopted if we are interested in determining the allowed range of
$p_i$.

Scanning over this large number of parameters we focus on regions where
the 2HDM is known to be consistent \cite{WahabElKaffas:2007xd}.
More details are given in Secs.~\ref{sec:med-mass} and \ref{sec:high-mass}.

It should be stressed that adopting the strategy described above, some solutions
could be missed. However, as we {\it do find} interesting regions of parameter space
that are allowed, we do not insist on exhausting the whole parameter space 
what would imply a dramatic increase of the CPU running time.

\subsection{Positivity and unitarity}
For the IDM, it was found \cite{LopezHonorez:2006gr} that certain domains in the $M_S$--$m_\eta$ plane (denoted $M_{H_0}$ and $\mu_2$ in \cite{LopezHonorez:2006gr}) are forbidden by positivity. For the present model, because we have more parameters (in particular, the $\alpha_i$ parameters), we have not found any such domain excluded by positivity. 
However, if we restrict the scan to the CP-conserving limit $\alpha_2\to0$, $\alpha_3\to0$, then some parts of parameter space are actually excluded, in particular for $M_{\eta^\pm}<M_S$. But that region is of course not interesting, since we want the DM candidate to be electrically neutral.

If we impose also unitarity in addition to positivity, the allowed parameter space in $M_{\eta^\pm}$--$m_\eta$ starts to shrink, as indicated in Fig.~\ref{hi-0550-pos-uni} for $M_S=550~\text{GeV}$ and $M_1=120~\text{GeV}$.
For the lower range of $M_S$-values, we do not find any such forbidden region within a ``reasonable'' range of parameters. For example, with $(M_S, M_A)=(75,110)~\text{GeV}$, there is no forbidden region within $70~\text{GeV}\leq M_{\eta^\pm}\leq 150~\text{GeV}$ and $0~\text{GeV}\leq m_{\eta}\leq150~\text{GeV}$.
Imposing next the experimental constraints discussed in Sec.~\ref{subsec:expt-constraints}, we find rather dramatic reductions of the allowed parameter space, as will be discussed in Secs.~\ref{sec:med-mass} and \ref{sec:high-mass}.

\FIGURE[ht]{
\let\picnaturalsize=N
\def\picsize{6cm}
\ifx\nopictures Y\else{
\let\epsfloaded=Y
\centerline{\hspace{4mm}{\ifx\picnaturalsize N\epsfxsize \picsize\fi
\epsfbox{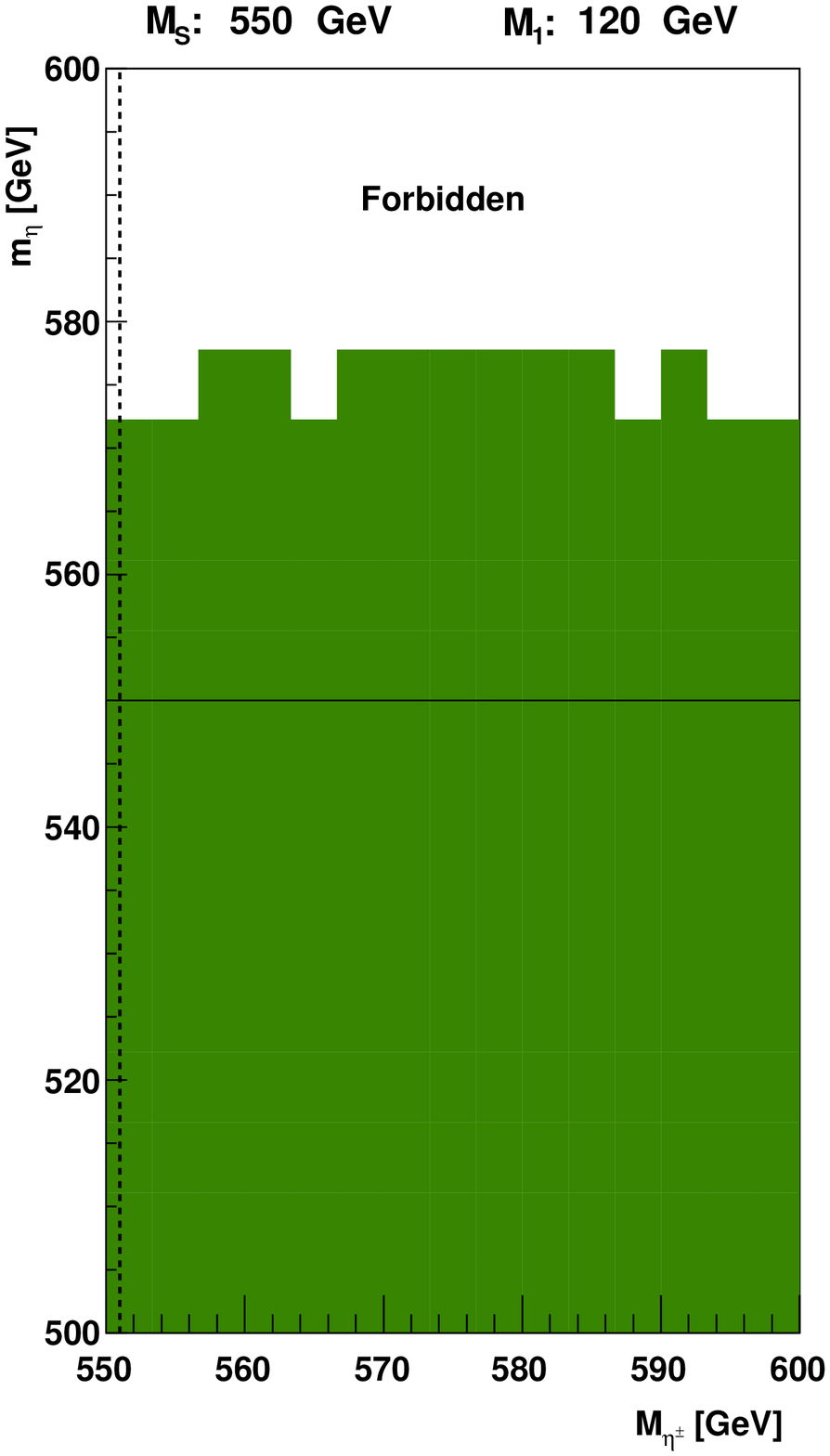}} }
}\fi
\vspace{-6mm}
\caption{\label{hi-0550-pos-uni} Regions in the $M_{\eta^\pm}-m_\eta$ plane that are allowed by positivity and unitarity, for DM mass $M_S=550~\text{GeV}$ and lightest Higgs mass $M_1=120~\text{GeV}$. } }

\section{Overview}
\label{sec:overview}
\setcounter{equation}{0}

Before going into a detailed discussion of allowed parameter regions, we here give a brief summary, comparing with the IDM, for which three mass regions were established:  light ($M_S \ll m_W$) \cite{Andreas:2008xy,Hambye:2007vf}, medium ($M_S\sim m_W$) \cite{Barbieri:2006dq,LopezHonorez:2006gr} or heavy ($M_S\gsim535~\text{GeV}$) \cite{Cirelli:2005uq,LopezHonorez:2006gr,Hambye:2009pw}. A recent analysis finds a further substructure of those regions\cite{Dolle:2009fn}. Finally, a very recent study \cite{LopezHonorez:2010tb} reports a new viable region for masses in the range $m_W\lsim M_S\lsim 150~\text{GeV}$.

We found two regions of allowed $M_S$ values, a low-to-medium region, $5~\text{GeV}\lsim M_S \lsim 100~\text{GeV}$ and a high region, $M_S \gsim 545~\text{GeV}$. The allowed $M_S$-ranges of these two models are compared in Fig.~\ref{Fig:overview}. The other parameters are chosen such that the allowed regions are maximized. In the low-to-medium DM-mass region, Early-Universe annihilation via the lightest neutral Higgs boson, $H_1$ plays an important role for obtaining the observed value of $\Omega_\text{DM}$. 
In the high DM-mass region, it is instead annihilation to two on-shell gauge bosons or two Higgs bosons that provide the correct amount of DM. In the forbidden intermediate range of $M_S$, $\Omega_\text{DM}$ is too low.
While the present model has a few more parameters than the IDM, we do not find solutions at as low values of $M_S$ as were found for the IDM. A main restriction on the model at low $M_S$ is the more tight up-to-date constraint on $\Omega_\text{DM}$ imposed here, the value comes out too high. In addition, our constraint on $M_A$ is different.
Likewise, the new viable region of the IDM \cite{LopezHonorez:2010tb} is only partly reproduced in our model, again mainly because of our more tight constraint on $\Omega_\text{DM}$.

\FIGURE[ht]{
\let\picnaturalsize=N
\def\picsize{17cm}
\ifx\nopictures Y\else{
\let\epsfloaded=Y
\centerline{\hspace{4mm}{\ifx\picnaturalsize N\epsfxsize \picsize\fi
\epsfbox{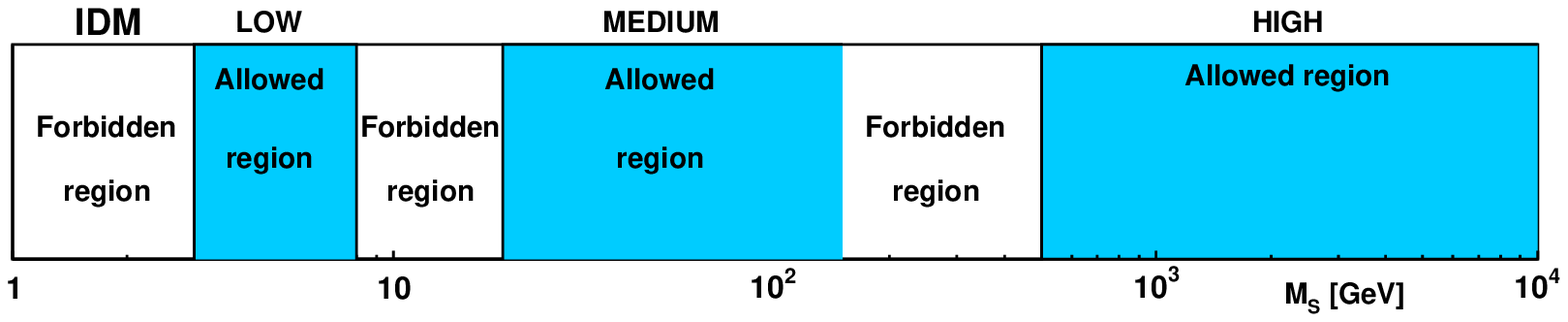}} } 
\centerline{\hspace{4mm}{\ifx\picnaturalsize N\epsfxsize \picsize\fi
\epsfbox{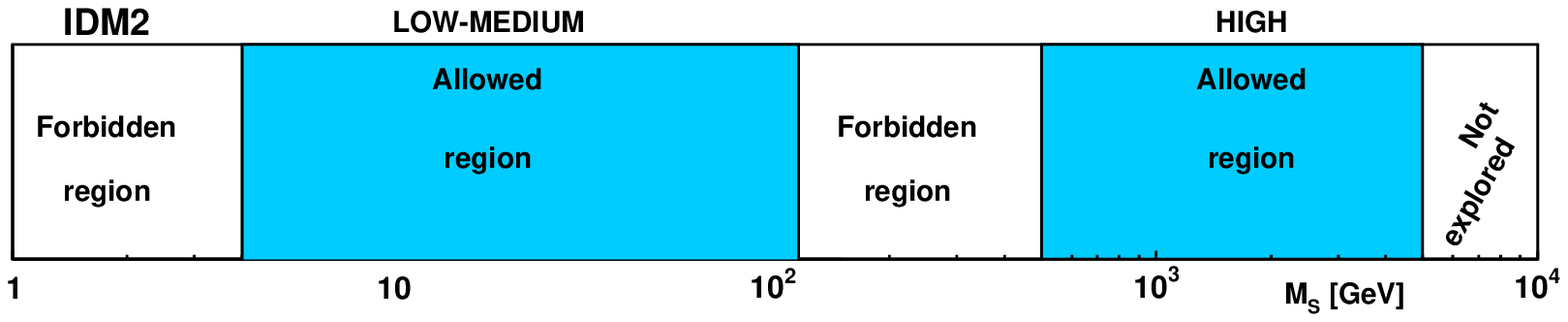}} } 
}\fi
\vspace{-6mm}
\caption{\label{Fig:overview} Top panel: Allowed $M_S$ ranges in the IDM \cite{Andreas:2008xy,Hambye:2007vf,Barbieri:2006dq,LopezHonorez:2006gr,Cirelli:2005uq,Hambye:2009pw,LopezHonorez:2010tb}. Bottom: Allowed $M_S$ ranges in the present model, IDM2.  Corresponding $M_1$ ranges are given in Secs.~\ref{sec:med-mass} and \ref{sec:high-mass}.} }

\section{Low--Medium DM Mass Regime}
\label{sec:med-mass}
\setcounter{equation}{0}

For a range of DM-masses, $M_S<{\cal O}(100~\text{GeV})$, we have explored the range of lightest neutral-Higgs-boson masses, $M_1\gsim120~\text{GeV}$, for which we find consistent solutions. Selected results are shown in Figs.~\ref{low-006-008}--\ref{med-075}. In these figures we display, for a given set of ($M_S,M_1$) values, the allowed region(s) in the $M_{\eta^\pm}$--$m_\eta$ plane, obtained by a scan over the ranges $70~\text{GeV}<M_{\eta^\pm}\leq150~\text{GeV}$ and $0\leq m_\eta\leq160~\text{GeV}$.

We note that for a given set of inert-sector masses, ($M_S$, $M_A$, $M_{\eta^\pm}$), one may think of $m_\eta$ (denoted $\mu$ in \cite{Barbieri:2006dq,Lundstrom:2008ai,LopezHonorez:2006gr}, and not to be confused with the parameter normally denoted $\mu$ in the 2HDM) as basically determining the trilinear coupling $SSH_1$, see Eqs.~(\ref{Eq:SShiggs}) and (\ref{Eq:lambda_L}). Thus, the vertical axis in these plots is a measure of how strongly the DM particles annihilate via a virtual Higgs. Indeed, along the right-hand edge of these plots, we also indicate some values of $\lambda_L$.
\subsection{Scanning strategy}

For this low-to-medium range of $M_S$, we start out with a fixed value of $M_A=110~\text{GeV}$ (approximately the lower limit compatible with LEP data  \cite{Lundstrom:2008ai}), and then scan over $M_{\eta^\pm}$ and $m_\eta$ as indicated in Sec.~\ref{subsec:-scan-strategy}. As mentioned above, for $M_{\eta^\pm}$, we impose the constraint $M_{\eta^\pm}>70~\text{GeV}$, adopted from the LEP searches for charginos \cite{Pierce:2007ut}. No particular hierarchy is assumed, we may have $M_S<M_A\leq M_{\eta^\pm}$ or $M_S<M_{\eta^\pm}\leq M_A$. If no solution is found for $M_A=110~\text{GeV}$, the scan is repeated for $M_A=115~\text{GeV}$. If still no solution is found, we declare there to be no solution for the chosen set $(M_S,M_1)$. (Only in one case was a solution found for $M_A=115~\text{GeV}$ and nothing for $110~\text{GeV}$.)

For fixed $M_A$, the scanning over $M_2$ and $\mu$ is organized as follows. For $M_1<300~\text{GeV}$, we first consider $M_2=300~\text{GeV}$ and $\mu=200~\text{GeV}$. If nothing is found, we increment $\mu$ to $350~\text{GeV}$ and $500~\text{GeV}$. If still nothing is found, we increment $M_2$ to 400 and $500~\text{GeV}$, repeating the values of $\mu$. For higher values of $M_1$, correspondingly higher values of $M_2$ and $\mu$ are adopted, but with less splitting with respect to $M_1$. This choice is inspired by the knowledge of allowed regions in the 2HDM parameter space \cite{Kaffas:2007ei,WahabElKaffas:2007xd}. For example, for $M_1=300~\text{GeV}$, we take $M_2=350~\text{GeV}$, $425~\text{GeV}$, and $500~\text{GeV}$.

\FIGURE[ht]{
\let\picnaturalsize=N
\def\picsize{5.5cm}
\ifx\nopictures Y\else{
\let\epsfloaded=Y
\centerline{\hspace{4mm}{\ifx\picnaturalsize N\epsfxsize \picsize\fi
\epsfbox{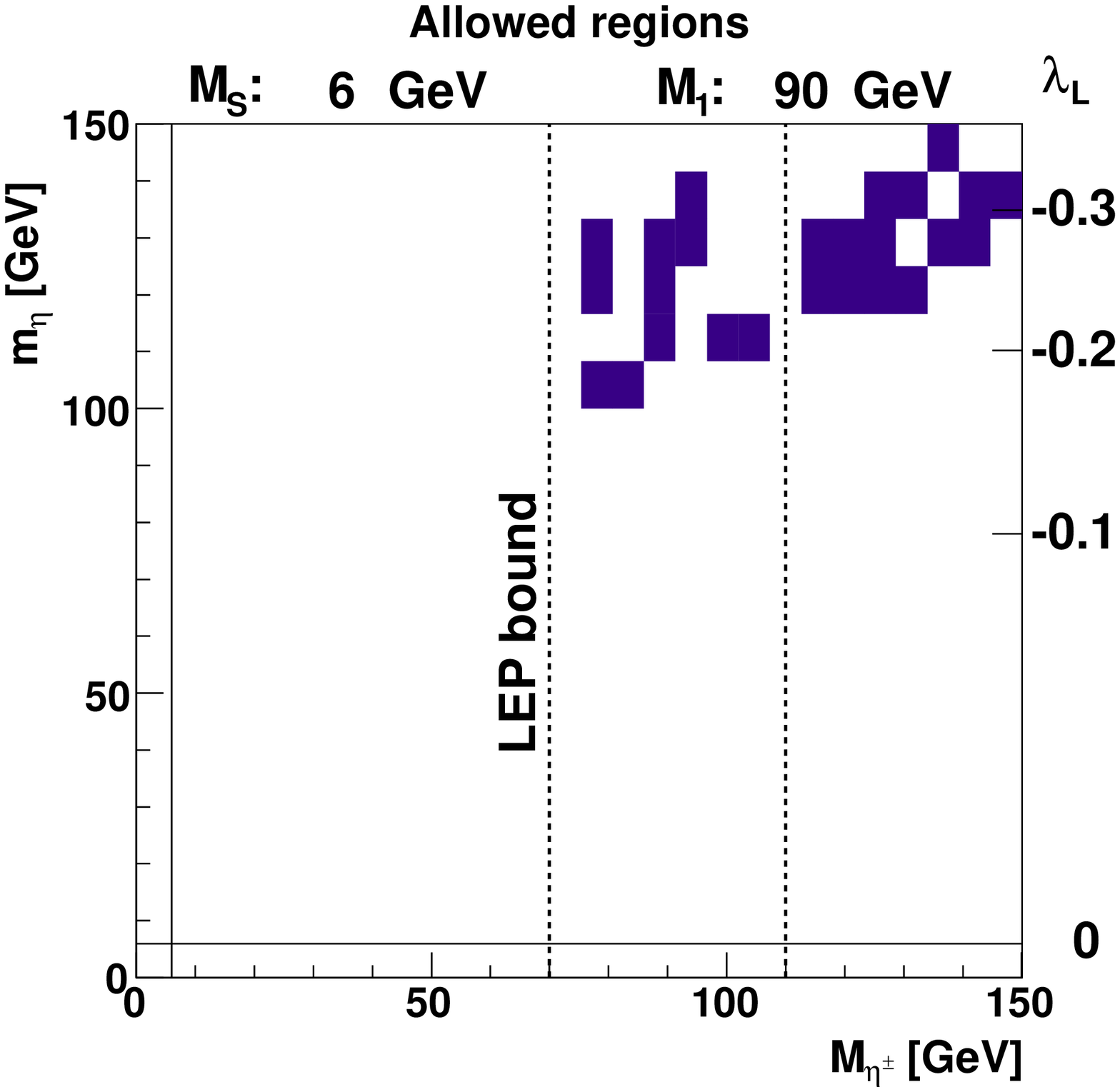}} 
\hspace{4mm}{\ifx\picnaturalsize N\epsfxsize \picsize\fi
\epsfbox{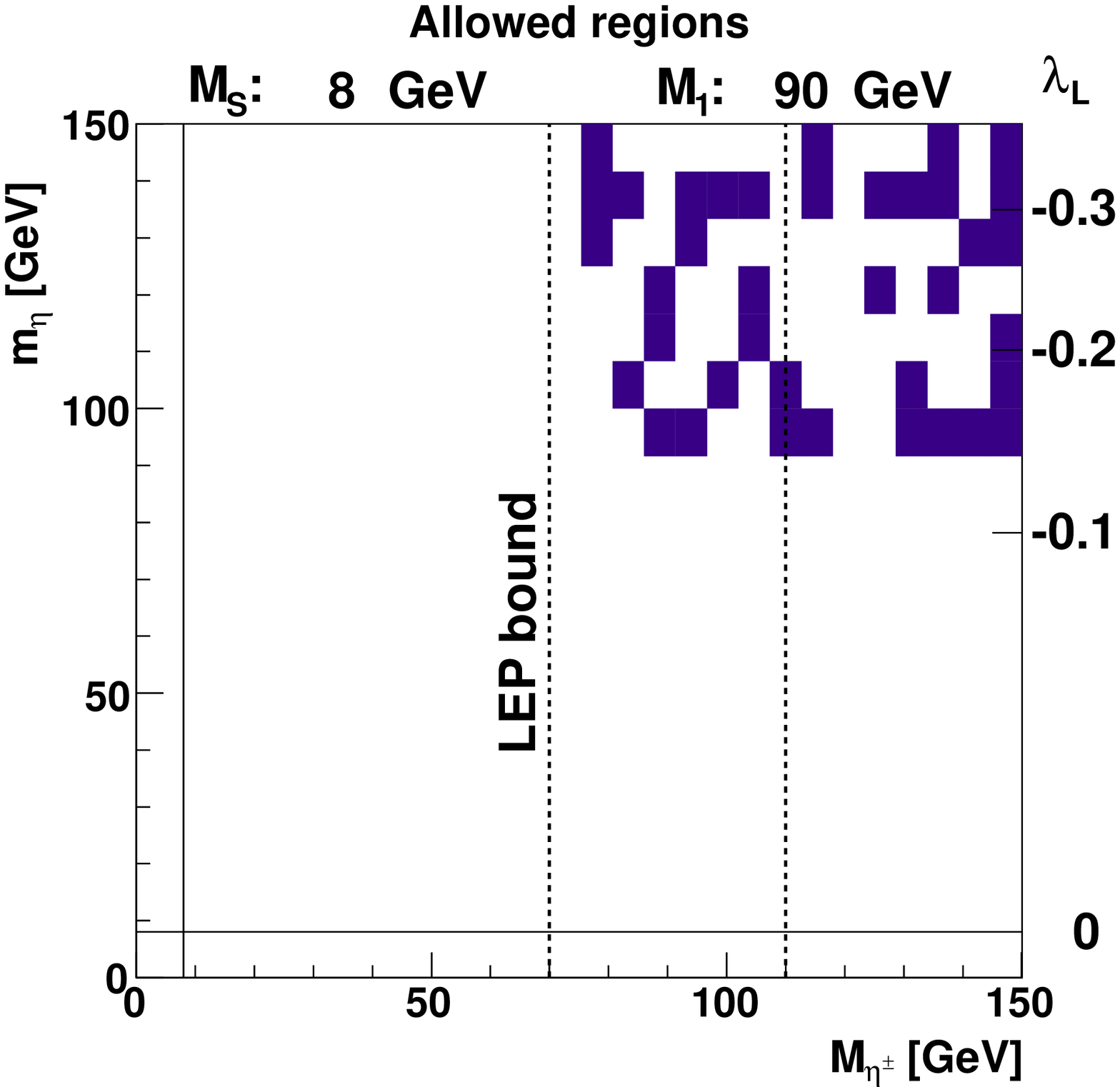}} } 
}\fi
\vspace{-6mm}
\caption{\label{low-006-008} Allowed regions (dark blue) in the $M_{\eta^\pm}-m_\eta$ plane, for DM mass $M_S=6~\text{GeV}$ and $8~\text{GeV}$, with lightest Higgs mass $M_1=90~\text{GeV}$. The thin solid lines indicate $M_{\eta^\pm}=M_S$ and $m_{\eta}=M_S$ (or $\lambda_L=0$), whereas the dashed line at $M_{\eta^\pm}=70~\text{GeV}$ (labelled ``LEP bound'') gives the adopted experimental bound. The right-most dashed line gives $M_{\eta^\pm}=110~\text{GeV}$, the default value for $M_A$.} }

\subsection{Results for $M_S\lsim100~\text{GeV}$}

Let us now comment on the main features of the plots showing allowed regions in the $M_{\eta^\pm}$--$m_\eta$ plane. As pointed out above, for fixed masses of the inert sector, ($M_S,M_A,M_{\eta^\pm}$), the ``soft'' parameter $m_\eta$, which represents the bilinear coupling in the inert sector, see Eq.~(\ref{v3}), will also represent the trilinear couplings between the inert sector and the non-inert one, as expressed by Eqs.~(\ref{Eq:trilinear}), (\ref{Eq:lambda_L})  and (\ref{Eq:lambda-vs-splitting}). Explicitly, the trilinear coupling $SSH_j$ vanishes in the limit $m_\eta\to M_S$, but can become large when either $m_\eta\ll M_S$ or $m_\eta\gg M_S$.

\FIGURE[ht]{
\let\picnaturalsize=N
\def\picsize{5.5cm}
\ifx\nopictures Y\else{
\let\epsfloaded=Y
\centerline{\hspace{4mm}{\ifx\picnaturalsize N\epsfxsize \picsize\fi
\epsfbox{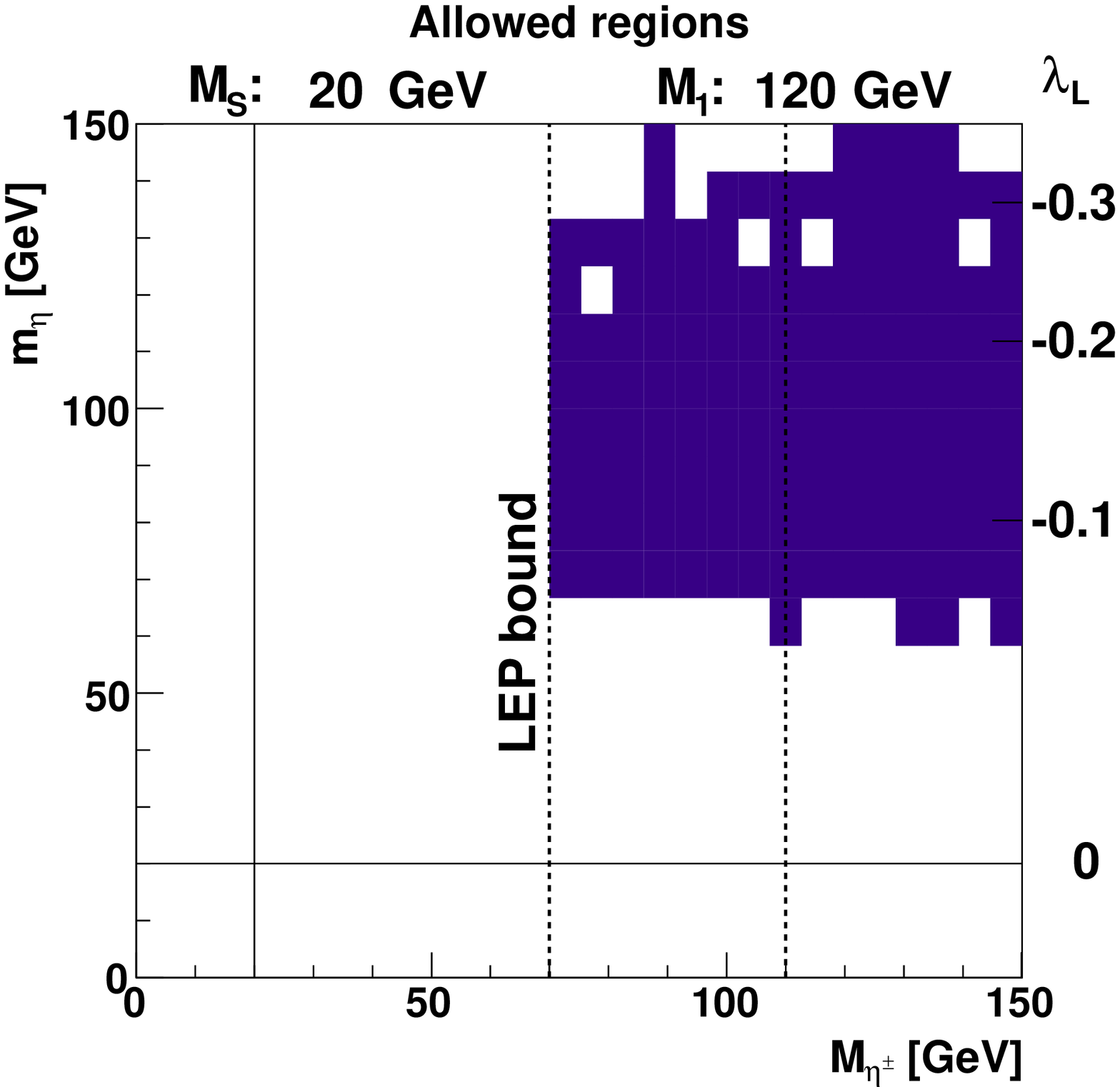}} 
\hspace{4mm}{\ifx\picnaturalsize N\epsfxsize \picsize\fi
\epsfbox{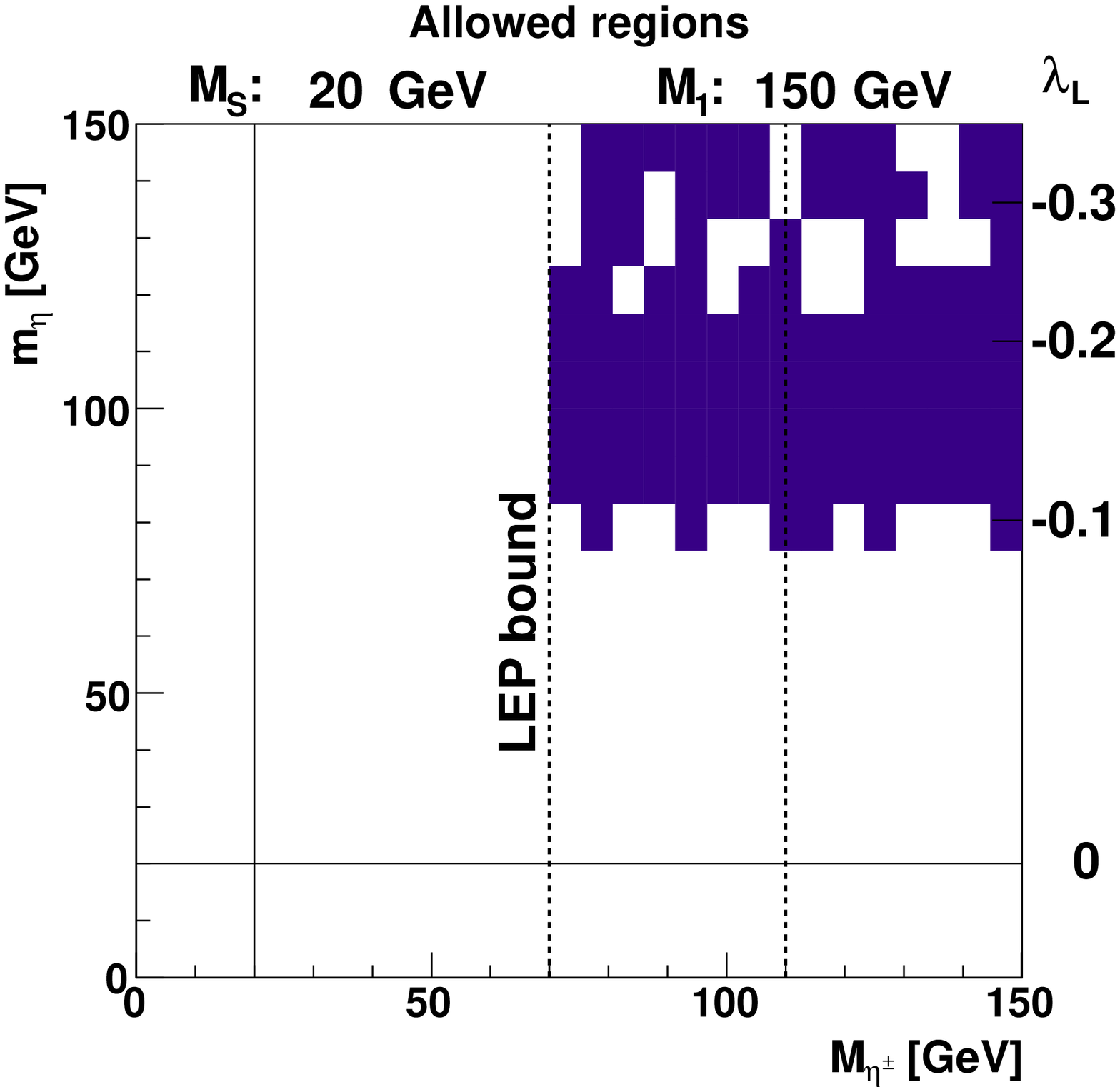}} } 
\centerline{\hspace{4mm}{\ifx\picnaturalsize N\epsfxsize \picsize\fi
\epsfbox{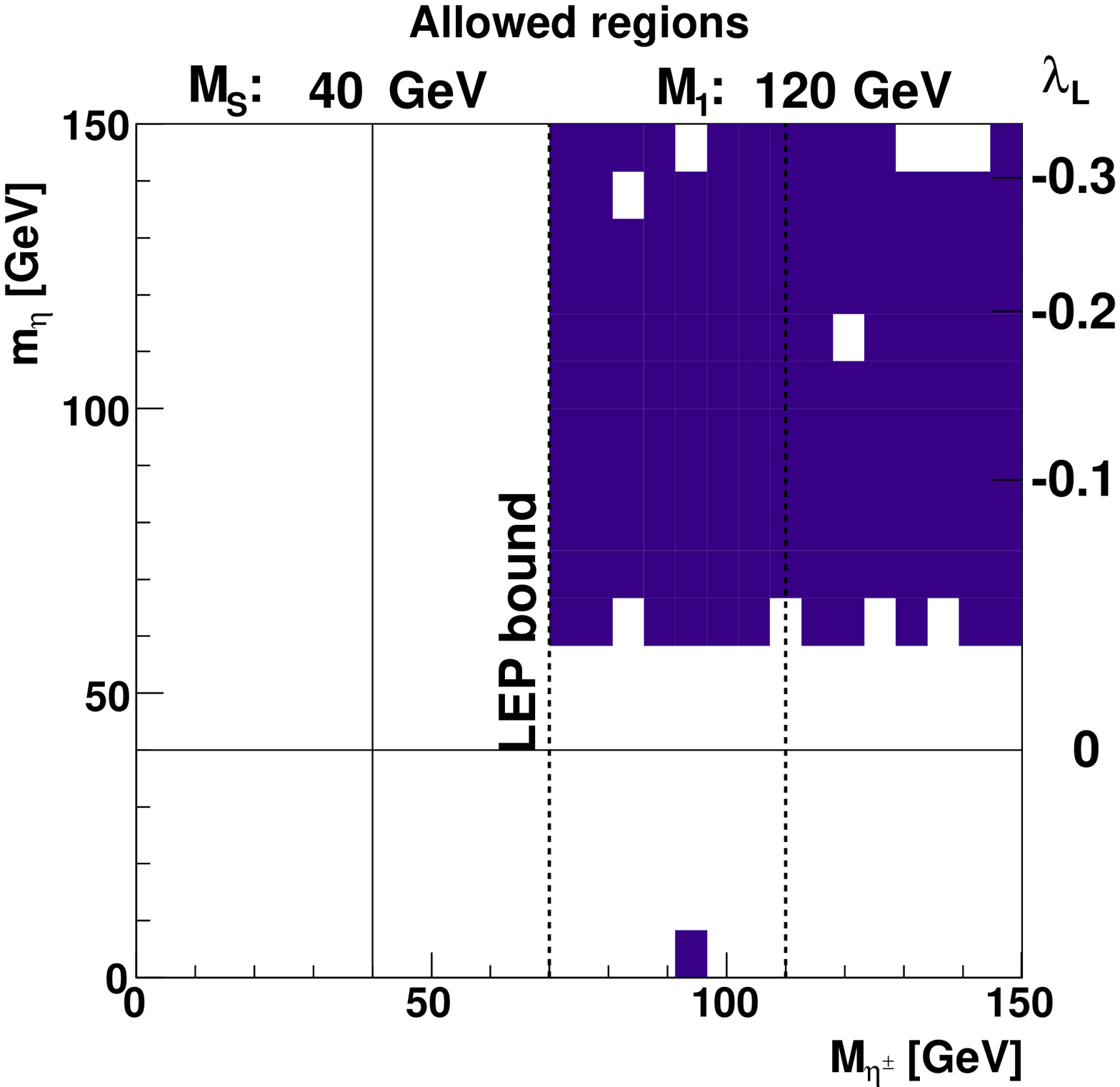}} 
\hspace{4mm}{\ifx\picnaturalsize N\epsfxsize \picsize\fi
\epsfbox{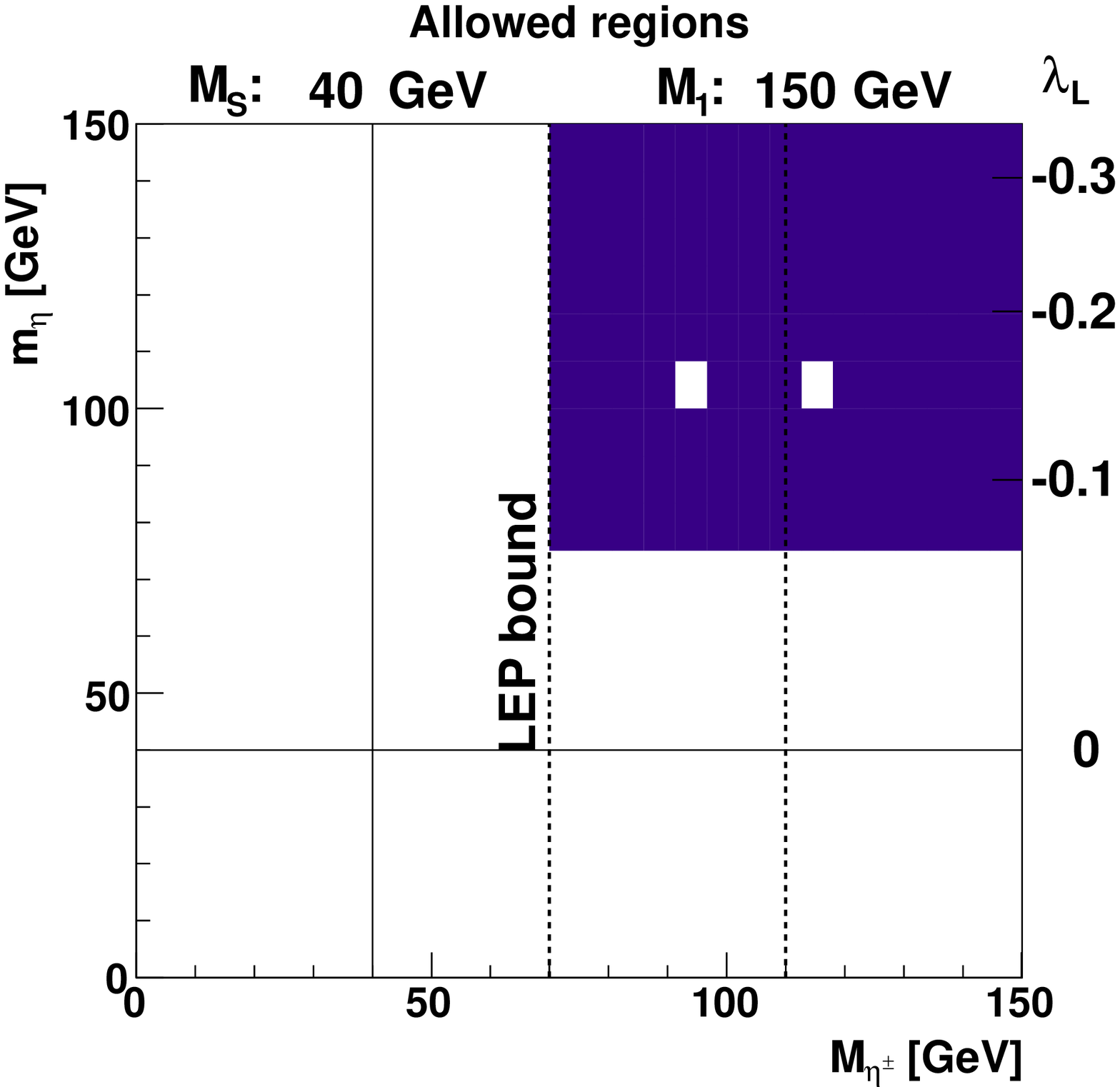}} } 
}\fi
\vspace{-6mm}
\caption{\label{med-020-040} Allowed regions in the $M_{\eta^\pm}-m_\eta$ plane, for DM mass $M_S=20~\text{GeV}$ and $M_S=40~\text{GeV}$, with lightest Higgs mass $M_1=120~\text{GeV}$ and 150~GeV.} }

In the case of a low- to medium-mass DM particle ($M_S\lsim100~\text{GeV}$), annihilations via a virtual Higgs boson
play an important role for obtaining the correct DM density. We recall that this coupling is proportional to $\lambda_L$, which in turn is proportional to $M_S^2-m_\eta^2$. This feature is the same as for the IDM \cite{LopezHonorez:2006gr}, and has implications for whether or not the parameter region $m_\eta \sim M_S$ is allowed.
However, in distinction from the IDM, here the coupling also contains a factor $F_j$, satisfying $0\leq|F_j|\leq1$ (see Appendix~A), which depends on $\tan\beta$ and the mixing angles of the neutral Higgs sector. Thus, for a given set of inert-sector parameters, one can always find non-inert-sector parameters for which this coupling is turned off (but positivity and unitarity constraints may make such points uninteresting).

If the DM particle is very light, $2M_S<M_1$, the intermediate Higgs particle will be off mass shell, and a heavier one (larger value of $M_1$) will be less efficient in mediating annihilations, because of the propagator suppression. In order to have sufficient Early-Universe annihilation, the $SSH_1$-coupling $\lambda_L$ must therefore be stronger, the higher $M_1$ is. Indeed, for the very lowest values of $M_S$, we do not find solutions for $M_1=120~\text{GeV}$, whereas $M_1=90~\text{GeV}$ and a large value of $m_\eta$ give acceptable solutions (see Fig.~\ref{low-006-008}). This is further illustrated in Fig.~\ref{med-020-040}, for $M_S=20~\text{GeV}$ and 40~GeV, where we see that indeed $|M_S^2-m_\eta^2|\propto |\lambda_L|$ must increase with $M_1$.

In this region, the cut-off towards higher values of $M_1$ is due to the $\Delta T$ and $\Omega_\text{DM}$ constraints. For example, at the edge of the forbidden region, for $M_S=40~\text{GeV}$ and $M_1=220~\text{GeV}$, more solutions are found if we either disregard the $\Delta T$ constraint, or allow a {\it higher} value of $\Omega_\text{DM}$. For this case of $M_1=220~\text{GeV}$, because of the mentioned propagator suppression, one must have $m_\eta\gsim110~\text{GeV}$.

\FIGURE[ht]{
\let\picnaturalsize=N
\def\picsize{5.5cm}
\ifx\nopictures Y\else{
\let\epsfloaded=Y
\centerline{\hspace{4mm}{\ifx\picnaturalsize N\epsfxsize \picsize\fi
\epsfbox{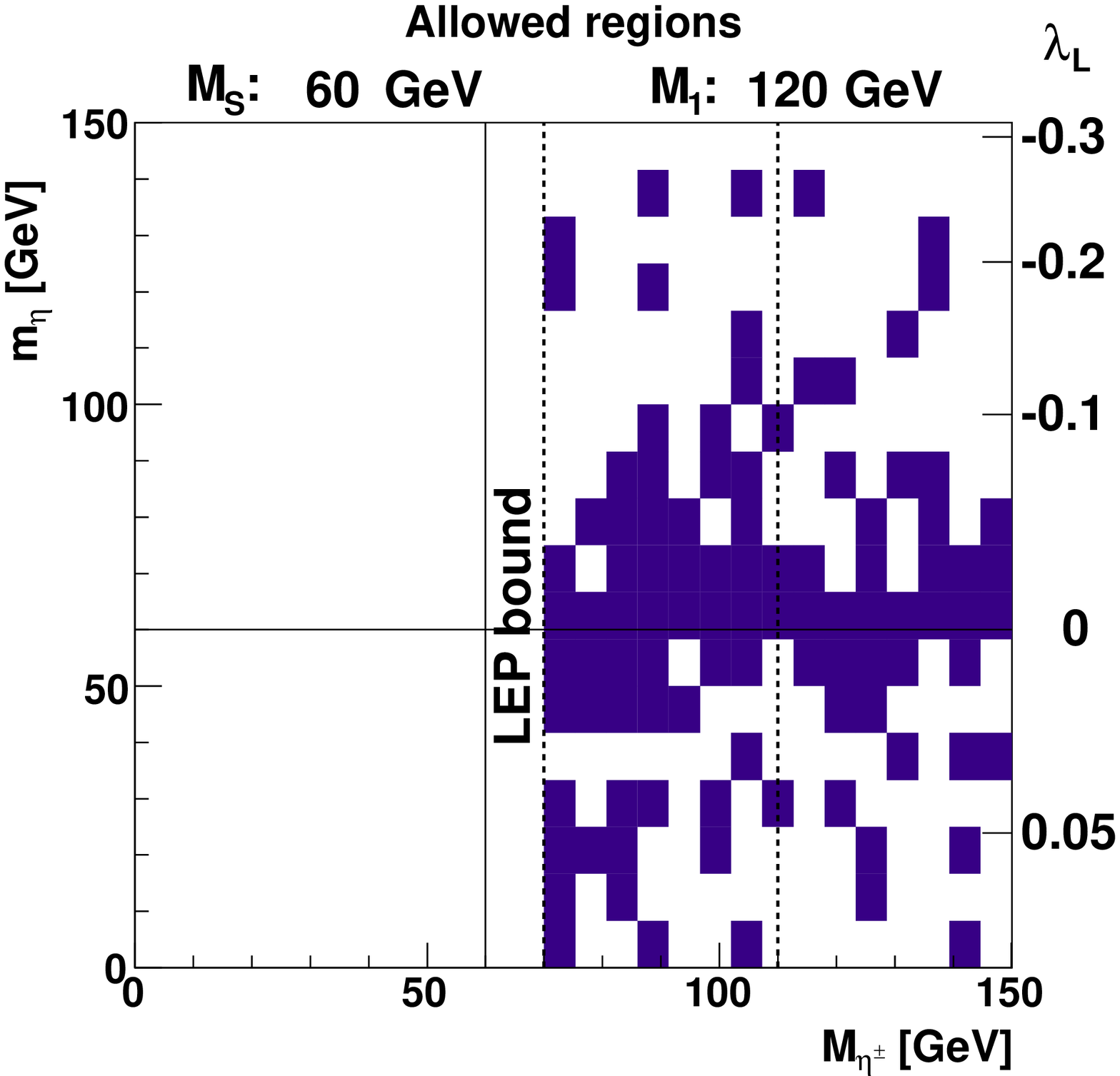}} 
\hspace{4mm}{\ifx\picnaturalsize N\epsfxsize \picsize\fi
\epsfbox{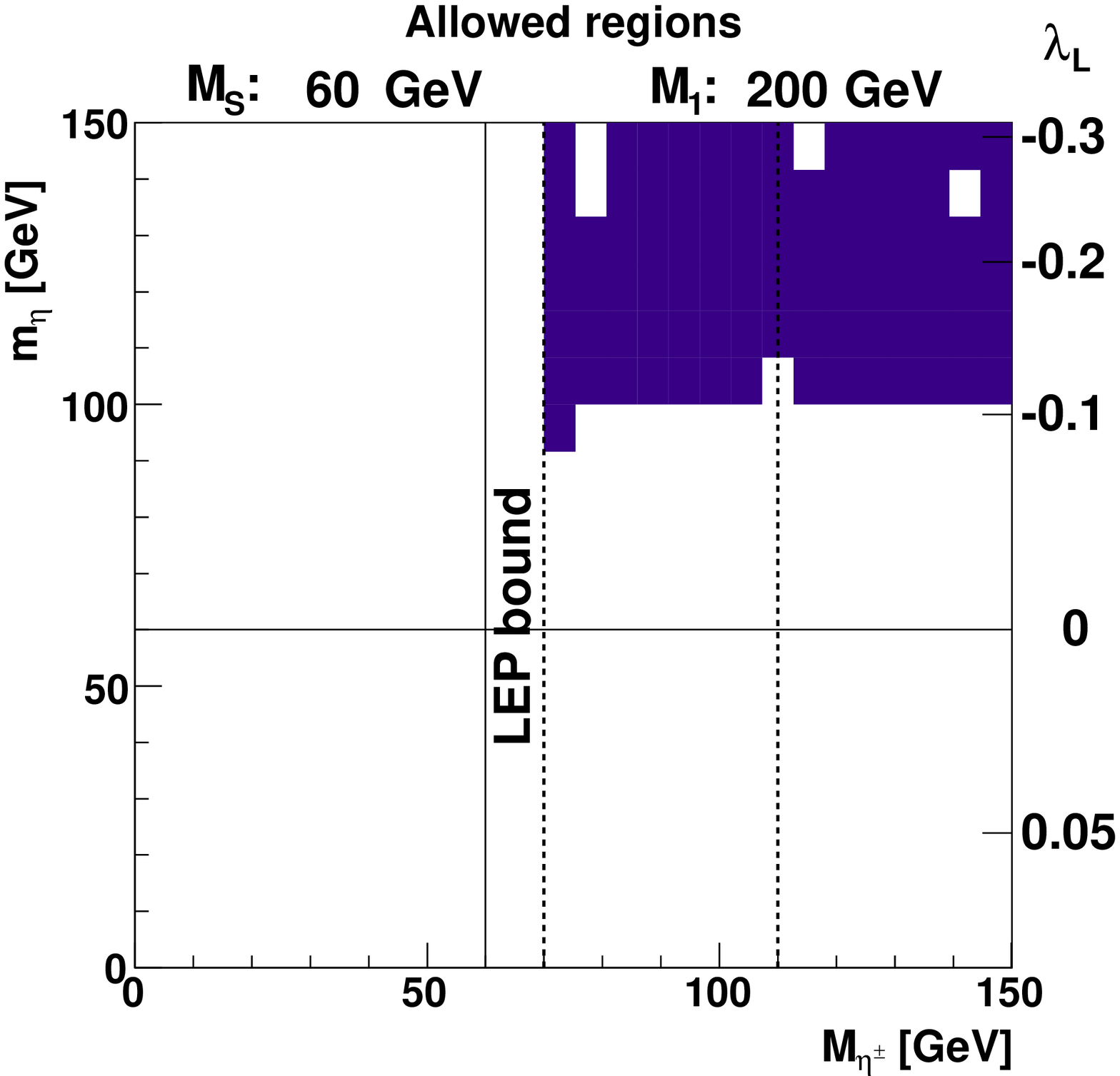}} } 
}\fi
\vspace{-6mm}
\caption{\label{med-060} Allowed regions in the $M_{\eta^\pm}-m_\eta$ plane, for DM mass $M_S=60~\text{GeV}$, with lightest Higgs mass $M_1=120~\text{GeV}$ and 200~GeV.} }

As $M_S$ is further increased, two things happen. First, as $M_S$ approaches the mass of the $W$, annihilation (in the Early Universe) via off-shell $W$'s and $Z$'s starts to play an important role, as illustrated quantitatively by Eq.~(\ref{Eq:80-annihilate}). Secondly, the lightest neutral Higgs can be produced {\it resonantly} via $SS$ annihilation. This is illustrated in Fig.~\ref{med-060}, where we consider $M_S=60~\text{GeV}$ and (left panel) $M_1=120~\text{GeV}$. Only small values of the trilinear coupling are allowed, reflected in the plane being populated by allowed solutions around $m_\eta=M_S$. (The ``holes'' are presumably due to the ``small'' number of points being scanned over.) For a larger value of $M_1$, $H_1$ is no longer produced resonantly, and a certain minimum value of $\lambda_L$ (or, equivalently, $|m_\eta^2-M_S^2|$) is required, as shown in the right panel of Fig.~\ref{med-060} for $M_1=200~\text{GeV}$.

\FIGURE[ht]{
\let\picnaturalsize=N
\def\picsize{5.5cm}
\ifx\nopictures Y\else{
\let\epsfloaded=Y
\centerline{\hspace{4mm}{\ifx\picnaturalsize N\epsfxsize \picsize\fi
\epsfbox{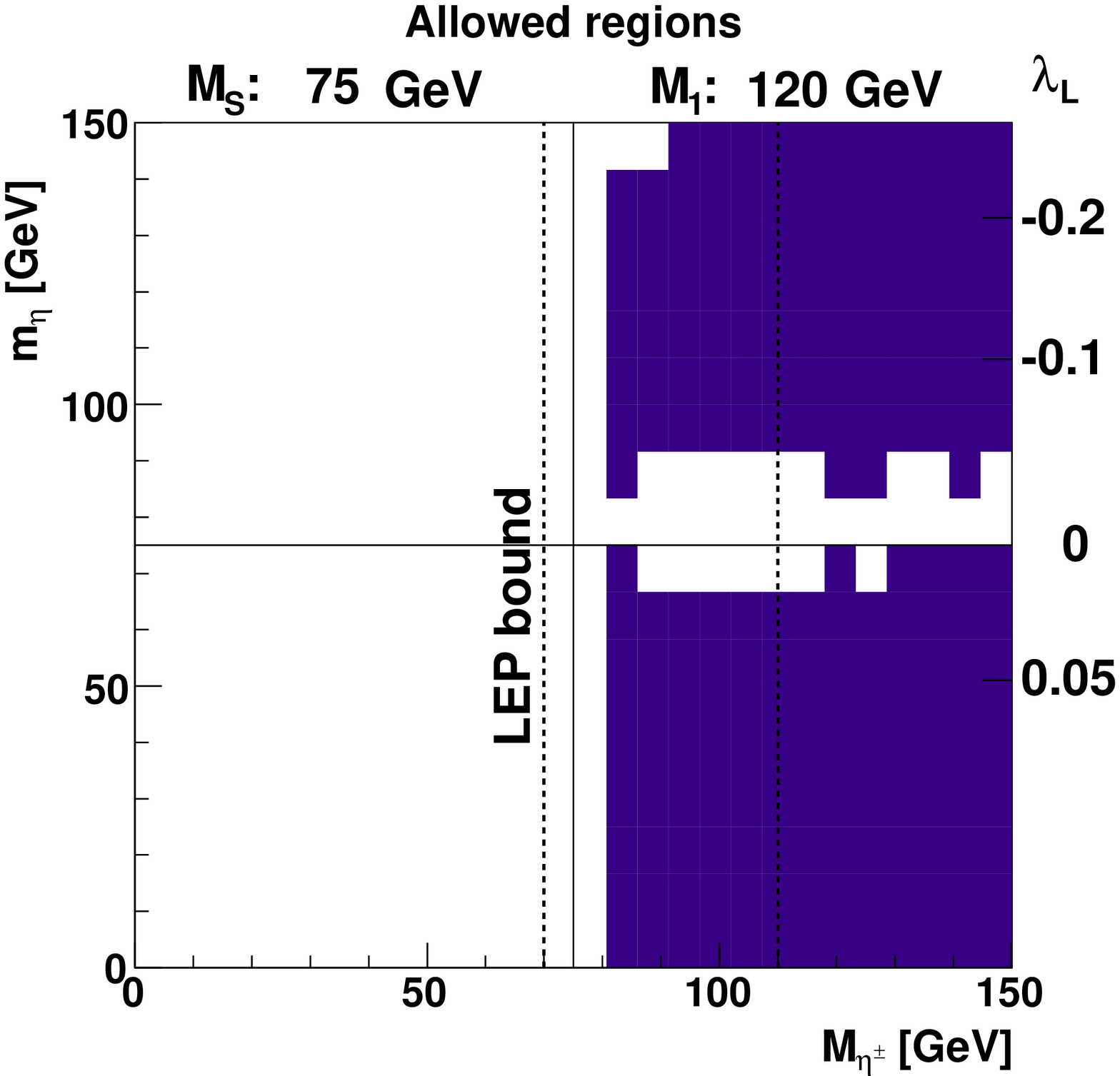}} 
\hspace{4mm}{\ifx\picnaturalsize N\epsfxsize \picsize\fi
\epsfbox{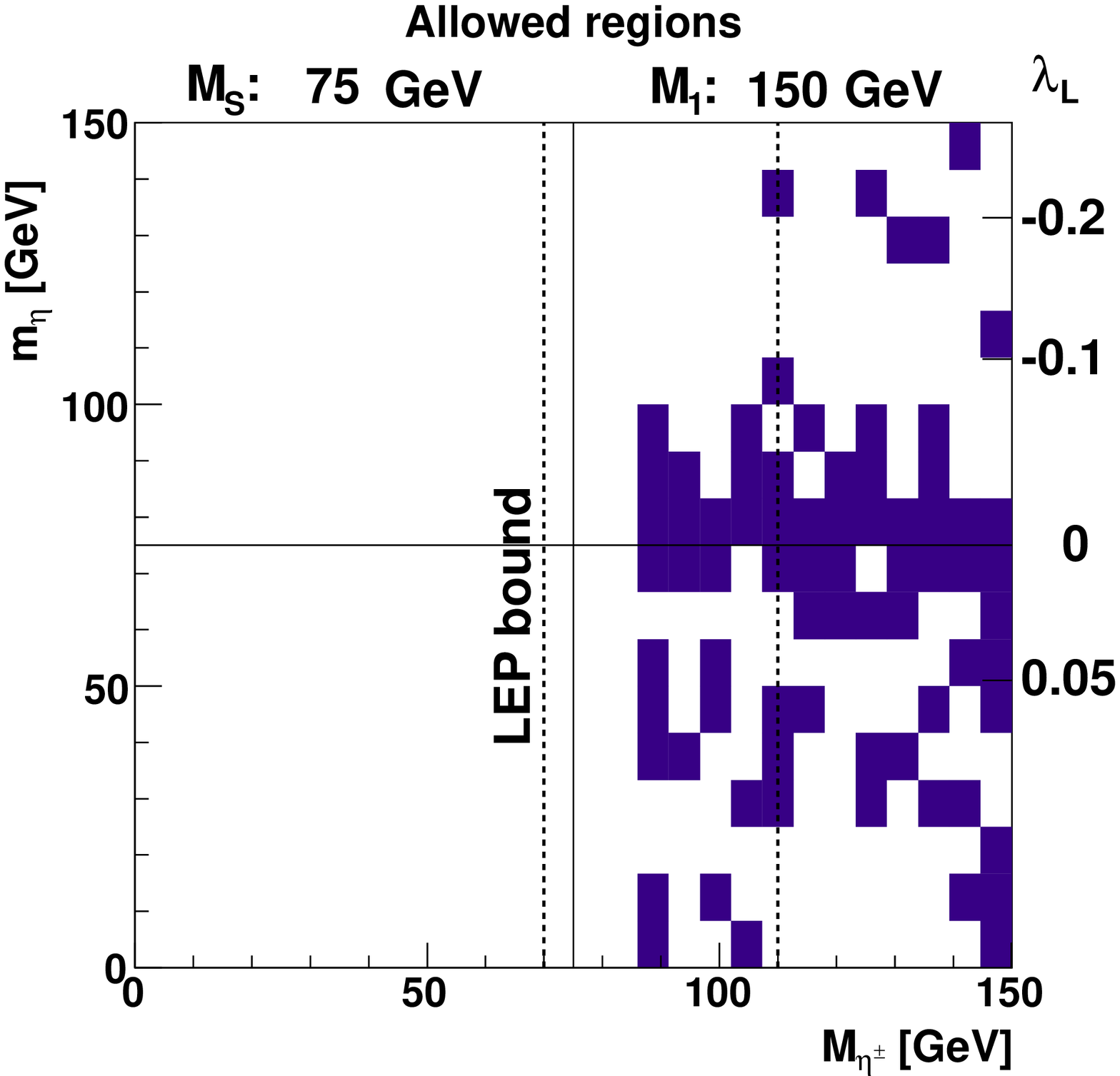}} } 
\centerline{\hspace{4mm}{\ifx\picnaturalsize N\epsfxsize \picsize\fi
\epsfbox{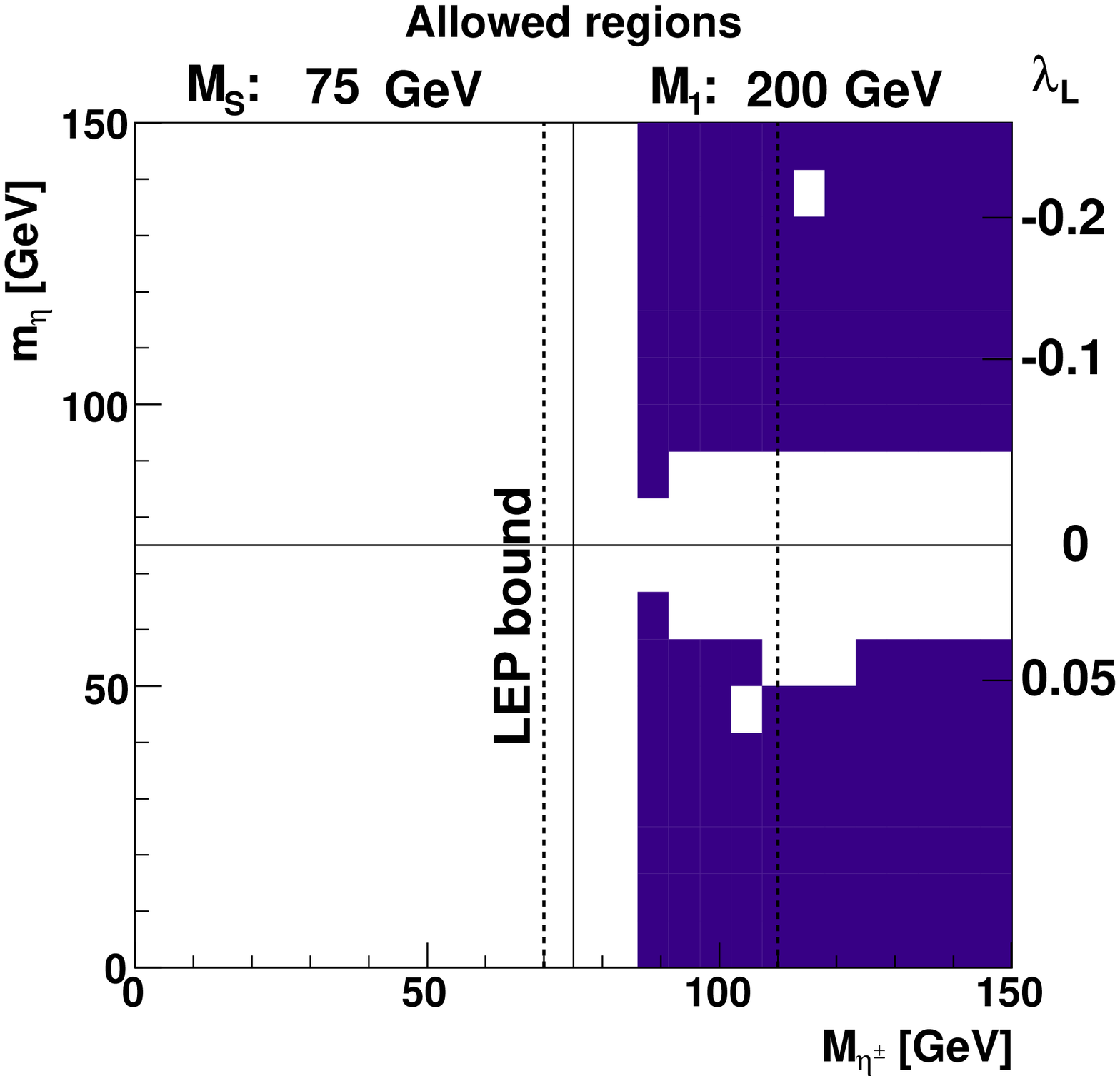}} 
\hspace{4mm}{\ifx\picnaturalsize N\epsfxsize \picsize\fi
\epsfbox{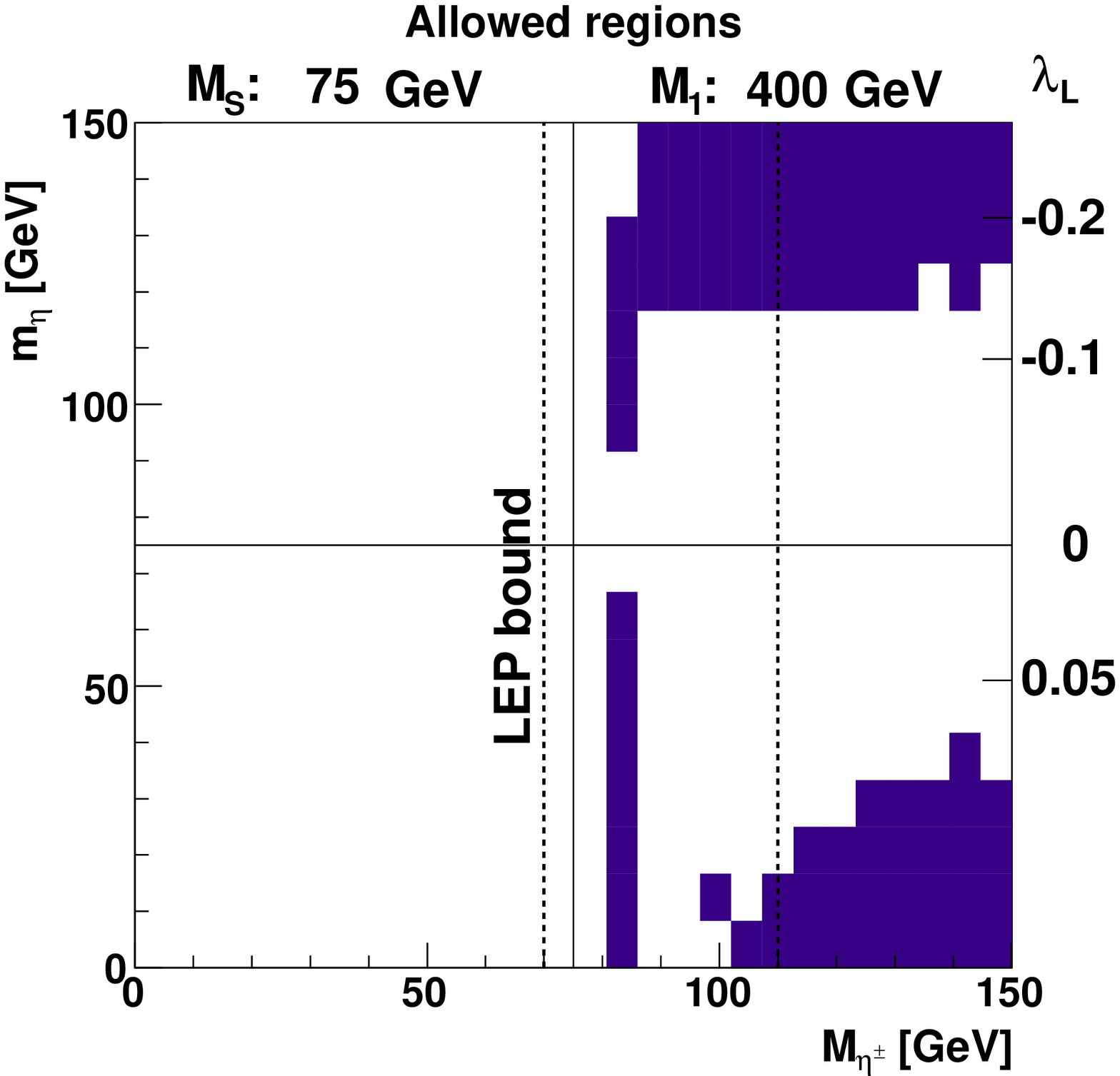}} } 
}\fi
\vspace{-6mm}
\caption{\label{med-075} Allowed regions (dark blue) in the $M_{\eta^\pm}-m_\eta$ plane, for DM mass $M_S=75~\text{GeV}$, with lightest Higgs mass $M_1=120~\text{GeV}$, 150~GeV, 200~GeV and 400~GeV.} }

\FIGURE[ht]{
\let\picnaturalsize=N
\def\picsize{5.5cm}
\ifx\nopictures Y\else{
\let\epsfloaded=Y
\centerline{\hspace{4mm}{\ifx\picnaturalsize N\epsfxsize \picsize\fi
\epsfbox{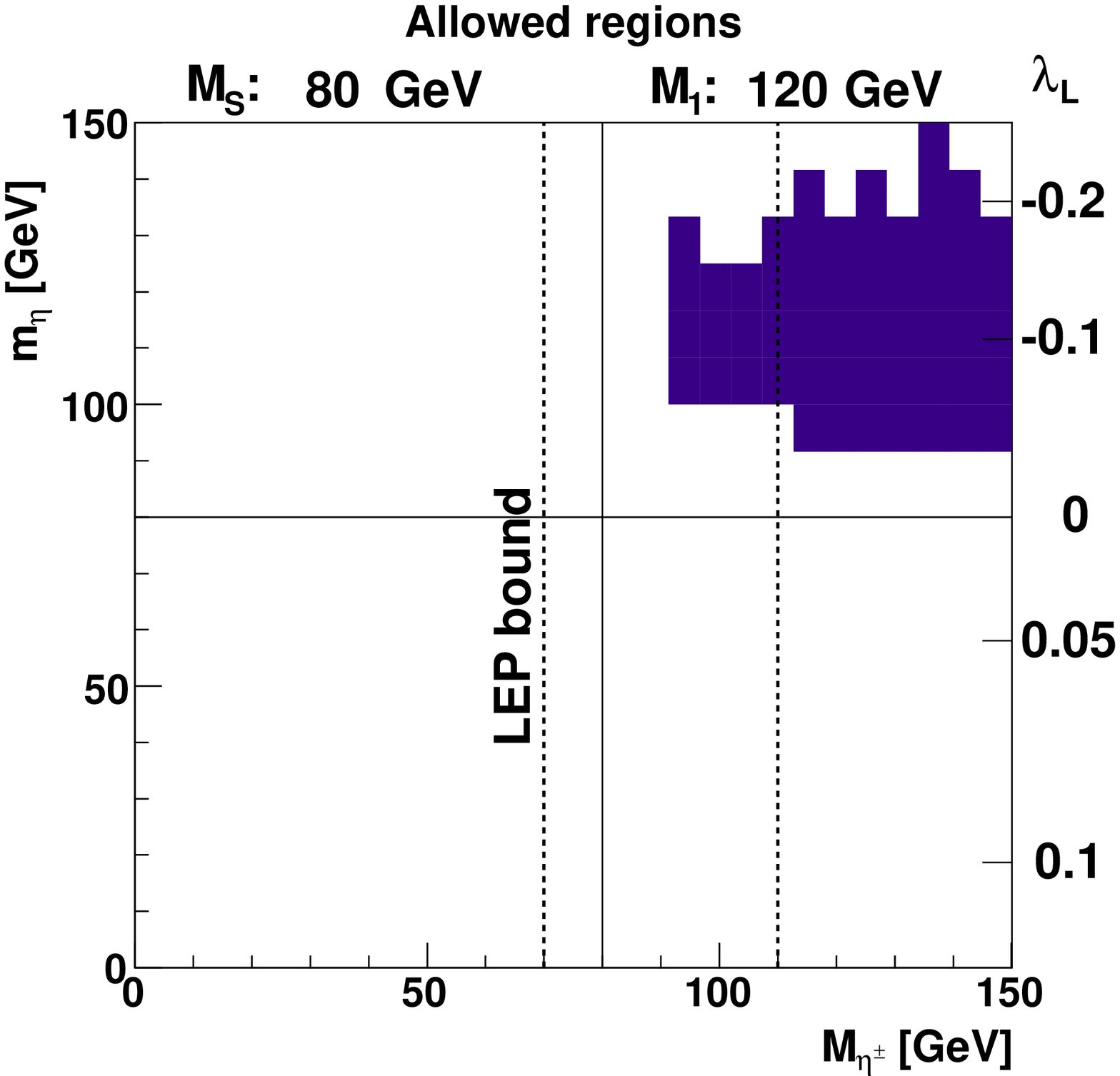}} 
\hspace{4mm}{\ifx\picnaturalsize N\epsfxsize \picsize\fi
\epsfbox{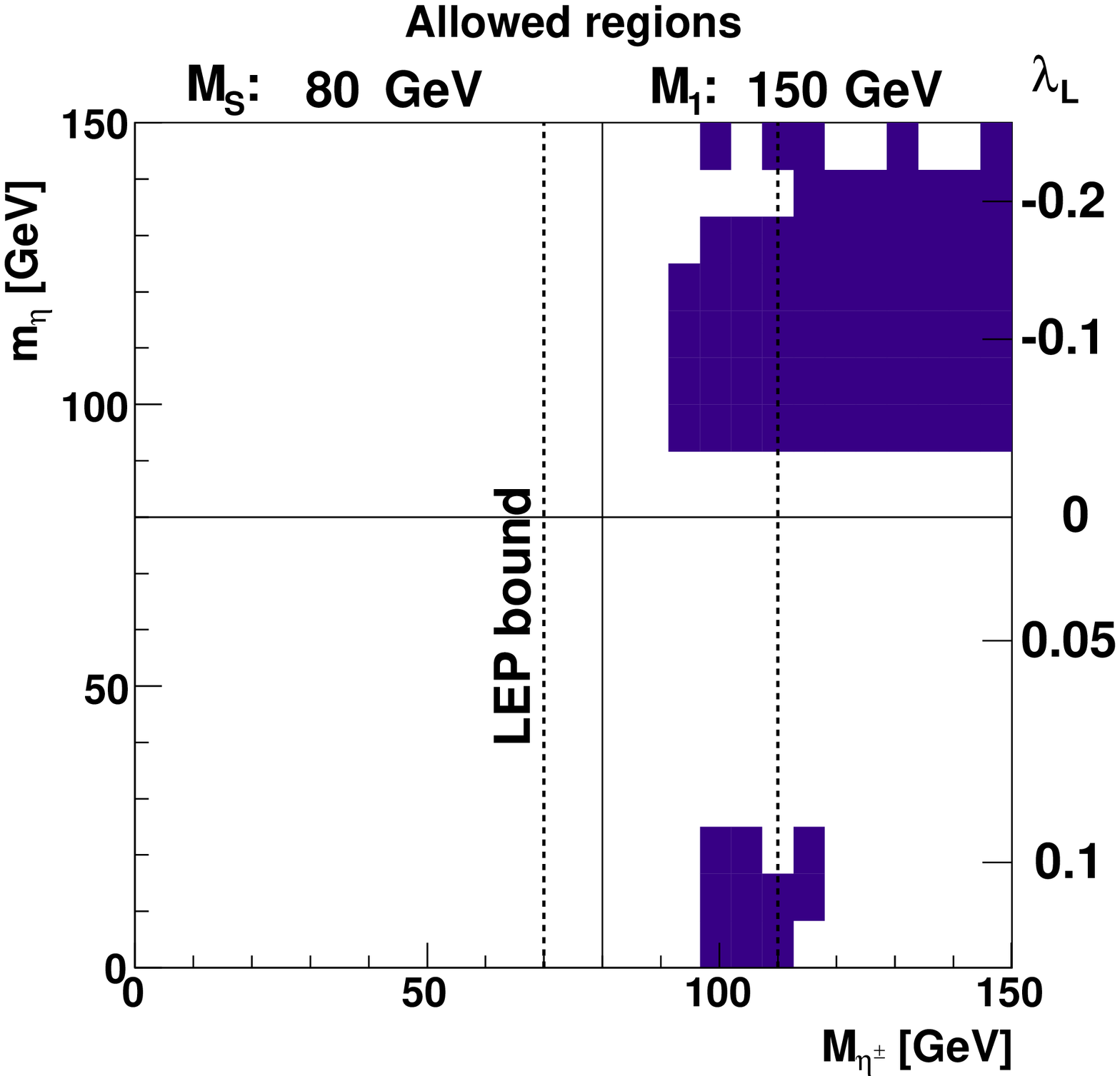}}
\hspace{4mm}{\ifx\picnaturalsize N\epsfxsize \picsize\fi
\epsfbox{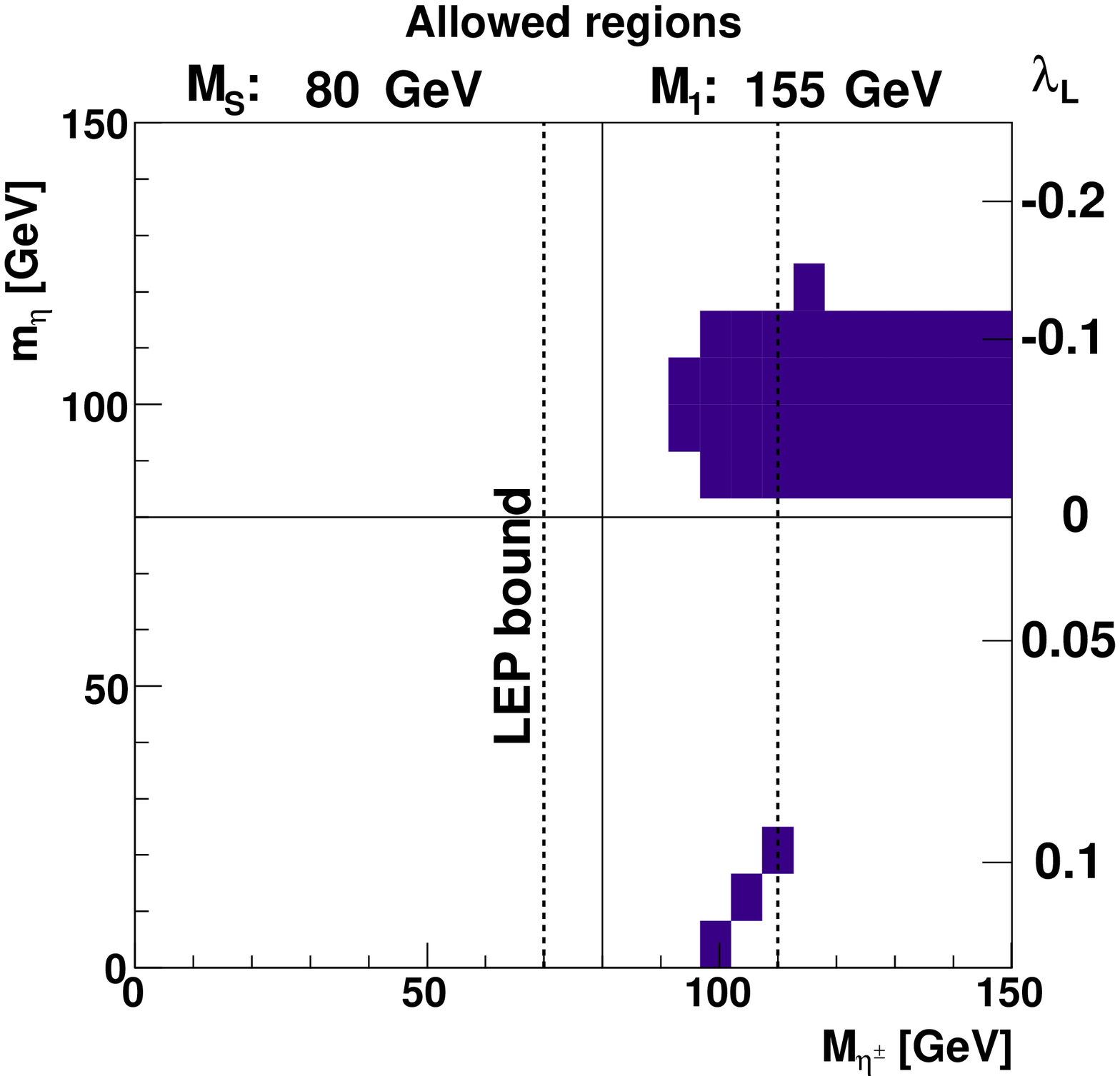}} } 
}\fi
\vspace{-6mm}
\caption{\label{med-080} Allowed regions in the $M_{\eta^\pm}-m_\eta$ plane, for DM mass $M_S=80~\text{GeV}$, with lightest Higgs mass $M_1=120$, 150~GeV,  and 155~GeV.} }

As $M_S$ reaches a value around $M_W$ or $M_Z$, annihilation becomes very easy via the $WW$ and $ZZ$ channels (the strength is given by the gauge coupling).  However, to reproduce the correct DM abundance
some annihilation must also take place via the Higgs (dominantly $H_1$) intermediate state. It is worth noticing that in order to obtain substantial contribution from that process, an increase of $M_1$ must be compensated by larger $\lambda_L$, therefore for a given $M_1$ only suitable values of $m_\eta$ and non-inert-sector parameters ($\tan\beta$, $R_{ij}$) provide an appropriate $\Omega_\text{DM}$. Some such cases are shown in Fig.~\ref{med-075}.
Of course, the annihilation via a neutral Higgs can be resonant, in which case $\lambda_L$ has to be small, as illustrated for $2M_S=M_1=150~\text{GeV}$ in the upper right panel in Fig.~\ref{med-075}.

In this region of $M_S\lsim M_W$, there is also another issue. As recently pointed out \cite{Honorez:2010re}, the annihilation could be enhanced by going via $WW^\ast$, where $W^\ast$ denotes a virtual $W$. Thus, three-body decay will set in below the two-body $WW$ threshold. However, we do not find an effect as large as reported in \cite{Honorez:2010re}. 
In this region, there are two important mechanisms: the four-point $SSWW$ gauge coupling, and the s-channel annihilation via an intermediate Higgs, the strength of which is given by $\lambda_L$ (and thus tunable via the model parameters). If the intermediate Higgs is close to its mass shell, the s-channel becomes more relevant. However in this region of the parameter space special care is required, as the Higgs resonance contribution  is very sensitive to the proper treatment of the Higgs width. This effect may be responsible for the reduced (as compared to \cite{Honorez:2010re}) effect caused by three-body final states that we have found.
As an example showing the importance of the three-body final states, we consider one of the ``good'' points, $(M_S,M_A,M_{\eta^\pm},m_\eta)=(75, 110, 86, 0)~\text{GeV}$, with $(M_1,M_2,\mu,M_{H^\pm})=(140, 300, 200, 389.7)~\text{GeV}$ and $\tan\beta=0.689$. Allowing a virtual $W$, we find a drop of $\Omega_\text{DM}$ from values around $0.11$ to below $0.07$. If we turn off the s-channel, with $\lambda_L=0$, or $m_\eta=M_S$, the corresponding values are $0.13$ and $0.05$ (a larger effect). In view of the excessive computational requirements, we have not pursued this.

For $M_S$ approaching $M_{\eta^\pm}$, as happens near the lower bound on $M_{\eta^\pm}$ ($70~\text{GeV}$), the $S$ and $\eta^\pm$ number densities in the early universe would have been similar. In this parameter region, the quartic $S\eta^\pm W^\mp\gamma$ gauge coupling becomes relevant, since the $W\gamma$ channel is kinematically open. For parameters in this region, there would be too much co-annihilation $S\eta^\pm\to W^\pm\gamma$, and $\Omega_\text{DM}$ would be too low. Thus, values of $M_{\eta^\pm}$ close to $M_S$ are not allowed. This is reflected as a forbidden band in Figs.~\ref{med-075} and \ref{med-080}. For higher values of $M_{\eta^\pm}$ (with respect to $M_S$) this is not a problem, because of the Boltzmann suppression of the $\eta^\pm$ number density.

For $M_S=75~\text{GeV}$ and $M_1=400~\text{GeV}$, there is a strip of allowed parameters around $M_{\eta^\pm}\sim85~\text{GeV}$, for almost all values of $m_\eta$ (see Fig.~\ref{med-075}, lower right panel). Along this strip, co-annihilation of $S\eta^\pm \to W^\pm\gamma$, as well as $S\eta^\pm \to W^{\pm\,\star} \to u\bar d,\; d\bar u,\; c \bar s, \; s \bar c$ play an important role. On the other hand, with such a heavy Higgs boson, the role of an intermediate $H_1$ is much reduced. Thus, for somewhat higher values of $M_{\eta^\pm}$, the co-annihilation is no longer effective (as discussed above), and the DM density would be too high.  The region of small $\lambda_L$ is thus not allowed beyond a narrow strip $M_{\eta^\pm}\gsim M_S$. 

Also, for $M_S=75~\text{GeV}$, if we relax the constraints, we find solutions at even higher values of $M_1$. For example, if we disregard the $\Delta T$ constraint, or allow a somewhat larger value of $\Omega_\text{DM}$, we also find solutions at $M_1=700~\text{GeV}$, but then at a somewhat higher value of $m_\eta$ or $\lambda_L$.

For $M_S\gsim M_W$, the annihilation $SS\to W^+W^-$ sets in, with a rate, which near threshold is controlled by 
\begin{equation} \label{Eq:SS-to-WW}
\langle\sigma\times v\rangle\propto \frac{\beta}{M_S^2},
\end{equation}
with the final-state $W^\pm$ velocity
\begin{equation}
\beta=\sqrt{1-\left(\frac{M_W}{M_S}\right)^2}.
\end{equation}
As soon as this becomes sizable, the annihilation rate exceeds the value that is compatible with $\Omega_\text{DM}$.  Apart from the region discussed in Sec.~\ref{sec:new-viable}, this annihilation mechanism thus provides an upper cut-off of the allowed region around $M_S=100~\text{GeV}$. Eventually, as we will see in Sec.~\ref{sec:high-mass}, for values of $M_S\gsim550~\text{GeV}$, the denominator in (\ref{Eq:SS-to-WW}) will bring the annihilation rate down again, to an acceptable level.

Below the cut-off, at $M_S\lsim100~\text{GeV}$, we again find an extension of the allowed range in $m_\eta$, as $M_1$ is increased towards $2M_S$, where the annihilation via a neutral Higgs boson is resonant. This is illustrated in Fig.~\ref{med-080}. For the particular case of $M_S=80~\text{GeV}$, a wider range of solutions is found for $M_1=150~\text{GeV}$ than for $M_1=120~\text{GeV}$. However, this has shrunk again at $M_1=155~\text{GeV}$, and nothing is found for $M_1=160~\text{GeV}$.
In this region of $M_S$ and $M_1$, the near-resonant annihilation via $H_1$ provides too much depletion of the DM.

\FIGURE[ht]{
\let\picnaturalsize=N
\def\picsize{5.5cm}
\ifx\nopictures Y\else{
\let\epsfloaded=Y
\centerline{\hspace{4mm}{\ifx\picnaturalsize N\epsfxsize \picsize\fi
\epsfbox{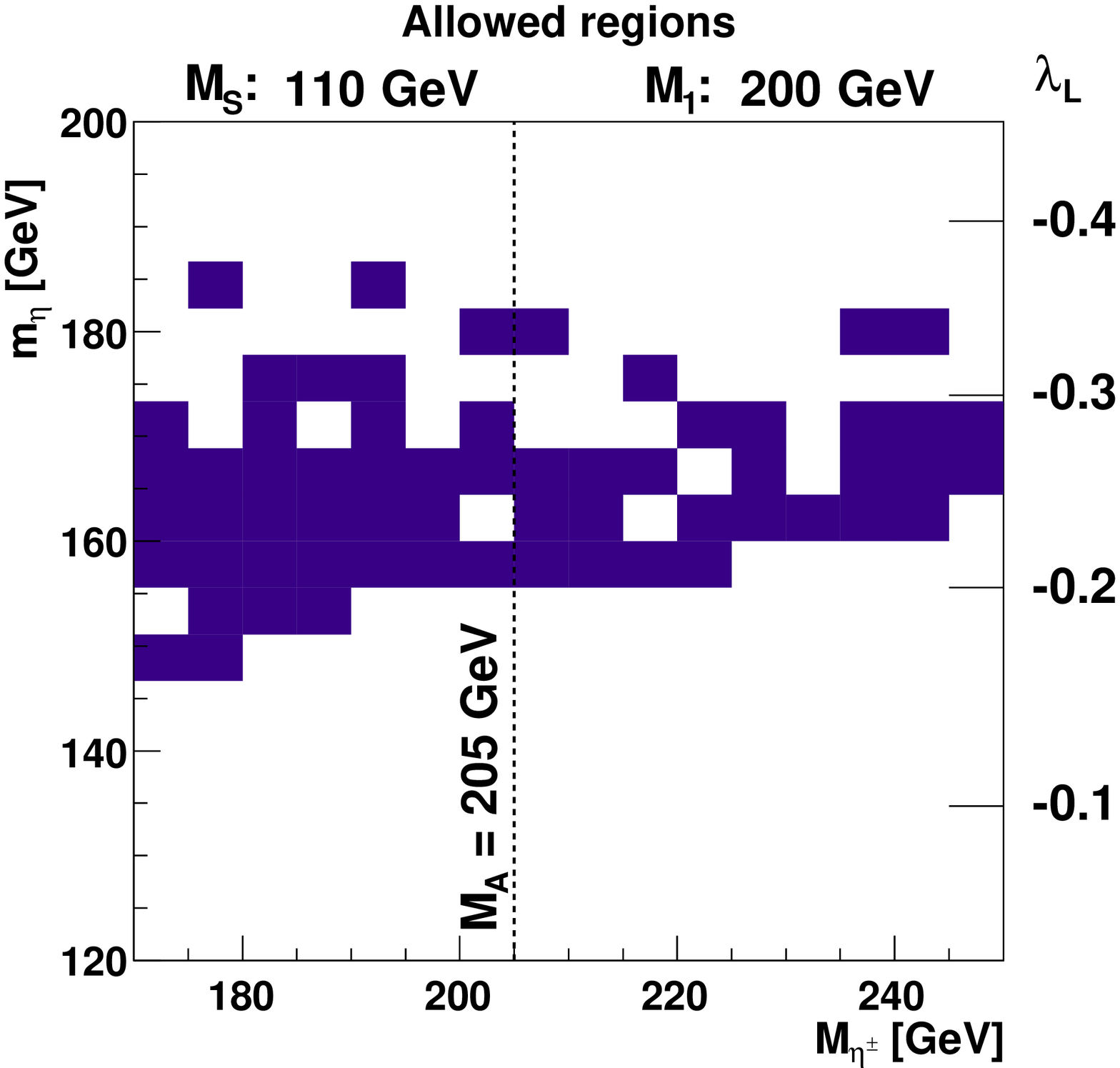}} 
\hspace{4mm}{\ifx\picnaturalsize N\epsfxsize \picsize\fi
\epsfbox{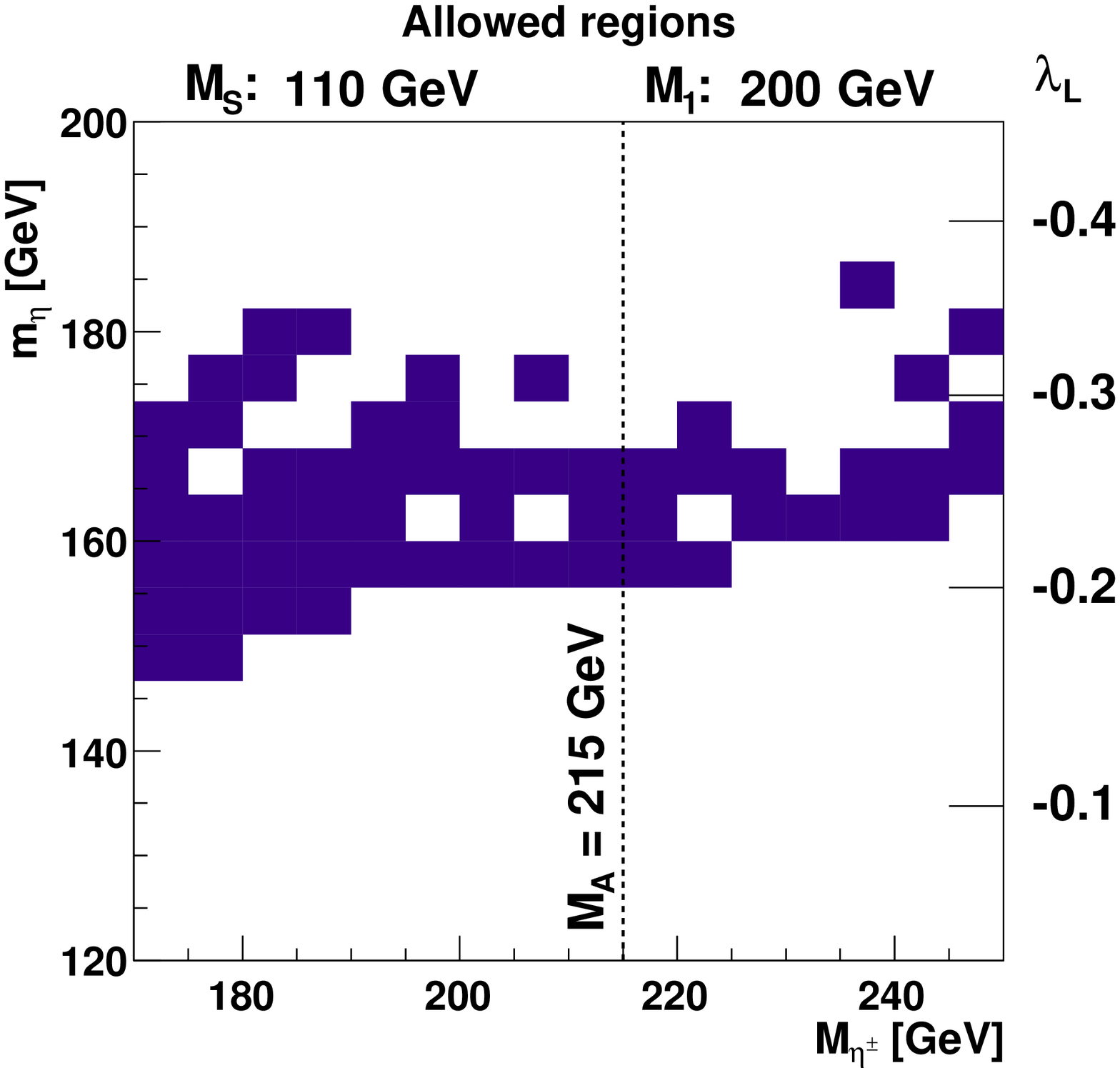}} } 
}\fi
\vspace{-6mm}
\caption{\label{good-viable} Allowed regions in the $M_{\eta^\pm}-m_\eta$ plane, for DM mass $M_S=110~\text{GeV}$, with two values of $M_A$ and lightest Higgs mass $M_1=200$.} }

\subsection{Results for ``new viable region''}
\label{sec:new-viable}

We confirm the existence of the new viable region identified by Lopez Honorez and Yamura for the IDM \cite{LopezHonorez:2010tb}. The existence of this region is due to cancellation between the four-point $SSW^+W^-$-coupling and the contribution via s-channel $H_1$ exchange, proportional to $\lambda_L$. In our notation, the cancellation condition becomes  \cite{LopezHonorez:2010tb}
\begin{equation}
\frac{1}{v^2}\left(M_S^2 - m_\eta^2\right)= \lambda_L\simeq-\frac{2}{v^2}\left[M_S^2-\left(\frac{M_1}{2}\right)^2\right],
\end{equation}
or
\begin{equation}
m_\eta^2=3M_S^2 -\half M_1^2.
\end{equation}

However, since we have adopted more tight constraints on $\Omega_\text{DM}$, we find a somewhat smaller allowed region. Examples are shown in Fig.~\ref{good-viable}, where we consider $M_S=110~\text{GeV}$ and $M_1=200~\text{GeV}$, with $M_A=205~\text{GeV}$ (left panel) and $M_A=215~\text{GeV}$ (right panel). The dominant loss mechanism here is $SS\to W^+W^-$ (four-point coupling, and $\eta^\pm$ t-channel exchange), together with a significant amount of $SS\to b\bar b$ (via $H_1$ s-channel exchange). We note a slight tendency for $|\lambda_L|$ to increase with $M_{\eta^\pm}$, indicating that the s-channel increases in significance as the t-channel decreases.

\FIGURE[ht]{
\let\picnaturalsize=N
\def\picsize{10cm}
\ifx\nopictures Y\else{
\let\epsfloaded=Y
\centerline{\hspace{4mm}{\ifx\picnaturalsize N\epsfxsize \picsize\fi
\epsfbox{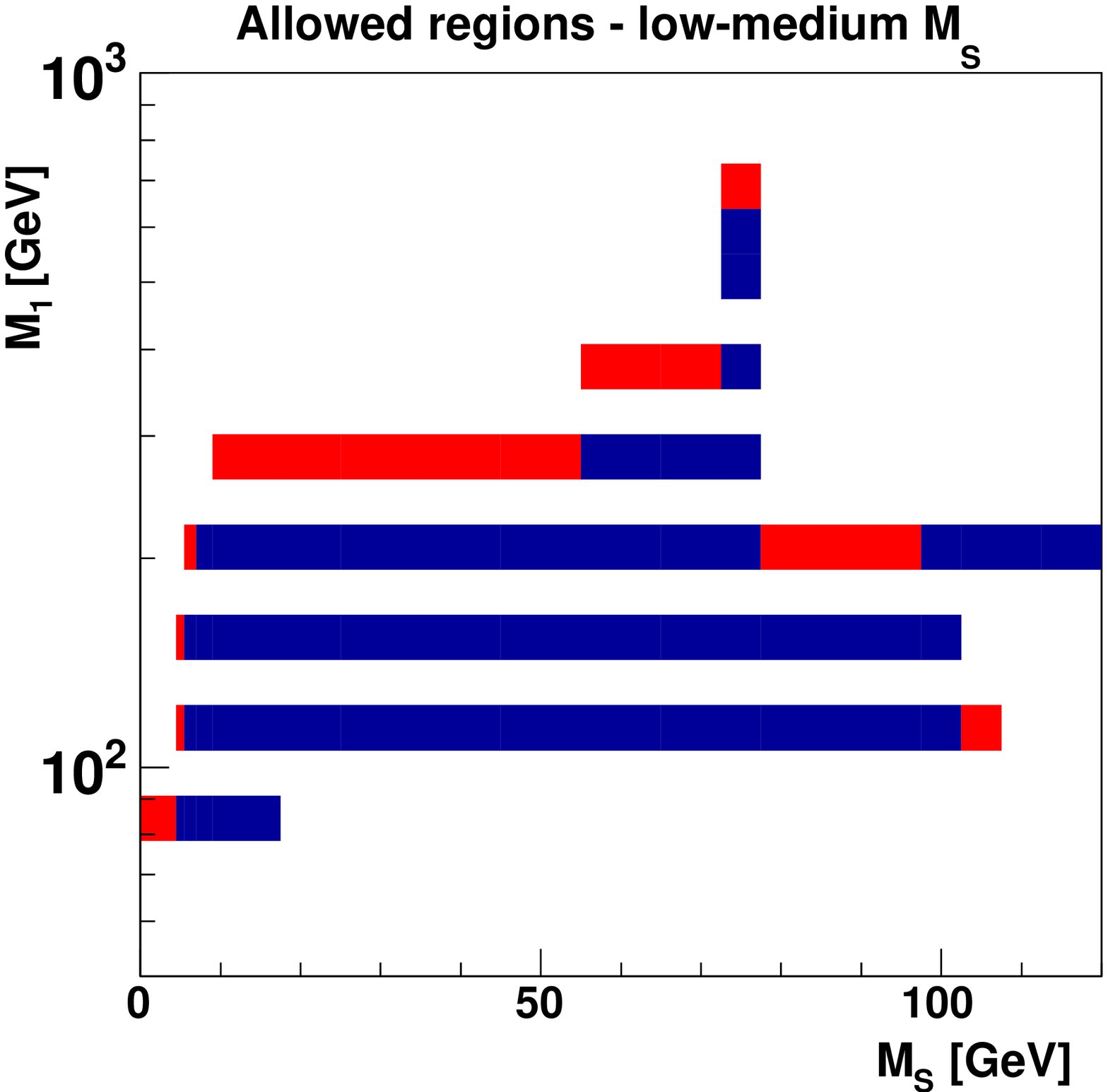}} } 
}\fi
\vspace{-6mm}
\caption{\label{ms-m1-plane} Allowed (blue) and not allowed (red) regions in the ($M_S,M_1$) plane. Discrete values of $M_1$ that are investigated, are represented as horizontal bars (note logarithmic scale in $M_1$).} }

\subsection{Summary of low-medium region}

For the low-to-medium mass region of $M_S$, an overview of allowed ranges of $M_1$ is given in Fig.~\ref{ms-m1-plane}. The model is viable from about $M_S\sim6~\text{GeV}$ up to about 120~GeV. As a default value, we have taken $M_1=120~\text{GeV}$, in the range favored by current electroweak SM fits \cite{Amsler:2008zz}. The low range of $M_S$ can be extended a bit by lowering $M_1$ to 90~GeV. More interestingly, the value of $M_1$ can be increased significantly, without any conflict with the data. It is typically restricted to $M_1\lsim300~\text{GeV}$, except for a narrow region around $M_S\sim75~\text{GeV}$, for which values up to $M_1\sim600~\text{GeV}$ are acceptable. Beyond 75~GeV, it rapidly falls again (except for the region discussed in Sec.~\ref{sec:new-viable}). This region  around $M_S=75~\text{GeV}$ is a very interesting one, since the little hierarchy can be considerably alleviated~\cite{Barbieri:2006dq}, and, as we shall see in Sec.~\ref{Sec:direct-detection}, the direct-detection cross sections are in this case within experimental reach.

The cut-offs in $M_S$ and $M_1$ can be summarized as follows:
\begin{itemize}
\item
In the low range of $M_S$ ($\sim 5~\text{GeV}$), we can extend it a bit (lower $M_S$, higher $M_1$) by allowing a {\it lower} value of $\Omega_\text{DM}$. 
\item
In the high range of $M_S$ ($\sim 100~\text{GeV}$), the cut-off is mainly due to too much DM annihilation in the Early Universe, via the gauge coupling.
\item
The upper cut-off of $M_1$ is mainly determined by the $\Omega_\text{DM}$ and $\Delta T$ constraints.
\end{itemize}

\section{High DM Mass Regime}
\label{sec:high-mass}
\setcounter{equation}{0}

We next study the model-parameter space when the DM is heavy. In analogy with the results for the simpler IDM \cite{LopezHonorez:2006gr}, solutions are found for $M_S\gsim545~\text{GeV}$. In this mass range, annihilation via a single Higgs boson is not very efficient, whereas annihilations to two gauge bosons or two Higgs bosons are relevant. There is a lower cut-off around $M_S\sim540-550~\text{GeV}$, below which the two-body annihilation to two gauge bosons, scaling like $1/M_S^2$ (see Eq.~(\ref{Eq:SS-to-WW})), is too fast to accommodate the observed value of $\Omega_\text{DM}$.

\subsection{Scanning strategy}
In this high-mass region, the scanning is done differently from that of the low-medium-mass region. The main difference is that, for a viable model, we need $M_A$, $M_{\eta^\pm}$ and $m_\eta$ all to be close to $M_S$. There are actually three reasons for this:
\begin{itemize}
\item
A significant splitting would lead to ``large'' values of $\lambda_a$, $\lambda_b$ or $\lambda_c$, and the Early-Universe annihilation to one or two Higgs bosons would be too fast, leaving too low a value for $\Omega_\text{DM}$.
\item
A significant splitting would require ``large'' values of $\lambda_a$, $\lambda_b$ or $\lambda_c$, and positivity or unitarity would be violated.
\item
A significant splitting would lead to a value for $\Delta T$ in violation of the LEP data.
\end{itemize}
The former constraints are stronger, such that the $\Delta T$ constraint has a negligible impact.

Unless otherwise specified, in this section we consider $M_A=M_S+1~\text{GeV}$. However, as we will discuss in Sec.~\ref{Sec:LHC}, this splitting could be much smaller.

\FIGURE[ht]{
\let\picnaturalsize=N
\def\picsize{5.4cm}
\vspace{-6mm}
\ifx\nopictures Y\else{
\let\epsfloaded=Y
\centerline{\hspace{4mm}{\ifx\picnaturalsize N\epsfxsize \picsize\fi
\epsfbox{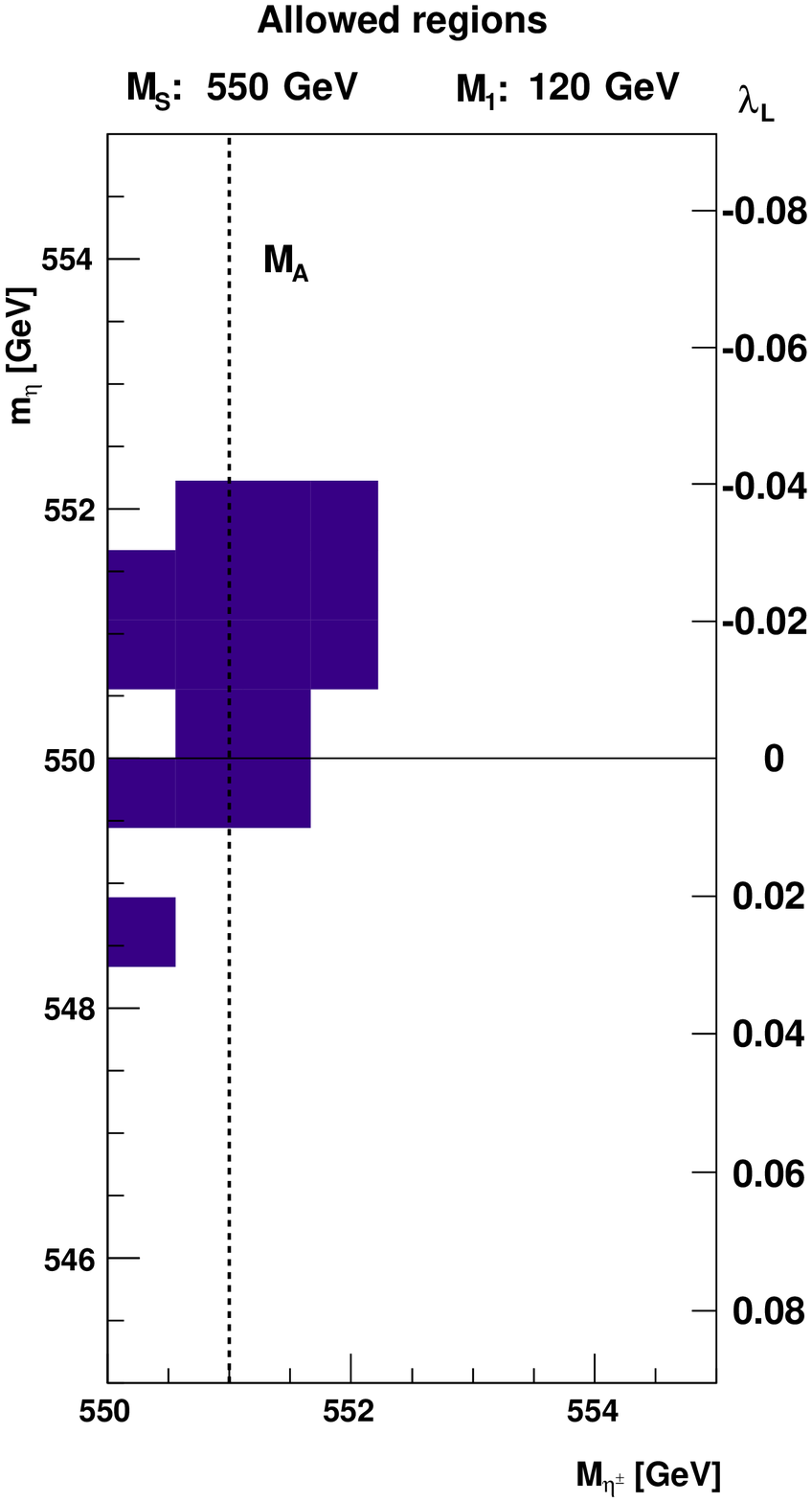}} 
\hspace{4mm}{\ifx\picnaturalsize N\epsfxsize \picsize\fi
\epsfbox{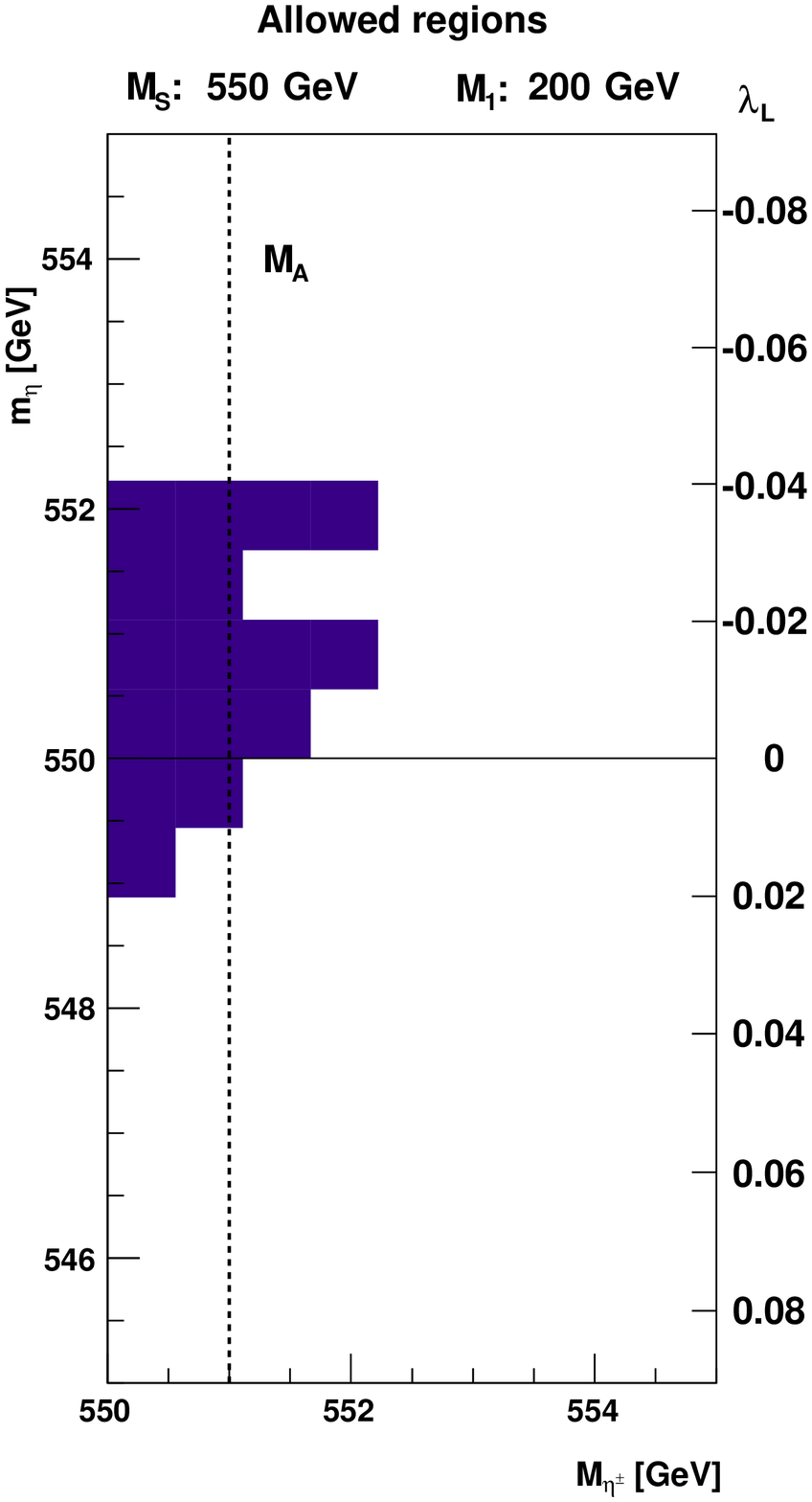}} } 
}\fi
\vspace{-8mm}
\caption{\label{hi=0550} Allowed regions (dark blue) in the $M_{\eta^\pm}-m_\eta$ plane, for DM mass $M_S=550~\text{GeV}$, $M_A=551~\text{GeV}$ with lightest Higgs mass $M_1=120~\text{GeV}$ and 200~GeV. The thin solid line indicates $m_\eta=M_S$ (and $\lambda_L=0$), whereas the dashed line gives $M_{\eta^\pm}=M_A$.} }

\subsection{Results}
Selected results on allowed regions in the $M_{\eta^\pm}$--$m_\eta$ plane are displayed in Figs.~\ref{hi=0550}--\ref{hi=3000}. As indicated above, in this high-mass region, all inert-sector masses are rather degenerate, including the soft-mass parameter $m_\eta$ (implying that $\lambda_L$ is small). However, the way a correct DM density is obtained, is a bit different in the lower end of this high-mass region, where $M_S\sim550~\text{GeV}$, from that of the higher-mass region, exemplified by $M_S=3000~\text{GeV}$ in Fig.~\ref{hi=3000}.

Figure~\ref{hi=0550} is devoted to $M_S=550~\text{GeV}$. This is representative of the lower allowed value (in this high-mass region). Characteristic of this region is the annihilation via four-point gauge couplings, as illustrated by Eq.~(\ref{Eq:low-edge-of-high-region}). These ``loss'' mechanisms involve not only $SS$ annihilation, but also $\eta^+\eta^-$, $AA$, $S\eta^\pm$ and $A\eta^\pm$ annihilations to gauge bosons. The cut-off around $540-550~\text{GeV}$ is due to the too high annihilation rate (at lower masses) \cite{Cirelli:2005uq}, scaling by Eq.~(\ref{Eq:SS-to-WW}).

\FIGURE[ht]{
\let\picnaturalsize=N
\def\picsize{5.4cm}
\vspace{-6mm}
\ifx\nopictures Y\else{
\let\epsfloaded=Y
\centerline{\hspace{4mm}{\ifx\picnaturalsize N\epsfxsize \picsize\fi
\epsfbox{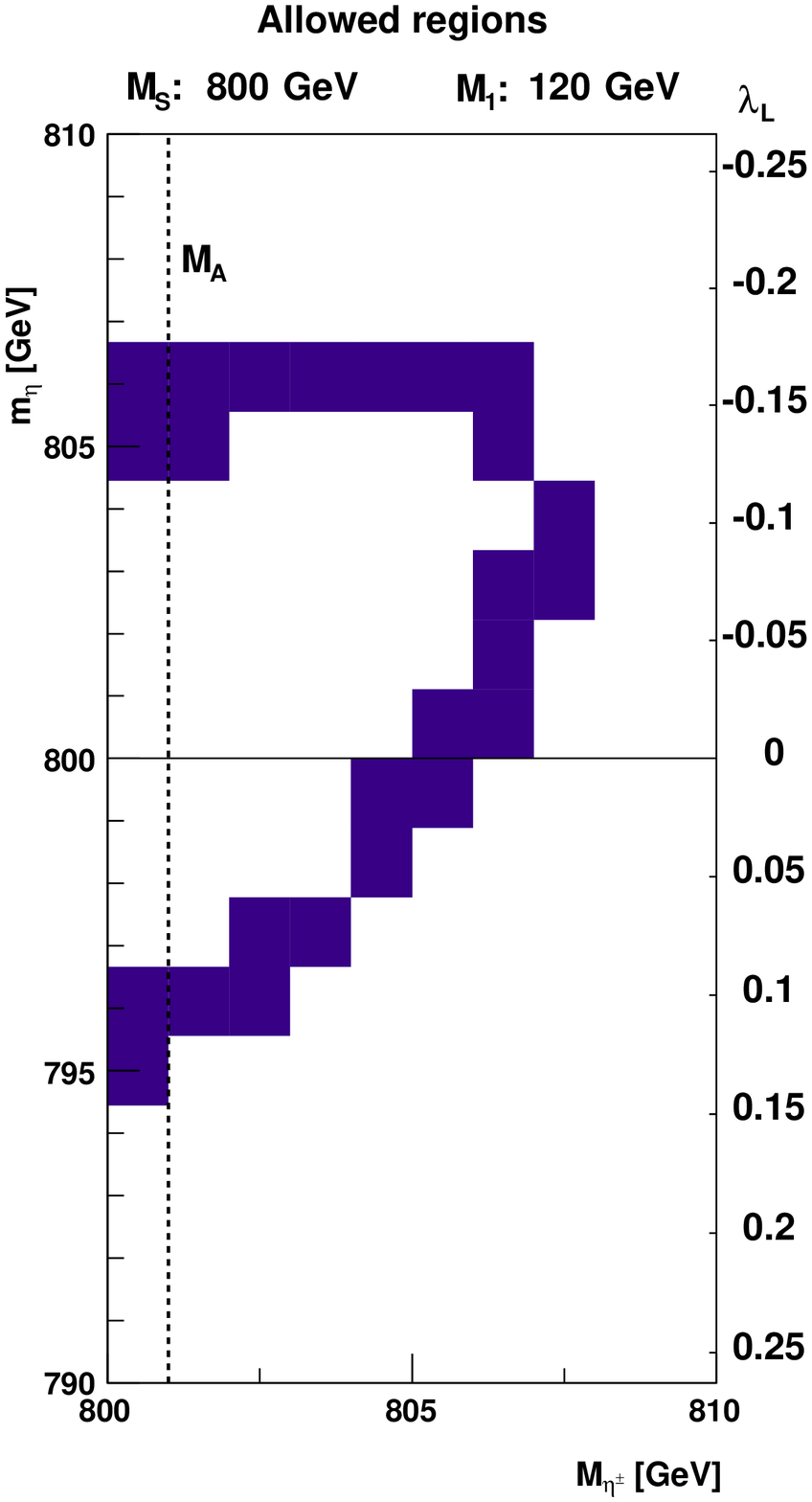}} 
\hspace{4mm}{\ifx\picnaturalsize N\epsfxsize \picsize\fi
\epsfbox{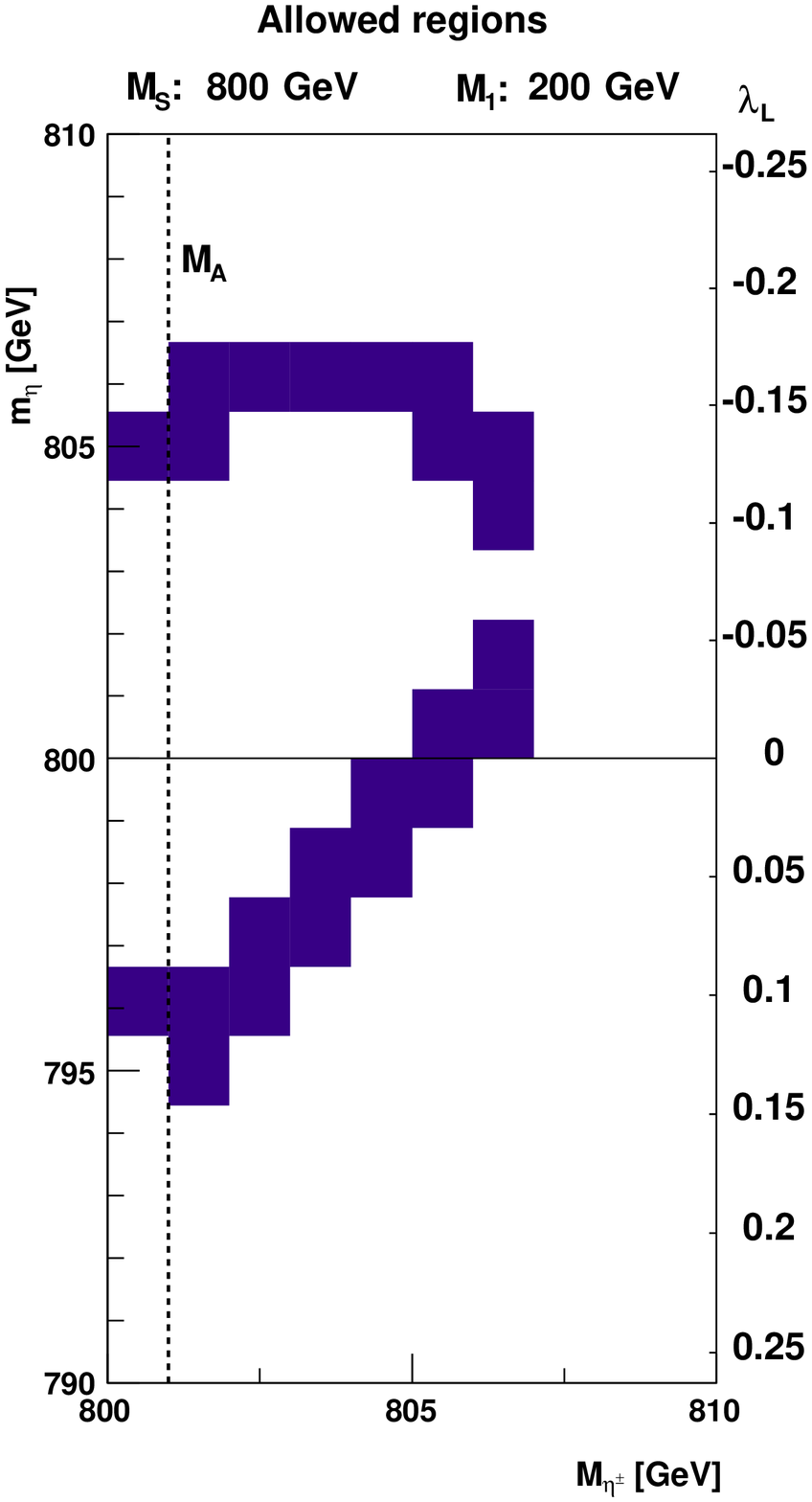}} } 
}\fi
\vspace{-8mm}
\caption{\label{hi=0800} Similar to Fig.~\ref{hi=0550}, for $M_S=800~\text{GeV}$.} }

\FIGURE[ht]{
\let\picnaturalsize=N
\def\picsize{5.4cm}
\vspace{-6mm}
\ifx\nopictures Y\else{
\let\epsfloaded=Y
\centerline{\hspace{4mm}{\ifx\picnaturalsize N\epsfxsize \picsize\fi
\epsfbox{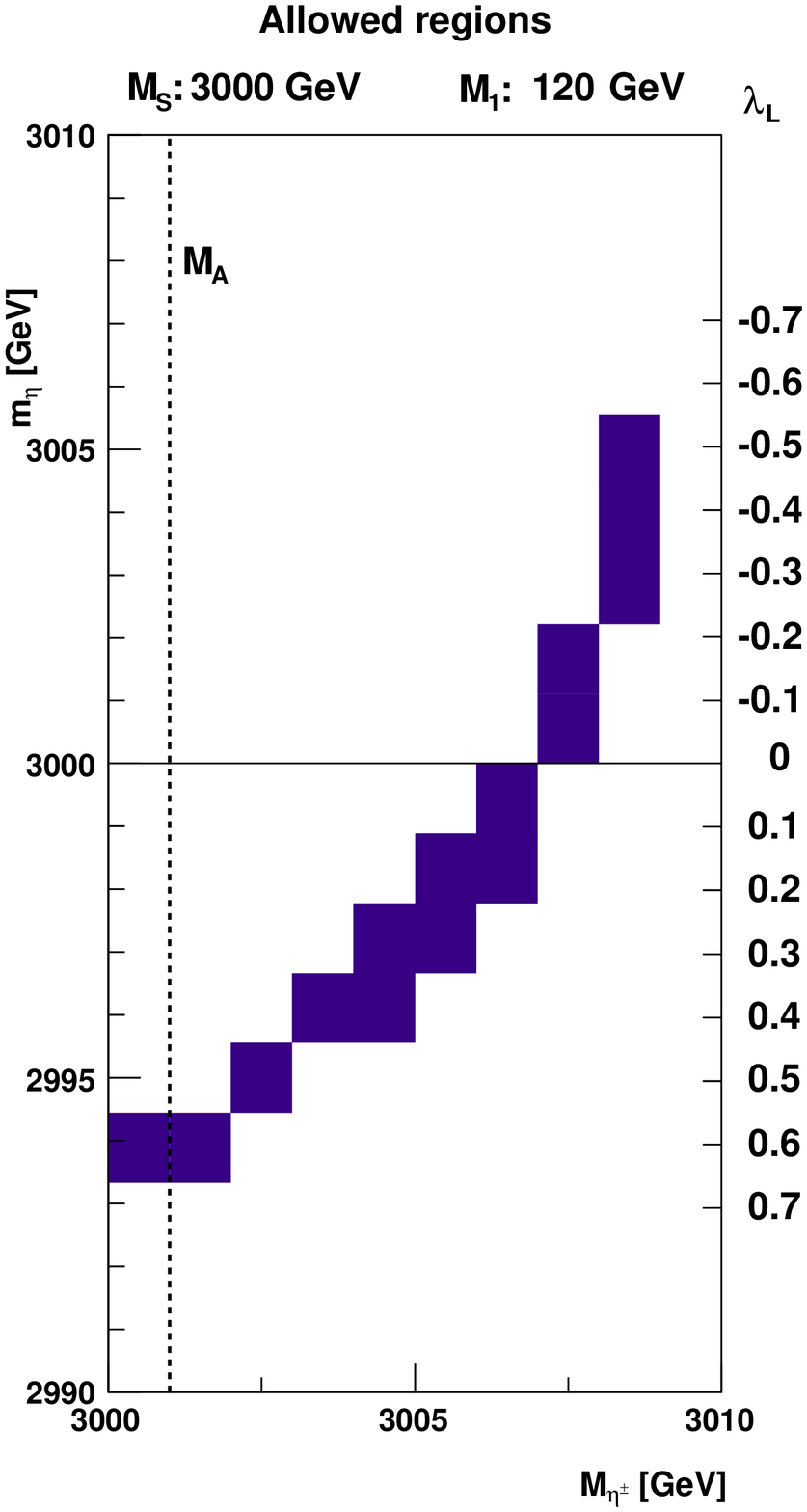}} 
\hspace{4mm}{\ifx\picnaturalsize N\epsfxsize \picsize\fi
\epsfbox{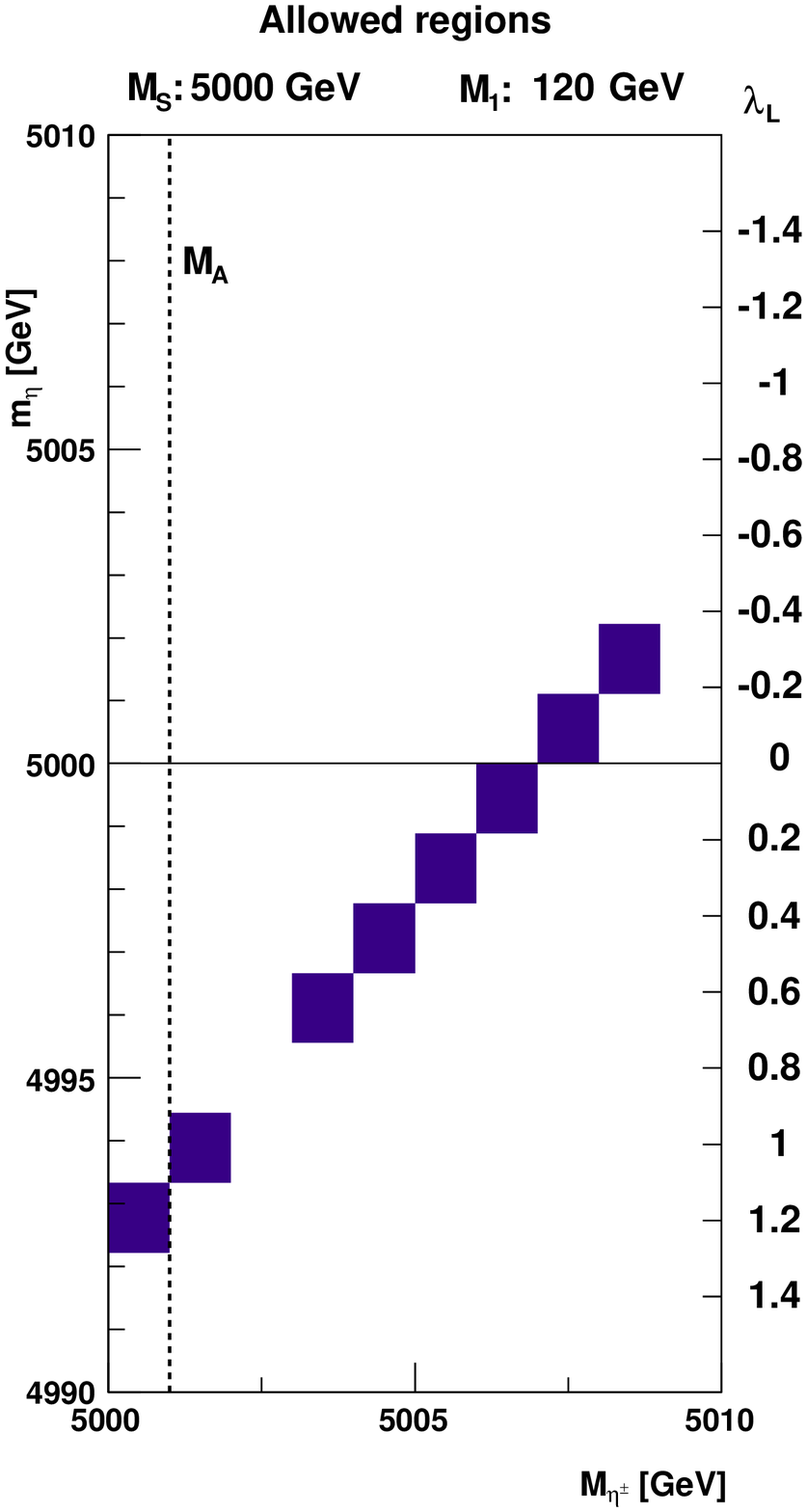}} } 
}\fi
\vspace{-8mm}
\caption{\label{hi=3000} Similar to Fig.~\ref{hi=0550}, for $M_S=3000~\text{GeV}$ and 5000~GeV, both with $M_1=120~\text{GeV}$.} }

As $M_S$ is increased to higher values, the annihilation to gauge bosons will fall off, but annihilation to neutral and charged Higgs bosons start playing an important role, as shown by Eq.~(\ref{Eq:high-edge-of-high-region}) for the case of $M_S=3000~\text{GeV}$ and $M_1=120~\text{GeV}$. These annihilations are also due to four-point couplings, but originating from the potential, rather than being gauge couplings. Thus, their strengths can be adjusted by tuning the mass splittings in the inert sector. 

In Fig.~\ref{hi=0800} we display allowed regions in the $M_\eta^{\pm}$--$m_\eta$ plane, for $M_S=800~\text{GeV}$. Compared to the case of $M_S=550~\text{GeV}$, an allowed ``ear''-shaped region has now developed. The interior is characterized by a high degree of degeneracy among the inert-sector mass parameters, which in turn leads to values of $\lambda_L$ that are too small for an efficient annihilation of dark matter (or its partners) to final-state Higgs particles in the early universe, the value for $\Omega_\text{DM}$ would become too high. In fact, the dependence of $\Omega_\text{DM}$ on these couplings is quite steep: for the case $(M_S,M_1)=(800,120)~\text{GeV}$ (left panel of Fig.~\ref{hi=0800}) and the point $(M_{\eta^\pm},m_\eta)=(M_A,M_S)$, the value of $\Omega_\text{DM}$ is too high by a factor of two. On the other hand, outside the ``ear'', some $\lambda$'s will be large, the annihilation is too fast, and $\Omega_\text{DM}$ too small.

As a final example, we show in Fig.~\ref{hi=3000} the cases of $M_S=3000~\text{GeV}$ and $M_S=5000~\text{GeV}$. Here, only the lower part of the ``ear'' is allowed. The upper part is forbidden, mostly due to the unitarity constraint.

\section{CP violation}
\label{Sec-CPV}

In order to illustrate the amount of CP violation which is available in the model
we will plot so-called weak-basis-transformation invariants that have non-zero
imaginary part. The benefit of studying invariants relies on the fact that
they provide a {\it measure} of CP violation since  any CP-violating observable must be
a linear combination of the invariants (or their higher odd powers). 
The issue of the invariants was extensively discussed in the 
literature \cite{Branco:1999fs,Davidson:2005cw,Gunion:2005ja,Lavoura:1994fv,Botella:1994cs,Branco:2005em} in the context of the 2DHM. According to Gunion and Haber, there are three independent invariants $J_{1,2,3}$ that are sufficient to describe any CP-violating
phenomenon which has its roots in the scalar potential (if Yukawa couplings are neglected). However, in the presence of three Higgs doublets the situation is much more involved and a complete study has
not been performed up to date; for the existing attempts, see \cite{Lavoura:1994fv,Botella:1994cs,Branco:2005em}. In order to discuss the
invariants it is useful to write the potential in the following compact manner~\cite{Branco:1999fs,Davidson:2005cw}
\begin{equation}
V(\Phi_1,\Phi_2,\Phi_3)= Y_{a\bar b} \Phi_{\bar a}^\dagger \Phi_b +
\frac12 Z_{a\bar{b}c\bar{d}}(\Phi_{\bar a}^\dagger \Phi_b)(\Phi_{\bar c}^\dagger \Phi_d),
\label{comp_pot}
\end{equation}
where $\Phi_3\equiv \eta$, and $Y_{a\bar b}$ and $Z_{a\bar{b}c\bar{d}}$ specify mass terms and quartic couplings, respectively.
Formulae for $Y_{a\bar b}$ and $Z_{a\bar{b}c\bar{d}}$ in terms of the standard potential (\ref{Eq:fullpot}) parameters  are given in Appendix~B in (\ref{param}).

In general, for three doublets one should expect more invariants, however here for illustration, we discuss only $J_{1,2,3}$ defined in~\cite{Gunion:2005ja} for the two-doublet case.  We generalize them in a very straightforward manner just by extending
the range for indices from $1,2$ to $1,2,3$:
\begin{align}
J_1&=\hat{v}_{\bar{a}}^*\hat{v}_{\bar{e}}^*Z_{a\bar{b}e\bar{f}}Z_{b\bar{d}}^{(1)}\hat{v}_d\hat{v}_f, \nonumber\\
J_2&=\hat{v}_{\bar{b}}^*\hat{v}_{\bar{c}}^*\hat{v}_{\bar{g}}^*\hat{v}_{\bar{p}}^*Z_{b\bar{e}g\bar{h}}Z_{c\bar{f}p\bar{r}}Z_{e\bar{a}f\bar{d}}\hat{v}_a\hat{v}_d\hat{v}_h\hat{v}_r, \nonumber\\
J_3&=\hat{v}_{\bar{b}}^*\hat{v}_{\bar{c}}^*Z_{b\bar{e}}^{(1)}Z_{c\bar{f}}^{(1)}Z_{e\bar{a}f\bar{d}}\hat{v}_a\hat{v}_d.
\end{align}
Of course, there exist additional invariants, however their determination is not necessary here.
We have calculated the three invariants in the basis adopted in this paper (specified by the 
vacuum expectation values) for the general potential (\ref{Eq:fullpot}). The results are 
complicated quadratic and cubic polynomials in quartic coupling constants, shown in Appendix B,
eqs.~(\ref{invJ1})--(\ref{invJ3}). However, if the dark democracy (\ref{Eq:DarkDemocracy}) is imposed
the results simplify considerably:
\begin{align}
\Im J_1&=-\frac{v_1^2v_2^2}{v^4}(\lambda_1-\lambda_2)\Im \lambda_5,
\label{Eq:ImJ_1}\\
\Im J_2&=-\frac{v_1^2v_2^2}{v^8}
\left[\left((\lambda_1-\lambda_3-\lambda_4)^2-|\lambda_5|^2\right) v_1^4
+2(\lambda_1-\lambda_2) \Re \lambda_5 v_1^2v_2^2\right. \nonumber\\
&\hspace*{1.5cm}\left.
-\left((\lambda_2-\lambda_3-\lambda_4)^2-|\lambda_5|^2\right) v_2^4\right]
\Im \lambda_5 \nonumber\\
&\hspace*{0.5cm}
+\frac{2v_1^4v_2^2v_3^2}{v^8}(\lambda_3+\lambda_4+\Re\lambda_5-\lambda_1)\lambda_c\Im \lambda_5 \nonumber\\
&\hspace*{0.5cm}
-\frac{2v_1^2v_2^4v_3^2}{v^8}(\lambda_3+\lambda_4+\Re\lambda_5-\lambda_2)\lambda_c\Im \lambda_5,
\label{Eq:ImJ_2}\\
\Im J_3&=\frac{v_1^2v_2^2}{v^4}(\lambda_1-\lambda_2)
(\lambda_1+\lambda_2+2\lambda_4+2\lambda_b)\Im \lambda_5.
\label{Eq:ImJ_3}
\end{align}
When finally we adopt the fact that $Z_2'$ is preserved by the vacuum (so $v_3=0$) then we obtain
\begin{align}
\Im J_1&=-\frac{v_1^2v_2^2}{v^4}(\lambda_1-\lambda_2)\Im \lambda_5,\\
\Im J_2&=-\frac{v_1^2v_2^2}{v^8}
\left[\left((\lambda_1-\lambda_3-\lambda_4)^2-|\lambda_5|^2\right) v_1^4
+2(\lambda_1-\lambda_2) \Re \lambda_5 v_1^2v_2^2\right.\nonumber\\
&\hspace*{1.5cm}\left.
-\left((\lambda_2-\lambda_3-\lambda_4)^2-|\lambda_5|^2\right) v_2^4\right]
\Im \lambda_5,\\
\Im J_3&=\frac{v_1^2v_2^2}{v^4}(\lambda_1-\lambda_2)
(\lambda_1+\lambda_2+2\lambda_4+2\lambda_b)\Im \lambda_5.
\label{inv3}
\end{align}
Note that, since $\lambda_{a,b,c}$ are real and the inert potential $V_3(\eta)$ is
CP-conserving all
the invariants are proportional to the same CP-violating parameter $\Im{\lambda_5}$
as in the case of the 2HDM. In addition the dark democracy and $\langle\eta\rangle=0$ imply
that $\Im J_{1,2}$ are identical to the corresponding invariants in the 2HDM, whereas $\Im J_3$ differs by a term proportional to $\lambda_b$, one of the quartic couplings between the non-inert and the inert sectors.

For $M_S=75~\text{GeV}$ and $M_{\eta^\pm}=90~\text{GeV}$ (and representative values for the other mass parameters), we show in Fig.~\ref{Im-J_i-075-090} the imaginary parts of $J_{1,2,3}$ for parameters which are consistent with both experimental and theoretical constraints described in Sec.~\ref{sect:constraints}. Note that only low values of $\tan\beta$ are allowed. Here, averages over sets of $\alpha$'s are shown. It turns out that for the parameters adopted in Fig.~\ref{Im-J_i-075-090}, the contribution of $\lambda_b$ is small, of the order of $\sim 5\%$. In the heavy DM case, where a high degree of degeneracy between inert masses is needed, the contribution from $\lambda_b$ is negligible. 

\FIGURE[ht]{
\let\picnaturalsize=N
\def\picsize{15cm}
\ifx\nopictures Y\else{
\let\epsfloaded=Y
\centerline{\hspace{4mm}{\ifx\picnaturalsize N\epsfxsize \picsize\fi
\epsfbox{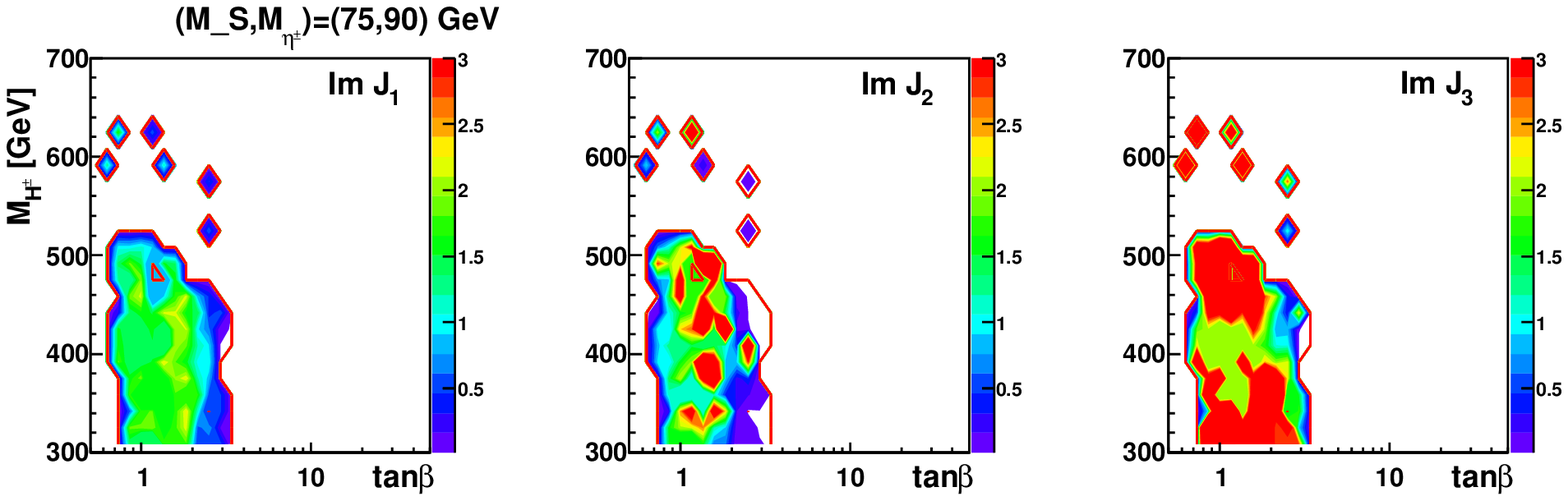}} } 
}\fi
\vspace{-6mm}
\caption{\label{Im-J_i-075-090} Contour plots for averaged (over $\alpha$'s) imaginary parts of the invariants, $\Im J_{1,2,3}$, illustrating the strength of CP violation, vs $\tan\beta$ and $M_H^\pm$, for $(M_S,M_A,M_{\eta^\pm},m_\eta)=(75,110,90,100)~\text{GeV}$, and $(M_1,M_2,\mu)=(120,300,200)~\text{GeV}$.} }

In Fig.~\ref{edm-075-090} the corresponding predictions for the electron electric dipole moment are presented, both as an average over the same data set, and maximum (over $\alpha$'s) values, in units of [$e$ $10^{-27}~\text{cm}$], which is also the 1-$\sigma$ bound (see Eq.~(\ref{Eq:edm-bound})). While typical values (left panel) are within 1~$\sigma$, the maxima (right panel) approach the cut-off, which is taken at 2~$\sigma$.

\FIGURE[ht]{
\let\picnaturalsize=N
\def\picsize{15cm}
\ifx\nopictures Y\else{
\let\epsfloaded=Y
\centerline{\hspace{4mm}{\ifx\picnaturalsize N\epsfxsize \picsize\fi
\epsfbox{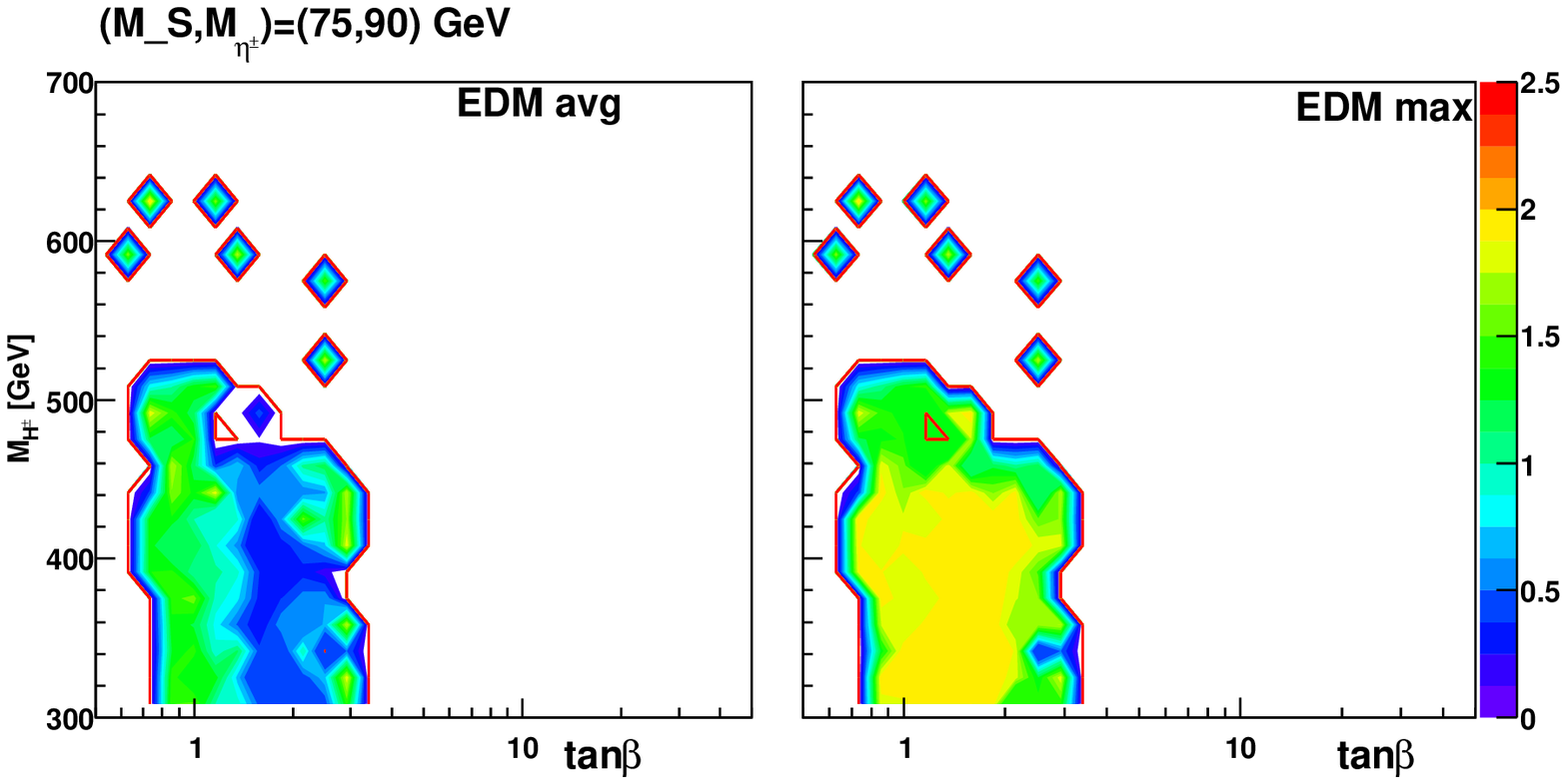}} } 
}\fi
\vspace{-6mm}
\caption{\label{edm-075-090} Contour plots for averaged and maximal (over $\alpha$'s) electron electric dipole moment (in units [$e$ $10^{-27}~\text{cm}$]), vs $\tan\beta$ and $M_H^\pm$, for $(M_S,M_A,M_{\eta^\pm},m_\eta)=(75,110,90,100)~\text{GeV}$, and $(M_1,M_2,\mu)=(120,300,200)~\text{GeV}$.} }

The presence of CP violation is also reflected in the distribution of $\alpha_2$ and $\alpha_3$, two of the three angles which determine the rotation matrix of the neutral Higgs sector, $R$. There are three limits of {\it no} CP violation, all identifiable in this plane \cite{ElKaffas:2007rq}:
\begin{alignat}{2}
&H_1\text{ odd:} &\quad \alpha_2&=\pm\pi/2,  \alpha_3\text{ arbitrary},\nonumber \\
&H_2\text{ odd:} &\quad \alpha_2&=0, \alpha_3=\pm\pi/2, \nonumber \\
&H_3\text{ odd:} &\quad \alpha_2&=0, \alpha_3=0. 
\end{alignat}
For the case $M_S=75~\text{GeV}$, $M_1=120~\text{GeV}$, we show in Fig.~\ref{fig-alphas} the populated parts of this plane. There is a broad distribution of values, with no particular accumulation point. Thus, for the majority of these model points, CP is violated by a non-negligible amount (but still within the limits imposed by the EDM constraint).

\FIGURE[ht]{
\let\picnaturalsize=N
\def\picsize{15cm}
\ifx\nopictures Y\else{
\let\epsfloaded=Y
\centerline{\hspace{4mm}{\ifx\picnaturalsize N\epsfxsize \picsize\fi
\epsfbox{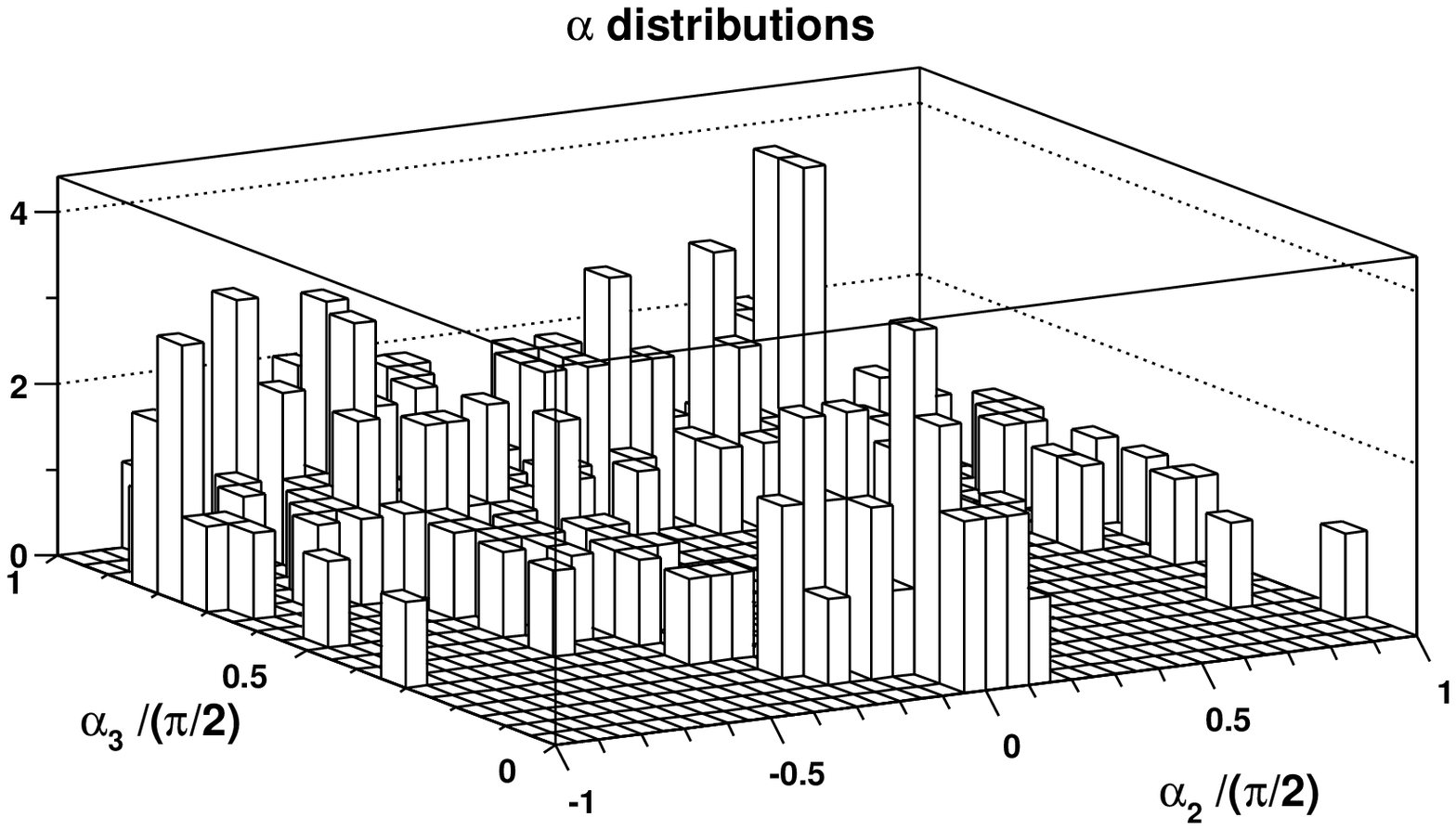}} } 
}\fi
\vspace{-6mm}
\caption{\label{fig-alphas} Populated regions in the ($\alpha_2,\alpha_3$) plane, for $M_S=75~\text{GeV}$, $M_1=120~\text{GeV}$.} }

\section{Direct detection}
\label{Sec:direct-detection}
\setcounter{equation}{0}

The parameter points which give models compatible with particle-physics and DM constraints, also give specific predictions for what signal should be observed in direct-detection experiments. In this regard, we have chosen to compare with the recent CDMS-II \cite{Ahmed:2009zw} and XENON100 \cite{Aprile:2010um} results for spin-independent scattering.

\FIGURE[ht]{
\let\picnaturalsize=N
\def\picsize{10cm}
\ifx\nopictures Y\else{
\let\epsfloaded=Y
\centerline{\hspace{4mm}{\ifx\picnaturalsize N\epsfxsize \picsize\fi
\epsfbox{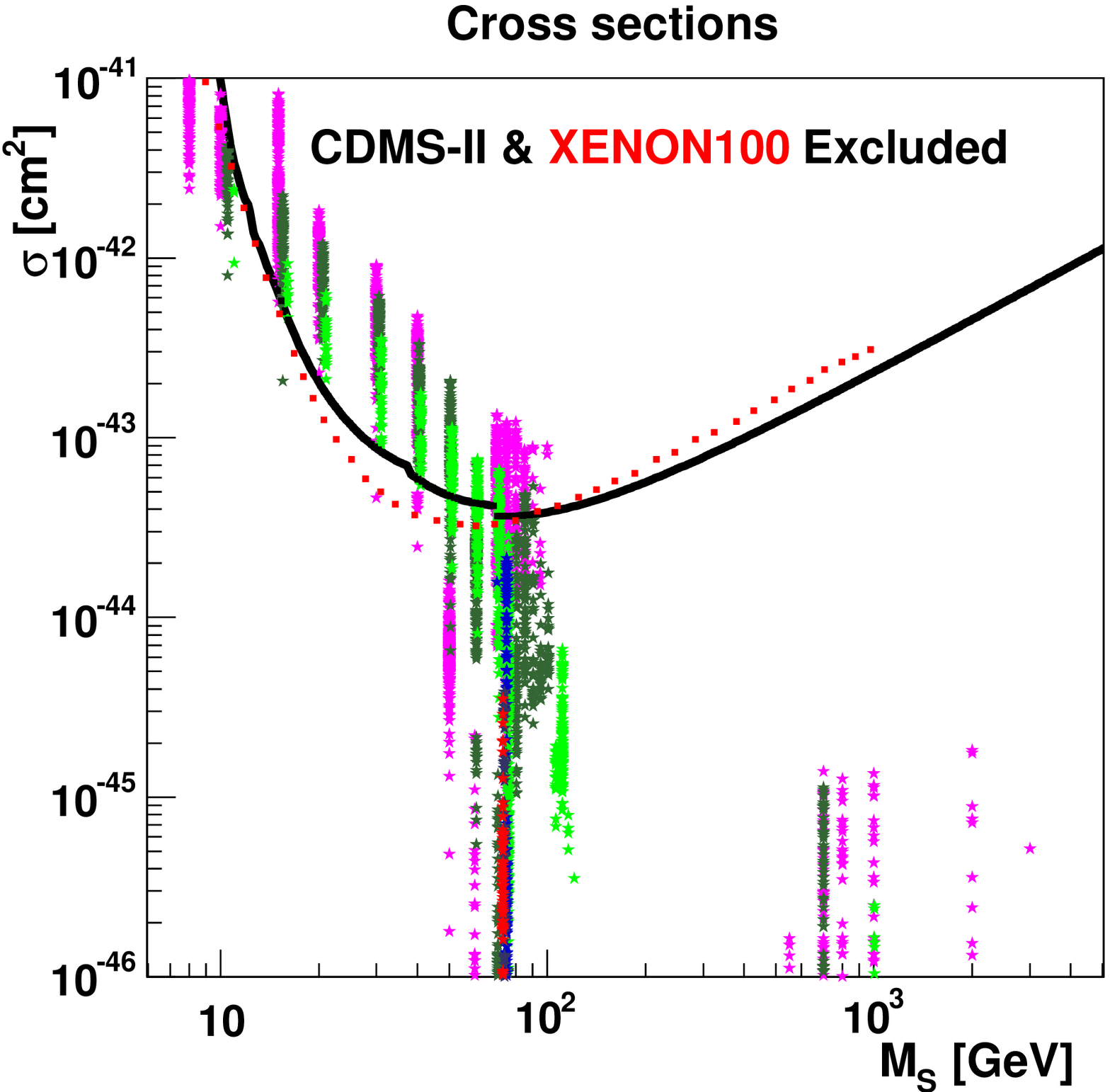}} } 
}\fi
\vspace{-6mm}
\caption{\label{fig:sigma} Direct-detection cross sections compared with the CDMS-II (solid, \cite{Ahmed:2009zw}) and XENON100 (dashed, \cite{Aprile:2010um}) bounds. Magenta: $M_1\leq120~\text{GeV}$, green: $150~\text{GeV}\leq M_1\leq230~\text{GeV}$, blue: $300~\text{GeV}\leq M_1\leq400~\text{GeV}$, red: $M_1\geq500~\text{GeV}$.} }

In Fig.~\ref{fig:sigma} we compare the cross sections for our otherwise acceptable model points with these recent constraints. For each studied value of $M_S$, we show a ``column'' of cross section values corresponding to different values of the other parameters. The cross section falls steeply with increasing mass $M_S$. The different colors refer to the value of $M_1$, as specified in the caption. Values above the curves are excluded by the CDMS-II (solid curve) or the XENON100 (dashed, red) experiments.
We can here make a few observations:
\begin{itemize}
\item
For a given value of $M_S$, the cross section tends to get lower for higher values of $M_1$ (see the color coding).
\item
At low masses, $M_S\lsim10~\text{GeV}$, the model is compatible with the bounds.
\item
In the range $10~\text{GeV}\lsim M_S\lsim50~\text{GeV}$, most model points are excluded.
\item
Around $M_S\sim60-80~\text{GeV}$, much of the parameter space is again compatible with the bounds.
\item
In the ``high'' region, the predicted cross sections are very low, implying that it would be difficult to test (exclude) the model in the near future.
\end{itemize}

We should also stress that during the scanning, in the interest of covering as much of the ``interesting'' parameter space as possible, we did not evaluate the cross section. This is why some points violate the cross section constraint.

\section{LHC prospects}
\label{Sec:LHC}
\setcounter{equation}{0}
At the LHC, one could imagine all the inert-sector scalars being pair-produced,
\begin{equation}
pp\to SSX, AAX, SAX, S\eta^\pm X, A\eta^\pm X, \eta^+\eta^- X,
\end{equation}
followed by the decay of $A$ or $\eta^\pm$ to the lightest one, $S$.

In favorable situations, decays involving $\eta^\pm$ could lead to observable signals.
It was recently pointed out that in a related model \cite{Huitu:2010uc}, the combination of a small mass splitting between the charged scalar and the inert one, together with a small mixing angle, can lead to long-lived charged scalars that give displaced vertices in LHC detectors. In that model, the small mass splitting comes about from the assumption of unification at a high scale. In the model  considered here, the small splitting is required by the
appropriate prediction for the present DM abundance in the case of a heavy $S$ (the DM candidate).
It is therefore of interest to check whether similar experimental signals are expected here as well.
We split this discussion into two cases, according to the mass hierarchies.

\subsection{$M_S<M_{\eta^\pm}<M_A$}

The decay (via a virtual $W$)
\begin{equation} \label{Eq:eta-decay}
\eta^+ \to S \ell^+\nu_\ell
\end{equation}
has several similarities to the familiar muon decay. The main differences are that (i) a scalar-scalar-vector vertex replaces a fermion-fermion-vector vertex, and (ii) one of the invisible final-state particles is massive. For the case of interest, $M_{\eta^\pm}-M_S\ll M_{\eta^\pm}$, the decay rate can be written as
\begin{equation} \label{Eq:Gamma_eta}
\Gamma_{\eta^\pm}=\frac{G_\text{F}^2}{30\pi^3}\left(M_{\eta^\pm}-M_S\right)^5.
\end{equation}

\FIGURE[ht]{
\let\picnaturalsize=N
\def\picsize{7.5cm}
\ifx\nopictures Y\else{
\let\epsfloaded=Y
\centerline{\hspace{4mm}{\ifx\picnaturalsize N\epsfxsize \picsize\fi
\epsfbox{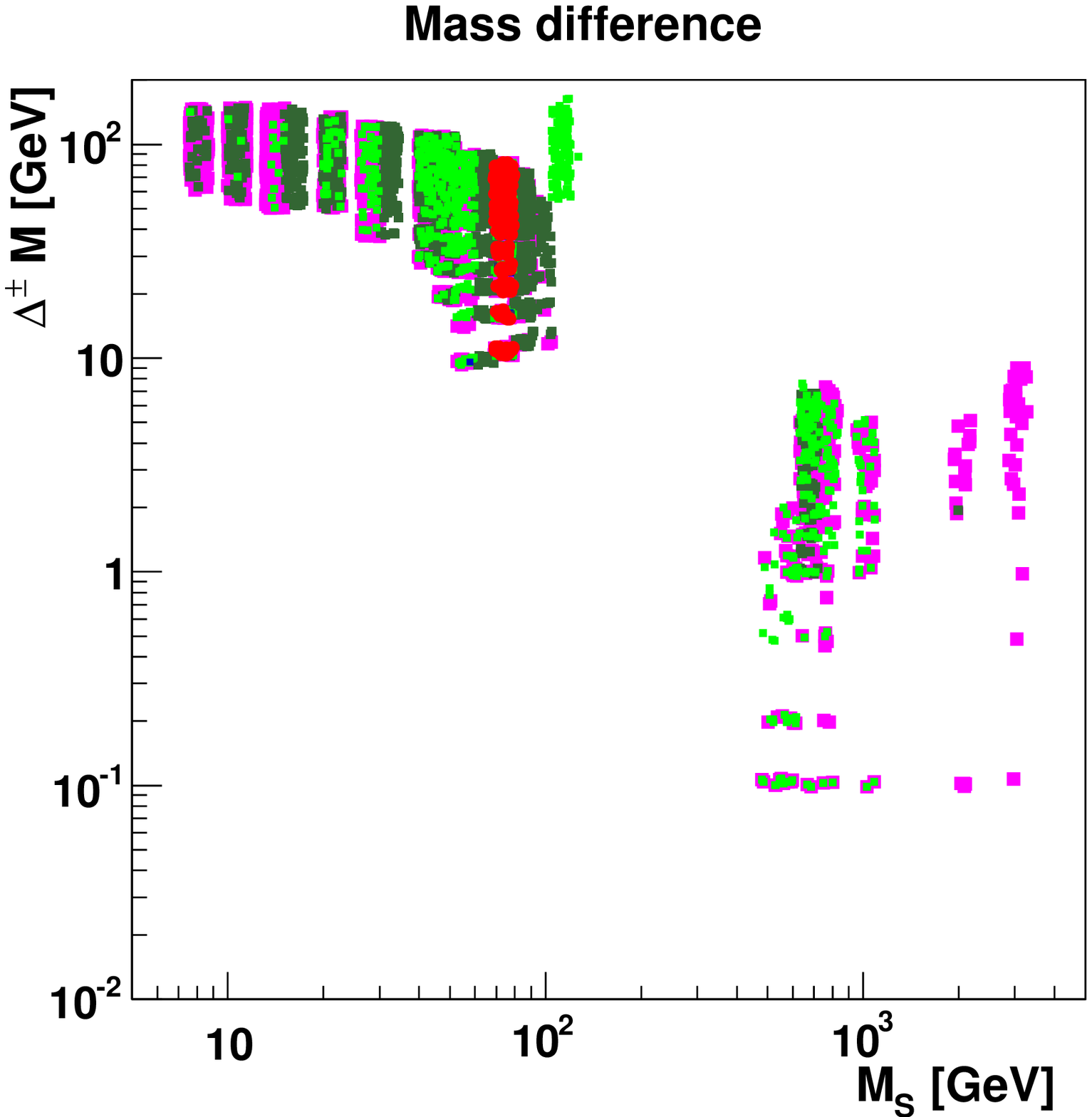}} 
\hspace{4mm}{\ifx\picnaturalsize N\epsfxsize \picsize\fi
\epsfbox{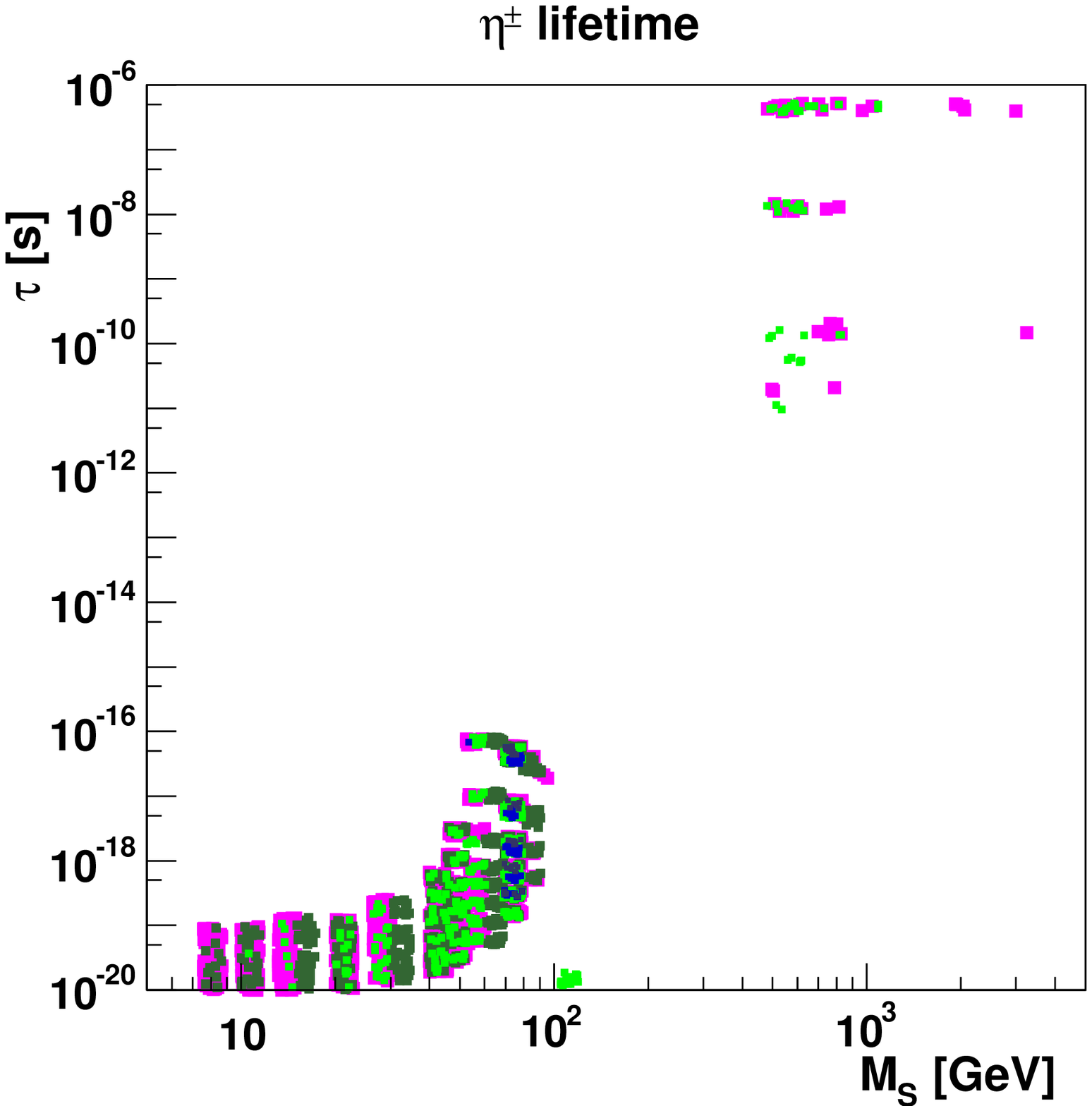}} } 
}\fi
\vspace{-6mm}
\caption{\label{fig:delta_mass} Left: Mass differences, $\Delta^\pm M\equiv M_{\eta^\pm}-M_S$ vs $M_S$.
Right: Corresponding $\eta^\pm$ lifetime.} }

We show in Fig.~\ref{fig:delta_mass} mass differences, $M_{\eta^\pm}-M_S$, for a random subset of the allowed solutions found in Secs.~\ref{sec:med-mass} and \ref{sec:high-mass}. In addition, we show some more degenerate solutions ($\Delta^\pm M\equiv M_{\eta^\pm}-M_S=0.1~\text{GeV}$) for the high-mass region. In the low-medium-mass range, the mass splitting is obviously too large to give any interesting lifetime for $\eta^\pm$, as can also be seen from Figs.~\ref{low-006-008}--\ref{med-080}. Even for $M_S$ above the lowest allowed values of $M_{\eta^\pm}$, the co-annihilation $S\eta^\pm\to W^\pm\gamma$ prevents a mass splitting less than a few GeV. However,
in the high-mass region, the mass splitting can be quite small, and longer lifetimes are possible. 

The experimental signature in this case would be the observation of an $\eta^\pm$ track from the production point up to the decay vertex ($\eta^+ \to S \ell^+\nu_\ell$) followed by a kink corresponding to the decay and a track of the
charged lepton. Although bosons of the inert doublet must be pair produced, such a kink does not
depend on the nature of the accompanying boson being produced; be it $\eta^\mp$, $A$ or $S$,
at least one kink is always there. Of course, in each case some missing energy (through the presence of $SS$ in the final state) will also be helpful. The displacement of the decay vertex depends on the mass splitting and $\eta^\pm$
velocity, for example, for $\Delta^\pm M=0.1~\text{GeV}$, the lifetime is $4.5\times10^{-7}~\text{s}$, and with a velocity of $\beta=0.1$, the decay length would be $L=13.6~\text{m}$. This could easily be measured, and thus, e.g.\ 
for $\beta=0.1$ even a splitting as large as $\Delta^\pm M = {\cal O}(1~\text{GeV})$ could lead to observable effects. 

\subsection{$M_S<M_A<M_{\eta^\pm}$}
In this case, if the mass splitting between $A$ and $S$ is small, $M_A-M_S\ll M_A$, so any
$A$ produced would decay relatively slowly through $A\to S Z^\star \to S f\bar f$ with the width
\begin{equation} \label{Eq:Gamma_A}
\Gamma_{A}=\frac{G_\text{F}^2}{30\pi^3\cos^2\theta_\text{W}}\left(M_A-M_S\right)^5,
\end{equation}
The experimental signature in this case would be a displacement between the production point
and the $A$ decay vertex, where the two tracks of the $l^+l^-$ start. Again $A$ would be accompanied by $A$, $S$ or $\eta^\pm$,
however at least one displaced vertex would be there, regardless what is the nature of the other inert boson.
In each case missing energy will be carried away by two $S$ bosons.
In this case the displacement of the $A$ decay vertex could be determined down to
$0.2$~cm, therefore the observation would be sensitive to values of
$M_A-M_S$ up to $\sim0.5~\text{GeV}$.

\section{Summary}
\label{Sec:summary}
\setcounter{equation}{0}
We have explored in some detail the properties of the ``IDM2'', an extension of the Inert Doublet Model (``IDM'') \cite{Deshpande:1977rw,Barbieri:2006dq} to the case of an additional non-inert doublet \cite{Grzadkowski:2009bt}, allowing for CP violation. As compared with our earlier paper \cite{Grzadkowski:2009bt}, we here implement the full positivity conditions, and run a full implementation of the model in {\tt micrOMEGAs}. Allowed regions in the $m_\eta$--$M_{\eta^\pm}$ and $M_S$--$M_1$ planes have been identified. Since our preliminary study \cite{Grzadkowski:2009bt} was restricted by $m_\eta<M_{\eta^\pm}$, only the lower-right parts of Figs.~\ref{low-006-008}--\ref{med-080} and Figs.~\ref{hi=0550}--\ref{hi=3000} were accessible. We now see that, in most cases, a similar region above that diagonal (i.e., $m_\eta>M_{\eta^\pm}$) is also allowed.
For some ranges of $M_S$, it was found that the little hierarchy problem can be significantly reduced.

In order to illustrate the strength of CP violation present in the model we generalize 
to the three-scalar-doublet case the three CP-sensitive weak-basis-transformation invariants $J_{1,2,3}$ (defined originally for two scalar doublets). It turns out that $\Im J_{1,2,3} \sim 0.5-3$, which is five orders of magnitude more than the corresponding invariant in the Standard Model.

Constraints on the model from direct-detection experiments have also been studied. In the region $10~\text{GeV}\lsim M_S\lsim50~\text{GeV}$, most model points are excluded by direct-detection experiments.

Finally, if inert scalars are produced in vector-boson fusion at the LHC, we discuss possible lepton signatures from the decays of the heavier partners of the DM candidate. It turns out the measurable displacement of their decay vertex could provide an efficient way to test the model, at least for the heavy DM case.

\vspace*{10mm} {\bf Acknowledgements.}  
We are grateful to M.~Tytgat for a discussion of the IDM, to C.~E. Yaguna for discussions of the three-body annihilation channel, to J. Cooley for providing the CDMS-II results. The research
of P.O. has been supported by the Research Council of Norway and by a stay at the Swedish Collegium for Advanced Study. The work of B.G. is supported in part by the Ministry of Science and Higher Education (Poland) as research project N~N202~006334 (2008-11). 

\section*{Appendix~A. Couplings of the inert sector}
\setcounter{equation}{0}
\renewcommand{\thesection}{A}
\label{App-couplings}


\subsection*{Gauge couplings}

The quartic couplings of two gauge fields and two inert fields will
have a ``trivial'' metric tensor factor, $g^{\mu\nu}$, coupling the
two gauge fields. In order to keep a ``light'' notation, we suppress
this factor. Furthermore, we denote the inert fields $S$, $A$ and
$\eta^\pm$. The heavy gauge fields are $W^\pm$ and $Z$, whereas the
photon will be referred to as $\gamma$ (not $A$). In this notation,
the quartic couplings involving two neutral inert fields are:
\begin{subequations}
\begin{alignat}{3}
&SSW^+W^-,&\quad 
&AAW^+W^-:&\quad
&\frac{ig^2}{2}, \\
&SSZZ,&\quad 
&AAZZ:&\quad
&\frac{ig^2}{2\cos^2\theta_W},
\end{alignat}
\end{subequations}
whereas those involving two charged inert fields are:
\begin{subequations}
\begin{alignat}{2}
&\eta^+\eta^-W^+W^-:&\quad 
&\frac{ig^2}{2}, \\
&\eta^+\eta^-ZZ:&\quad
&\frac{ig^2}{2\cos^2\theta_W}\cos^2(2\theta_W), \\
&\eta^+\eta^-\gamma\gamma:&\quad &2ig^2\sin^2\theta_W, \\
&\eta^+\eta^-Z\gamma:&\quad &ig^2\tan\theta_W\cos(2\theta_W),
\end{alignat}
\end{subequations}
(where for uniformity of notation we have not substituted $e=g\sin\theta_W$)
and finally the mixed ones are
\begin{subequations}
\begin{alignat}{2}
&S\eta^\pm ZW^\mp: &\quad &\frac{-ig^2\sin^2\theta_W}{2\cos\theta_W}, \\
&A\eta^\pm ZW^\mp: &\quad &\frac{\mp g^2\sin^2\theta_W}{2\cos\theta_W}, \\
&S\eta^\pm \gamma W^\mp: &\quad &\frac{ig^2}{2}\sin\theta_W, \\
&A\eta^\pm \gamma W^\mp: &\quad &\pm\frac{g^2}{2}\sin\theta_W.
\end{alignat}
\end{subequations}

There are no trilinear gauge couplings involving two identical inert
fields.  The non-vanishing trilinear gauge couplings (two inert fields
and one gauge field) are:
\begin{subequations}
\begin{align}
SAZ: &\quad \frac{-g}{2\cos\theta_W}(p_A-p_S), \\
S\eta^\pm W^\mp: &\quad \frac{\pm ig}{2}(p_S-p_\pm), \\
A\eta^\pm W^\mp: &\quad \frac{g}{2}(p_A-p^\pm), \\
\eta^+\eta^-Z: &\quad \frac{-ig}{2\cos\theta_W}\cos(2\theta_W)(p^+-p^-), \\
\eta^+\eta^-\gamma: &\quad -ig\sin\theta_W(p^+-p^-), 
\end{align}
\end{subequations}
where all momenta $p_S$, $p_A$ and $p^\pm$ are incoming, and carry the
Lorentz index of the vector field, in an obvious notation.
\subsection*{Scalar couplings}

The scalar couplings can be expressed in a compact manner if we
introduce the following notation:
\begin{subequations} \label{Eq:lambda_L-app}
\begin{align} 
\lambda_L&\equiv \half(\lambda_a+\lambda_b+\lambda_c)
=\frac{M_S^2-m_\eta^2}{v^2}, \\
\tilde \lambda_L&\equiv \half(\lambda_a+\lambda_b-\lambda_c)
=\frac{M_A^2-m_\eta^2}{v^2},
\end{align}
\end{subequations}
and
\begin{subequations}
\begin{align}
F_{j}&=\cos\beta R_{j1}+\sin\beta R_{j2}, \\
\tilde F_{j}&=\cos\beta R_{j2}-\sin\beta R_{j1}.
\end{align}
\end{subequations}
The latter quantities satisfy $|F_{j}|\leq1$, and $|\tilde F_{j}|\leq1$
since $R$ is unitary. In particular,
$F_{1}=\cos(\beta-\alpha_1)\cos\alpha_2$. 

\paragraph{Trilinear couplings.}
The trilinear scalar couplings are:
\begin{subequations}
\label{Eq:trilinear}
\begin{alignat}{2}
&SSH_j: &\quad &-2i\lambda_LvF_{j},  \\
&AAH_j: &\quad &-2i\tilde\lambda_L v F_j, \\
&SAH_j: &\quad &0, \\
&S\eta^\pm H^\mp: &\quad &0, \\
&A\eta^\pm H^\mp: &\quad &0, \\
&\eta^+\eta^-H_j: &\quad &-i\lambda_a vF_j, \\
&SSG^0: &\quad &0, \\
&AAG^0: &\quad &0, \\
&SAG^0: &\quad &-iv\lambda_c, \\
&S\eta^{\mp}G^{\pm}: &\quad &-\frac{i}{2}v(\lambda_b+\lambda_c), \\
&A\eta^{\mp}G^{\pm}: &\quad &\pm\frac{1}{2}v(\lambda_b-\lambda_c).
\end{alignat}
\end{subequations}
From Eqs.~(\ref{Eq:lambda_L-app}) we note that the splittings
$M_S^2-m_\eta^2$ and $M_A^2-m_\eta^2$ control the strengths of
important trilinear couplings of two inert neutral fields to a Higgs
field.  Likewise, $M_{\eta^\pm}^2-m_\eta^2$ controls the strength of
the inert charged fields to a Higgs field. If the dark democracy is
lifted, all these couplings (\ref{Eq:trilinear}) would be non-zero.

\paragraph{Quadrilinear couplings.}
The quadrilinear ones involving two neutral inert fields are
\begin{subequations}
\begin{alignat}{2}
\label{Eq:quadrilinear-neutr}
&SSH_jH_j:&\quad &-2i(\lambda_L-\lambda_c R_{j3}^2), \\
&SSH_jH_k:&\quad &2i\lambda_c R_{j3}R_{k3}, \quad j\ne k, \\
&SSH^+H^-:&\quad &-i\lambda_a, \\
&AAH_jH_j:&\quad &-2i(\tilde\lambda_L+\lambda_c R_{j3}^2),\\
&AAH_jH_k:&\quad &-2i\lambda_c R_{j3}R_{k3},  \quad j\ne k, \\
&AAH^+H^-:&\quad &-i\lambda_a, \\
&SAH_jH_j:&\quad &-2i\lambda_c R_{j3}\tilde F_j, \\
&SAH_jH_k:&\quad &-i\lambda_c 
(\tilde F_j R_{k3}+R_{j3}\tilde F_k),  \quad j\ne k, \\
&SSG^0G^0:&\quad &-2i\tilde{\lambda}_L, \\
&SSG^+G^-:&\quad &-i\lambda_a, \\
&SSH^{\pm}G^{\mp}:&\quad &0, \\
&SSG^0H_j:&\quad &0, \\
&AAG^0G^0:&\quad &-2i\lambda_L, \\
&AAG^+G^-:&\quad &-i\lambda_a, \\
&AAH^{\pm}G^{\mp}:&\quad &0, \\
&AAG^0H_j:&\quad &0, \\
&SAG^0G^0:&\quad &0, \\
&SAG^0H_j:&\quad &-i\lambda_cF_j, \\
\end{alignat}
\end{subequations}
and those involving two charged inert fields are
\begin{subequations}
\label{Eq:quadrilinear-ch}
\begin{alignat}{2}
&\eta^\pm\eta^\pm H^\mp H^\mp:&\quad &-2i\lambda_c, \\
&\eta^+\eta^-H_jH_j:&\quad &-i\lambda_a, \\
&\eta^+\eta^-H_jH_k:&\quad &0,  \quad j\ne k,
\label{Eq:quartic-eta-eta-j-k}\\
&\eta^+\eta^-H^+H^-:&\quad &-i(\lambda_a+\lambda_b), \\
&\eta^{\pm}\eta^{\pm}G^{\mp}G^{\mp}:&\quad &-2i\lambda_c, \\
&\eta^{\pm}\eta^{\pm}H^{\mp}G^{\mp}:&\quad &0, \\
&\eta^+\eta^-G^0G^0:&\quad &-i\lambda_a, \\
&\eta^+\eta^-G^0H_j:&\quad &0, \\
&\eta^+\eta^-G^+G^-:&\quad &-i(\lambda_a+\lambda_b), \\
&\eta^+\eta^-H^{\pm}G^{\mp}:&\quad &0,
\end{alignat}
\end{subequations}
and those involving one neutral and one charged charged inert field are
\begin{subequations}
\label{Eq:quadrilinear-neutr-ch}
\begin{alignat}{2}
&S\eta^\pm H^\mp H_j:&\quad 
&\frac{-i}{2}(\lambda_b+\lambda_c)\tilde F_j\pm\frac{1}{2}(\lambda_b-\lambda_c)R_{j3}, \\
&A\eta^\pm H^\mp H_j:&\quad 
&\frac{\mp 1}{2}(\lambda_b-\lambda_c)\tilde F_j-\frac{i}{2}(\lambda_b+\lambda_c)R_{j3}, \\
&S\eta^{\pm}G^{\mp}G^0:&\quad &\pm\frac{1}{2}(\lambda_b-\lambda_c), \\
&S\eta^{\pm}H^{\mp}G^0:&\quad &0, \\
&S\eta^{\pm}G^{\mp}H_j:&\quad &-\frac{i}{2}(\lambda_b+\lambda_c)F_j, \\
&A\eta^{\pm}G^{\mp}G^0:&\quad &-\frac{i}{2}(\lambda_b+\lambda_c), \\
&A\eta^{\pm}H^{\mp}G^0:&\quad &0, \\
&A\eta^{\pm}G^{\mp}H_j:&\quad &\mp\frac{1}{2}(\lambda_b-\lambda_c)F_j.
\end{alignat}
\end{subequations}
In Eq.~(\ref{Eq:trilinear})--(\ref{Eq:quadrilinear-neutr-ch}), the
couplings listed as zero, would be non-zero if the dark democracy is
lifted. On the other hand, couplings not listed, are absent also when
the dark democracy is lifted.

Couplings involving four fields, all from the inert doublet, are
\begin{subequations}
\begin{alignat}{2}
&SSSS:&\quad 
&-3i\lambda_\eta, \\
&AAAA:&\quad 
&-3i\lambda_\eta, \\
&SSAA:&\quad 
&-i\lambda_\eta, \\
&SS\eta^+\eta^-:&\quad 
&-i\lambda_\eta, \\
&AA\eta^+\eta^-:&\quad 
&-i\lambda_\eta, \\
&\eta^+\eta^+\eta^-\eta^-:&\quad 
&-2i\lambda_\eta.
\end{alignat}
\end{subequations}

\section*{Appendix~B. Invariants \boldmath{$\Im J_{1,2,3}$}}
\setcounter{equation}{0}
\renewcommand{\thesection}{B}
\label{App-invariants}

The tensors $Z_{a\bar{b}c\bar{d}}$ adopted in (\ref{comp_pot})
can be expressed through the standard parameters used in (\ref{Eq:fullpot})
as follows:
\begin{alignat}{3}
Z_{1\bar{1}1\bar{1}}&=\lambda_1, &\quad
Z_{2\bar{2}2\bar{2}}&=\lambda_2, &\quad
Z_{3\bar{3}3\bar{3}}&=\lambda_\eta, \nonumber\\
Z_{1\bar{1}2\bar{2}}=Z_{2\bar{2}1\bar{1}}&=\lambda_3, &\quad
Z_{1\bar{2}2\bar{1}}=Z_{2\bar{1}1\bar{2}}&=\lambda_4, &\quad
Z_{1\bar{1}3\bar{3}}=Z_{3\bar{3}1\bar{1}}&=\lambda_{1133}, \nonumber\\
Z_{2\bar{2}3\bar{3}}=Z_{3\bar{3}2\bar{2}}&=\lambda_{2233}, &\quad
Z_{1\bar{3}3\bar{1}}=Z_{3\bar{1}1\bar{3}}&=\lambda_{1331}, &\quad
Z_{2\bar{3}3\bar{2}}=Z_{3\bar{2}2\bar{3}}&=\lambda_{2332}, \nonumber\\
Z_{1\bar{2}1\bar{2}}&=\lambda_5, &\quad
Z_{2\bar{1}2\bar{1}}&=\lambda_5^*, &\quad
Z_{1\bar{3}1\bar{3}}&=\lambda_{1313}, \nonumber\\
Z_{3\bar{1}3\bar{1}}&=\lambda_{1313}^*, &\quad
Z_{2\bar{3}2\bar{3}}&=\lambda_{2323}, &\quad
Z_{3\bar{2}3\bar{2}}&=\lambda_{2323}^*.
\label{param}
\end{alignat}
We also have
\begin{equation}
\hat{v}_1=v_1/v,\quad
\hat{v}_2=v_2/v,\quad
\hat{v}_3=v_3/v.
\end{equation}
The invariants $\Im J_{1,2,3}$ expressed in the basis adopted here then read:
\begin{align}
\Im J_1&=-\frac{v_1^2}{v^4}(\lambda_1-\lambda_2+\lambda_{1331}-\lambda_{2332})(v_2^2\Im \lambda_5+v_3^2\Im\lambda_{1313}),
\label{invJ1}\\
\Im J_2&=-\frac{v_1^2}{v^8}(v_2^2\Im \lambda_5+v_3^2\Im\lambda_{1313})\nonumber\\
&\times
\biggl[\left((\lambda_1-\lambda_3-\lambda_4)^2-|\lambda_5|^2\right) v_1^4
+2(\lambda_1-\lambda_2) \Re \lambda_5 v_1^2v_2^2
-\left((\lambda_2-\lambda_3-\lambda_4)^2-|\lambda_5|^2\right) v_2^4\nonumber\\
&+2v_1^2v_3^2\left[(\lambda_{1133}+\lambda_{1331}-\lambda_{2233}-\lambda_{2332})(\lambda_{1133}+\lambda_{1331}-\lambda_3-\lambda_4+\Re\lambda_{1313})\right.\nonumber\\
&\hspace*{1.5cm}\left.-\Re\lambda_{1313}(\lambda_3+\lambda_4-\lambda_1)-\Re\lambda_{2323}\Re\lambda_5\right]\nonumber\\
&+2v_2^2v_3^2\left[(\lambda_{1133}+\lambda_{1331}-\lambda_{2233}-\lambda_{2332})(\lambda_{2233}+\lambda_{2332}-\lambda_3-\lambda_4+\Re\lambda_{2323})\right.\nonumber\\
&\hspace*{1.5cm}\left.+\Re\lambda_{2323}(\lambda_3+\lambda_4-\lambda_2)+\Re\lambda_{1313}\Re\lambda_5+\Im\lambda_{1313}\Im\lambda_5\right]\nonumber\\
&+v_3^4\left[|\lambda_{1313}|^2-(\lambda_{1133}+\lambda_{1331})^2\right.\nonumber\\
&\hspace*{1.5cm}\left.+(\lambda_{2233}+\lambda_{2332}-\Re\lambda_{2323})(\lambda_{2233}+\lambda_{2332}+\Re\lambda_{2323})\right.\nonumber\\
&\hspace*{1.5cm}\left.+2(\lambda_{1133}+\lambda_{1331}-\lambda_{2233}-\lambda_{2332})\lambda_\eta\right]\nonumber\\
&-\frac{2v_1^4v_3^2}{v_2^2}\Im\lambda_{1313}\Im\lambda_5-\frac{v_1^4v_3^4}{v_2^4}(\Im\lambda_{1313})^2\biggr],
\label{invJ2}\\
\Im J_3&=\frac{v_1^2}{v^4}(\lambda_1-\lambda_2+\lambda_{1331}-\lambda_{2332})
(\lambda_1+\lambda_2+2\lambda_4+\lambda_{1331}+\lambda_{2332})(v_2^2\Im \lambda_5+v_3^2\Im\lambda_{1313}).
\label{invJ3}
\end{align}
where, in order to retain generality we kept $v_3/\sqrt{2}\equiv \langle \eta \rangle \neq 0$, and made use of the relation
\begin{equation}
v_1^2\Im\lambda_{1313}+v_2^2\Im\lambda_{2323}=0,
\end{equation}
which emerges from the minimization conditions for $v_3\neq0$.

In the dark democracy case, these results simplify to those given in Eqs.~(\ref{Eq:ImJ_1})--(\ref{Eq:ImJ_3}).



\end {document}